\documentclass[11pt,a4paper]{article}
\usepackage[T1]{fontenc}
\usepackage[utf8]{inputenc} % use utf8 to display umlaute etc.
\usepackage[english]{babel} % set document language: english

\usepackage{mathtools} % math package (includes amsmath)
\usepackage{amssymb} % mathematical symbols (technically not imprtant)
\usepackage[version=4]{mhchem} % display chemical reaction equations
\usepackage{amsthm} %For theorem s and proofs
\usepackage{physics}

\usepackage{float} % environment for figures and tables
\usepackage{subcaption} % include subfloat-environments
\usepackage{graphicx} % include pictures etc
\usepackage[hidelinks]{hyperref} % include labels and references

\usepackage[usenames,dvipsnames]{xcolor} % change textcolors
\usepackage{geometry} % change page geometry
\usepackage{setspace} % cange linespacing

\usepackage{authblk} % add affiliations to authors
\usepackage{lipsum} % include lipsum text
\usepackage[textsize=tiny]{todonotes} % add todonotes to text
\usepackage{tikz-cd}
\usepackage{nicematrix}
\usepackage{tikz}
\usetikzlibrary{arrows.meta, positioning}
\usepackage[final,defaultcolor=red,commandnameprefix=ifneeded]{changes}
\usepackage{verbatim}
\title{Thermodynamic ranking of pathways in reaction networks}
\author[1]{Praful Gagrani}
\author[2]{Nino Lauber}
\author[3,4,5]{Eric Smith}
\author[2]{Christoph Flamm}
\date{\today}
\affil[1]{\small Institute of Industrial Science, The University of Tokyo, Tokyo, Japan}
\affil[2]{\small Institute for Theoretical Chemistry, University of Vienna, Vienna, Austria}
\affil[3]{\small Department of Biology, Georgia Institute of Technology, Atlanta, GA, USA }
\affil[4]{\small Earth-Life Science Institute, Tokyo Institute of Technology, Tokyo, Japan}
\affil[5]{\small  Santa Fe Institute, Santa Fe, NM, USA}
\newcommand\mbf[1]{\mathbf #1}
\newcommand\mbb[1]{\mathbb #1}
\newcommand\mcl[1]{\mathcal #1}
\newcommand\mfk[1]{\mathfrak #1}

\newcommand{\PG}[1]{{\color{teal}#1}}

\newcommand{\NL}[1]{{\color{Fuchsia}#1}}

\newcommand{\bbcal}[3]{\mbb{#1}(\mcl{#2}#3)}

\def\St{\mathbb{S}}

\def\vext{v_\text{ext}}
\def\supp{{\mathrm{supp}}}
\def\eq{\underline{q}}
\def\e{\mathbf{e}}

\def\fracqeq{\left(\frac{q}{\eq}\right)}
\def\fracqeqme{\left(\frac{q}{\eq}-\e\right)}

\begin{document}

    %----------------------------------------------------------------------------------
    \maketitle
    %----------------------------------------------------------------------------------

    %----------------------------------------------------------------------------------
    \begin{abstract}

   \added{ One of the puzzles left open by energetic analyses of
  irreversible stochastic processes is that boundary conditions that
  prevent the performance of work or the dissipation of heat make no
  contribution to an entropy-production budget; yet we see
  ubiquitously in both engineered and living systems that both
  transient and persistent energy costs are paid to create and
  maintain such boundaries.  We wish to know whether there are
  inherent limits for the costs of such phenomena, and common units in
  which those can be traded off against more familiar costs measured
  in terms of entropy production and heat dissipation.  We give this
  problem a concrete framing in the context of Chemical Reaction
  Networks (CRNs), for the problem of extracting a topologically
  restricted pathway from a larger distributed network, through the
  activation of some reactions and the selective elimination of
  others.}  We define a thermodynamic cost function for pathways
derived from the large-deviation theory of stochastic CRNs, which
decomposes into two components: an ongoing maintenance cost to sustain
a non-equilibrium steady state (NESS), and a restriction cost,
quantifying the ongoing improbability of neutralizing reactions
outside the specified pathway.  Applying this formalism to
detailed-balanced CRNs in the linear response regime, we make use of
their formal equivalence to electrical circuits.  We prove that the
resistance of a CRN decreases as reactions are added that support the
throughput current, and that the maintenance cost, the restriction
cost, and the thermodynamic cost of nested pathways are bounded below
by those of their hosting network.  For four- and five-species example
CRNs, we show how catalytic and inhibitory mechanisms can drastically
alter pathway costs, enabling the unfavorable pathways to become
favorable and to approach the cost of the hosting pathway.  Our
results provide insights into the thermodynamic principles governing
open CRNs and offer a foundation for understanding the evolution of
metabolic networks.

    \end{abstract}
    %----------------------------------------------------------------------------------

    %----------------------------------------------------------------------------------
    \tableofcontents

    \section{Introduction}
    \label{sec:intro}

Both biological and engineered systems employ selected or assembled
designs and control loops in hierarchies, whereby a design choice or
controller at a higher level dictates or modulates the operating
conditions of processes at lower levels.
The advantages conferred by these modulations can have many sources.
Some are evident and offer easy engineering analyses: if a catalyst
reduces the transition barrier to a metabolic reaction driven away
from equilibrium, some chemical work can be saved from dissipation in
performing the reaction, and used to drive other actions needed by a
cell or organism (including costs to make, maintain, and replace
copies of the catalyst), while the reaction rate itself may also be
increased.

Other designs or feedbacks serve the function of \emph{preventing
things from happening} that otherwise would.  It is well
known~\cite{Khersonsky:promiscuity:10,Khersonsky:enzyme_families:11}
that enzyme families are often organized around a highly conserved
reaction center, which builds the transition state for a given
reaction mechanism.  The residues making up that center are under
intense stabilizing selection because the transition state requires
very precise positioning of specific atomic centers.  Diversification
in the family occurs concentrically outside the reaction center, and
it is there that substrate-specificity evolves, excluding unwanted
activities or partitioning (through sub-functionalization) a class of
related reactions so that each may be independently
regulated~\cite{Zhu:CBB_control:24}.  For these designs whose function
is \emph{restriction}, the straightforwardly-evident advantages may be
indirect and apparently contingent on systemic features: protection of
organic matter from conversion by side-reactions to unusable
forms~\cite{Orgel:cycles:08},
toxicity~\cite{Gudelj:coli_syntrophy:16}, or (yet more indirect)
simplification of some downstream control problem for the network.

However they come to exist, all of the designs or control loops that
do not arise spontaneously, and therefore require selection or active
synthesis, incur costs.  These include costs to implement designs and
costs to operate the resulting processes.  An important class of
questions that can be raised for processes of selection and assembly
in hierarchical systems, which both incur costs and confer benefits,
is whether the costs and benefits can be quantified in any common
denomination, how both can be ranked among alternative choices, and
whether the tradeoff of costs and benefits is subject to fundamental
limits.
\added{
A first-principles theory of such limits would enable comparisons
among subsystems (\textit{e.g.}~the physiology catalyzed by specific
enzymes and the biochemistry and bioenergetics synthesizing them) or
across scales (the physiology catalyzed versus the overgrowth to
compensate for selective deaths to incorporate the information
required for specificity or to maintain it against mutational
degradation~\cite{Kimura:sel_info:61,Iwasa:free_fitness:88,Mustonen:fitness_flux:10}),
even absent the detailed knowledge of how these are structurally
connected that would be needed for an \textit{ad hoc} engineering
analysis.
}

\subsubsection*{\added{An illuminating instance of a more
    general question about optimal frontiers in non-equilibrium
    systems}} 

\added{
The paradigms for impossibility theorems and limiting horizons, which
may be applied no matter what our states of incomplete knowledge, are
the first and second laws of equilibrium
thermodynamics~\cite{Fermi:TD:56}.  Energy is a constraining value
determining what states may be jointly occupied in a multi-component
system, and its conservation (the first law) therefore limits all
possible distributional entropies no matter what the dynamical state .
Then, the absence of spontaneous mean increase of equilibrium entropy
(the second law) implies that no engine cycle is more efficient than a
Carnot cycle, one of a few fundamental impossibility theorems in
physics (because its impossibility comes from the negation of the
constructive result of conservation of entropy flux in reversible
transformations~\cite{Smith:gaps:25}).
}

\added{
Conserved quantities, including volume and particle number, but
especially energy because of its generality across types of
excitations, are privileged extensive state variables for equilibrium
thermodynamics, because their conservation follows from very general
properties of matter and geometry such as symmetries.
For driven non-equilibrium systems, fewer of the relations that may
provide encapsulating interfaces between subsystems or across scales
are accounted for by dynamically conserved quantities, and we find
ourselves asking what other relations might give rise to common
measures of costs and benefits in multi-scale systems.
}

A long-studied example of
\added{
a cost derived inherently from the operation of a non-equilibrium
system, which illustrates both the uses and the limitations of energy
denomination,} is the problem of optimizing engine efficiency at
maximum
power~\cite{Odum:max_power:55,vanDenBroeck:eff_max_power:05,Esposito:copol_eff:10,Allahverdyan:engines:13}.
Here the cost is the excess heat dissipation per work performed
relative to the ideal (zero) dissipation at Carnot efficiency.
\added{
Its non-equilibrium aspect is the way energy flow is partitioned
between work and dissipation, which depends on model class but can
still be quite general.
Its use as a bound follows from essentially equilibrium constraints,
which are tight only to the extent that potential work unavailable to
other processes equals the energy lost to dissipation.  
The foregoing analysis omits \emph{by construction} any indication why
there should be costs to aspects of system operation not measured in
realized event rates, and thus any way to assign inherent cost-values
to the boundary conditions that make a system an engine.
}

In this paper we will construct what we argue to be natural measures
of cost for the operation of a process,
\added{
acknowledging the wider range of system interfaces that can partition
non-equilibrium phenomena.
A benchmark for our approach will be the assignment of costs both to
events that are performed and those that are prevented, from common
first principles and in common units.
In contrast to the example of engine efficiencies, our costs are not
predicated upon energy conservation and thus are not limited to heat
dissipation rates and not naturally denominated in energies.
We will be able to recover familiar cost measures such as the entropy
production rate in heat units~\cite{Peliti:ST_intro:21} when energy
conservation is the relevant system-partitioning constraint, while
obtaining other costs for hierarchical \emph{control relations} such
as inhibition when those are the binding constraints.
}

\subsubsection*{\added{The Chemical Reaction Networks as a
    study class, combining concreteness and generalizability}}

We study the problem in the model class of stochastic stoichiometric
population processes~\cite{Smith:three_level:24}, familiar from their
most widespread use as models of chemical reaction networks (CRNs).
CRNs, \added{well-known and} widely developed for
chemical~\cite{Horn:mass_action:72,Feinberg:notes:79,Feinberg:def_01:87}
and
metabolic~\cite{schilling1998underlying,Palsson:systems_bio:06,BarEven12}
modeling and engineering,
\added{have many other applications, including} ecological
interaction and
succession~\cite{Sterner:eco_stoich:02,peng2020ecological},
epidemiology~\cite{avram2024advancing} and population
biology~\cite{Smith:evo_games:15}, and cellular
translation~\cite{cuevas2023modular} and realization of genomic
lifecycles~\cite{Smith:BFI_Lifecycles:23,Smith:BFII_Information:24}.
\added{
The availability of a uniform model abstraction for so many modes of
biological organization makes them particularly well-suited for
framing the cross-scale cost comparisons that concern us.
}

\added{Because we are particularly interested to compare
  costs associated with non-equilibrium driving and with restriction
  by specificity,} we will consider \textit{open}
CRNs~\cite{Polettini:open_CNs_I:14,rao2016nonequilibrium,Wachtel:transduction:22},
which exchange their species as well as heats with an environment.
\added{
Classes of such graphs, which differ in their internal reaction
redundancy, will be capable of carrying out a fixed chemical
conversion between the boundary species, which then defines the
separating interface for cost comparisons.
}
The theory of open networks driven by time-varying chemostatted
concentrations was developed in~\cite{rao2016nonequilibrium}.  In this
work, we focus on open CRNs where certain species flow in and out of
the system at fixed rates.  A category-theoretic formulation of such
networks is presented in \cite{baez2017compositional}, and analytical
tools developed to enumerate all pathways that facilitate exchange
between a CRN and its environment can be found in
\cite{anderesen:2016,andersen:2019}.

The abstraction we use to study alternative processes interacting with
the same boundary conditions is that of \textit{pathways}, roughly as
the term is understood in metabolism.  A pathway can be any collection
of reactions in a CRN capable of performing a given conversion between
inputs and outputs.  In realistic systems modeled with CRNs, there are
generally multiple alternative pathways.  Much is known descriptively
about pathways in metabolism, but many questions about their
possibilities and evolution remain: What determines the thermodynamic
favorability of one pathway over another?  Why has evolution favored
specific pathways catalyzed by particular enzymes?
\added{When pathways are projected down from large
  combinatorial networks by catalytic specificity, by how much does
  their topological restriction increase the chemical potential drop
  required to drive them, and how does the resulting cost in rate of
  work delivery relate to other costs of specificity?}  The cost
definition and examples presented in this paper are chosen to explore
these questions.

\subsubsection*{\added{A bridge from heat energy to
    system-state improbability: the fluctuation theorems of stochastic
    thermodynamics}}

Stochastic CRNs are naturally modeled as Markov jump processes on
directed multi-hypergraphs.  The large deviation theory for jump
processes has been extensively developed~\cite{barato2015formal,
  touchette2009large, peliti2021stochastic}, and is applied
specifically to CRNs in~\cite{lazarescu2019large}.  A variety of cost
measures arise within this framework (in large-deviation approximation
and more generally as exact results) from the constructions known as
\textit{fluctuation
  theorems}~\cite{Harris:fluct_thms:07,Esposito:fluct_theorems:10,Seifert:stoch_thermo_rev:12,Peliti:ST_intro:21},
which extend notions of minimum cost for a transformation in terms of
work~\cite{Jarzynski:neq_FE_diffs:97,Crooks:NE_work_relns:99} from
their familiar forms in equilibrium thermodynamics to a limited class
of non-equilibrium situations.  These costs, denominated in energy
units, carry the interpretations of the minimum work required to
prepare a system in an improbable fluctuation state, and thus
conversely, the maximum work extractable from its relaxation back to
equilibrium~\cite{kolchinsky2021work}.  A similar result (for certain
idealized computational rather than chemical process descriptions) is
Landauer’s principle~\cite{plenio2001physics},
\[ \frac{E}{k_B T} \geq \log 2, \]
stating that the energy cost to irreversibly erase a one-bit variable,
normalized by the thermal scale $k_B T$, is bounded below by the
information content of the bit.

The common assumption underlying all these constructions is the
separation-of-timescales idealization leading to
\added{Prigogine's \textit{local equilibrium
    approximation (LEA)}~\cite{Nicolis:fluct_NEQ:71,Glansdorff:structure:71,Prigogine:MT:98}
  for thermal baths and chemical reservoirs between reactions, and the
  coupling between these known as the \textit{local detailed balance}
  (LDB) approximation~\cite{rao2016nonequilibrium},} enabling the
interpretation of chemical affinities in terms of heat entropies of
fast-relaxing thermal baths.
\added{
The dissipative costs following from the LEA are defined in terms of
what are fundamentally equilibrium state variables, including their
dependence on conserved quantities such as energy, volume, or particle
numbers.}\footnote{
\added{Oono~\cite{Oono:nonlin_world:13} has aptly termed
these collections of equilibrated local distributions, cordoned off
from each other by physical partitions -- but equally well by
transition-state barriers during almost-all the time when reactions
are not occurring -- \textit{compartmented quasi-equilibria}, to
distinguish the order their partitions carry, from order inherently
carried in boundary-layer phenomena such as pattern-forming
reaction-diffusion fronts, the archetypes of ``dissipative
structures''\cite{Nicolis:fluct_NEQ:71,Prigogine:MT:98}.}}

\subsubsection*{\added{An approach to cost measures
    denominated directly in path entropies and entropy rates}}

\added{
To generalize to system-partitioning relations that can include but
extend beyond conservation of energy or particle number, we will
denominate costs directly in terms of path entropies for ensembles
conditioned on the parameters that define the separating boundaries.
Our approach can be understood as a version of the method of Maximum
Caliber
(MaxCal)~\cite{Jaynes:MEPP:80,Presse:max_cal:13,Yang:NEQ_net_flows:25}.
Extending one of the early relations from MaxCal, that the
unlikelihood of an unlikely state is equal at leading order to the
cumulative unlikelihood along the least-unlikely path to reach that
state, our elementary cost measures will be path-entropy rates with
the interpretation of likelihood differences for patterns of
\emph{event sequences}; where state-likelihoods arise, they will be
derived from these more basic distributions over events.  
}

\added{
To understand the construction that follows, it is important in the
presentation to distinguish the cost that is inherent in
\emph{operating} in a certain way, both from the cost of assembling
machinery that may dictate those operating parameters as a boundary
condition, and from whatever benefit elsewhere in the system
\emph{pays that cost} through the enhancement of some other operation.
Boundary conditions, in this accounting, \emph{supply} the
unlikelihood costs to realize event sequences away from those at
equilibrium.  (To make this intuitive: for the energetic problem of
driving a non-equilibrium flow, the cost will be the measure of
dissipation and the payment of this cost will be non-equilibrium
chemical work supply by the environment, under the work-dissipation
accounting identity~\cite{Wachtel:transduction:22}.)
Both the costs to realize the boundary conditions
capable of dictating non-equilibrium event conditions (for example,
evolving and supplying enzymes), and the benefits that in turn pay
those costs (fitness benefits accruing from selectivity), are separate
problems that are outside the scope of this treatment.  
}

        \begin{figure}[!htb]
            \centering
            \begin{subfigure}[c]{0.2\textwidth}
                \centering
            \includegraphics[width=\textwidth]{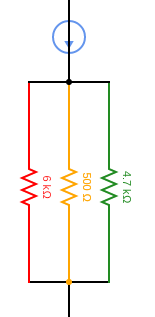}
            \end{subfigure}
            \hfil
            \begin{subfigure}[c]{0.45\textwidth}
                \centering
                \includegraphics[width=\textwidth]{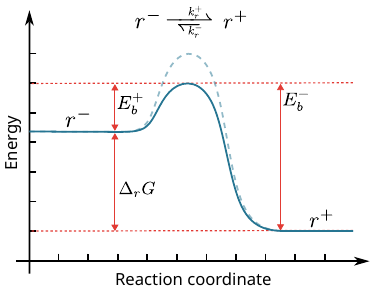}
            \end{subfigure}
            \hfil
            \begin{subfigure}[c]{0.2\textwidth}
                \centering
                \includegraphics[width=\textwidth]{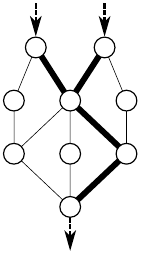}
            \end{subfigure}
            \caption{An analogy between electrical networks and CRNs. In the electrical network, the middle resistor (orange) has the lowest resistance, causing most of the current to flow through it. Similarly, in the open CRN shown on the right, the energy landscape can be modified so that most of the throughput is supported by the shaded pathway (thick black lines).
            }
            \label{fig:motivation}
        \end{figure}

\added{
For simplicity -- and importantly, as a model for the problem of
evolutionary pruning of reaction networks generated combinatorially by
sets of reaction mechanisms, as in~\cite{Smith:three_level:24} -- we
will consider only two states for a reaction: being driven in a
mass-action flow, or going un-used to reflect its active exclusion
from the prior combinatorial reaction network.  Therefore, the
}
cost of a pathway that we will propose comprises two
nonnegative contributions: (1) the minimum log-improbability to
sustain the pathway at a fixed exchange rate with the environment, and
(2) the minimum log-improbability to restrict flows to the reactions
in the pathway, excluding alternatives within a larger stochastic CRN.

For detailed-balanced CRNs operating in the linear response regime, we
demonstrate an analogy between electrical circuits and CRNs (see
Fig.\ \ref{fig:motivation}) that can be developed formally by
generalizing current and voltage to vectors, and conductance and
resistance to matrices.
In electrical circuits, it is well understood how serial and parallel
resistances affect the flow of current: adding resistors in parallel
reduces the effective resistance of the circuit and, for a fixed
current, decreases the power dissipated in the network.  Similarly,
for nested pathways in a detailed-balanced CRN operating close to
equilibrium, we prove that the thermodynamic cost of a smaller pathway
is always higher than that of any larger pathway of which it is a
part.  Interestingly, this relationship may not hold for networks
driven far from equilibrium,
\added{though in this work we find such cases only for
  unstable fixed points that represent the transition boundaries
  between high- and low-current attractors in multistable reaction
  systems, akin to the boundaries between coexisting phases in
  first-order phase transitions.}

    The outline of the paper is as follows. In Sec.\ \ref{sec:math_formalism}, we mathematically define the concept of a pathway and formalize its thermodynamic cost as an optimization problem. The cost function is derived from large-deviation theory, and its decomposition is presented. In Sec.\ \ref{sec:det_balanced}, we apply our formalism to detailed-balanced systems, deriving our main results for detailed-balanced CRNs operating in the linear response regime. The consequences of the framework far from equilibrium are also explained and illustrated using a simple example. In Sec.\ \ref{sec:methods}, we demonstrate our formalism on four- and five-species unimolecular CRNs and a multimolecular CRN involving competing autocatalytic cycles. Finally, we conclude in Sec.\ \ref{sec:discussion} with a summary of our contributions and an outlook for future research.
   
    %----------------------------------------------------------------------------------

    %----------------------------------------------------------------------------------
    \section{Mathematical formalism}
    \label{sec:math_formalism}

The objective of our study is to understand the thermodynamics of open
CRNs subject to fixed species currents. Such problems, involving
throughput currents in CRNs, are ubiquitous in nature and arise, for
instance, when an organism maintains a desired uptake current of
nutrients and a rejected waste current.  For a given CRN, there are
often multiple pathways through which the throughput current can be
realized.  The primary question we formulate and address in this
section is: what is a natural thermodynamic cost measure associated
with driving flow through one or more pathways while restricting to
zero flow through others that are alternatives?
  
\begin{figure}
        \centering
        \begin{subfigure}[c]{0.8\textwidth}
            \centering
            \includegraphics[width=\textwidth]{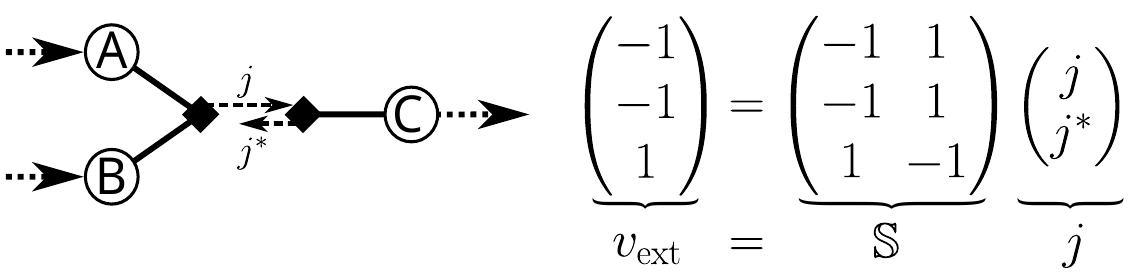}
            \caption{}
            \label{fig:ex1}
        \end{subfigure}
        
        \begin{subfigure}[c]{0.4\textwidth}
            \centering
            \includegraphics[width=\textwidth]{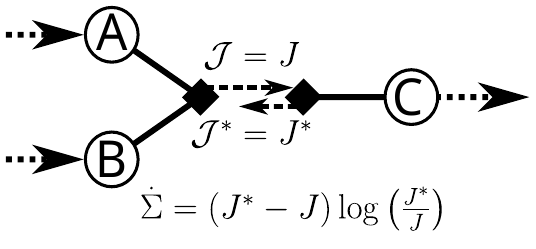}
            \caption{}
            \label{fig:ex2}
        \end{subfigure}
        \hfil
        \begin{subfigure}[c]{0.4\textwidth}
            \centering
            \includegraphics[width=\textwidth]{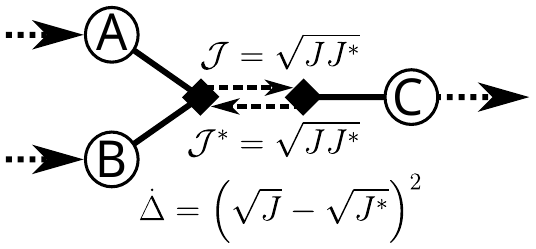}
            \caption{}
            \label{fig:ex3}
        \end{subfigure}
        \caption{
            \added{
                Scheme for a CRN involving a single, reversible reaction \ce{A + B <=> C} that serves as an example for the quantities given in Table~\ref{tab:symbol}.
                \textbf{(a)} Stoichiometric matrix $\mathbb{S}$ corresponding to the CRN as well as flux vector $j$ and throughput current $\vext$. 
                \textbf{(b)} Situation where the CRN runs normally with the mass action fluxes $J,J^{*}$, and maintanence cost $\dot{\Sigma}$
                \textbf{(c)} Situation where the reaction is blocked out and the blocking cost $\dot{\Delta}$.
                }
        }
        \label{fig:example}
        \end{figure}

In Sec.~\ref{sec:prelim}, we outline the basic setup of our work,
providing definitions for detailed-balanced CRNs, throughput currents,
pathways, and non-equilibrium steady states.  In
Sec.~\ref{sec:thermo_cost}, the thermodynamic cost of a pathway is
formulated as an optimization problem, and the natural decomposition
of this cost into two components, one for maintenance of the flow and
the other for restriction from alternatives, is also explained.  A
summary of defined quantities and their associated symbols introduced
in this section can be found in Table~\ref{tab:symbol} \added{and an example is shown in Fig.~\ref{fig:example}}.  Additional
remarks on the rationale behind our definitions and natural extensions
of our framework are given in Sec.~\ref{sec:discussion}

%Shall we make a table with symbols
\begin{table}[]
\centering
\begin{tabular}{ccc}
Variables & Symbol & Section introduced\\
\hline
Species in species set & $s \in \mcl{S}$ &\ref{sec:prel_CRN_partial}\\
Reaction in reaction set & $r \in \mcl{R}$&\ref{sec:prel_CRN_partial}\\
Species concentration & $q $&\ref{sec:prel_CRN_partial}\\
Species current & $\dot{q}$ &\ref{sec:prel_CRN_partial}\\
Reaction flux & $j $&\ref{sec:prel_CRN_partial}\\
Stoichiometric matrix & $\mbb{S} $ &\ref{sec:prel_CRN_partial}\\
Basis of conservation laws & $\mbb{L}$ &\ref{sec:prel_CRN_partial}\\
Reverse reaction & $r^*$ & \ref{sec:prel_detailed_balanced}\\
Reverse reaction flux &  $j_r^*$ & \ref{sec:prel_detailed_balanced}\\
Mass-action flux & $J(q)$ & \ref{sec:prel_detailed_balanced}\\
Detailed-balanced flux & $\Phi_r$& \ref{sec:prel_detailed_balanced}\\
Throughput current & $v_\text{ext}$& \ref{sec:prelim_thruput}\\
Net reaction flux of $j$ & $\mfk{N}(j)$& \ref{sec:prelim_thruput}\\
Partial mass-action flux & $\mcl{J}(\mcl{G}',q)$& \ref{sec:prelim_NESS}\\
Maintanence cost of a pathway & $\dot{\Sigma}(\mcl{G}',q)$& \ref{sec:thermo_cost_maintenence}\\
Restriction cost of a pathway & $\dot{\Delta}(\mcl{G}',q)$& \ref{sec:cost_block}\\
Restriction cost of a reaction & $\dot{\delta}(r,q)$& \ref{sec:cost_block}\\
Thermodynamic cost of a pathway & $\chi(\mcl{G}')$& \ref{sec:thermo_cost_thermo_cost}
\end{tabular}
\caption{Table of variables and their associated symbols.}
\label{tab:symbol}
\end{table}

    \subsection{Preliminaries: definitions and setup}
    \label{sec:prelim}
    \subsubsection{CRN and its partial networks}
    \label{sec:prel_CRN_partial}

A CRN $\mcl{G}$ is a pair $(\mcl{S},\mcl{R})$ where $\mcl{S}$ is a set
of species and $\mcl{R}$ is a set of reactions. We will index the set
$\mcl{S}$ by $s$ and denote a \textbf{species} in the set by $X_s$. A
\textbf{reaction} $r \in \mcl{R}$ is given as a pair of column vectors
$(r^-, r^+) \in \mbb{Z}^{\mcl{S}}_{\geq 0}$ and denoted with the
schema
    \[ r: r^- \to r^+.\]
    We use this notation as shorthand for the conventional representation of a reaction in chemistry 
    \[ r: \sum_{s \in \mcl{S}} (r^-)_s X_s \to \sum_{s \in \mcl{S}} (r^+)_s X_s\]
    where $r^-$ and $r^+$ are, respectively, the stoichiometries of the reactants and products. 

    A CRN specifies a list of rules by which the species \textbf{concentration}, denoted by $q \in \mbb{R}^{\mcl{S}}_{\geq 0}$, can change. 
    The net change in species due to a single firing of reaction $r$ is $\Delta r:= r^+ - r^-$. $\Delta r$ is called the reaction's \textbf{stoichiometric vector}, and the $\mcl{S}\times \mcl{R}$ matrix $\St$ whose columns are the stoichiometric vectors,
    \[ \St^r = \Delta r, \]
    is called the \textbf{stoichiometric matrix}.
    The event rate in each reaction is given by the \textbf{reaction flux} vector $j \in \mbb{R}^{\mcl{R}}_{\geq 0}$. 
     The rate of change of species concentrations, called \textbf{species current} $\dot{q} \in \mbb{R}^{\mcl{S}}$, due to a reaction flux $j$ is given as 
    \begin{equation}
        \dot{q} = \St j = \sum_r  \Delta r \phantom{\cdot} j_r. \label{eq:stoich_CRN}
    \end{equation}
Any species current resulting from a reaction flux must lie in
$\text{Im}(\St)$, also called the \textbf{stoichiometric subspace}.
It is the complement to $\text{Ker}(\St^T)$ in the linear space of
species concentrations.  Any vector $c \in \text{Ker}(\St^T)$
specifies a conservation law, meaning that its inner product along any
trajectory $q(t)$ is a conserved quantity, i.e.\ $c^T q(t) = c^T
q(0)$.  Let $\mbb{L}$ denote a matrix whose columns are any basis
for the generators of conservation laws,
    \begin{equation}
        \text{Cols}(\mbb{L}) = \text{Basis of Ker}(\St^T) \label{eq:conservation_law_matrix}
    \end{equation} 
A linear space of concentrations with fixed values for the conserved
quantities is called a \textbf{stoichiometric compatibility class}
(see Chap.\ 3, \cite{feinberg2019foundations}).

CRNs are directed multi-hypergraphs where the species are vertices and
reactions are hyperedges \cite{dai2023hypergraph}.  Borrowing
terminology from the hypergraph literature, for a CRN $\mcl{G} =
(\mcl{S},\mcl{R})$, any CRN $\mcl{G}'=(\mcl{S}',\mcl{R}')$ that
contains a subset of the reaction set such that $\mcl{R}'
\subseteq\mcl{R}$ and $\mcl{S}' = \mcl{S}|_{\mcl{R}'}$ will be called
a \textbf{partial network} of $\mcl{G}$, where $\mcl{S}|_{\mcl{R}'}$
denotes the set of species that participate in the reaction set
$\mcl{R}'$ in the CRN $\mcl{G}$. In contrast, a CRN
$\mcl{G}'=(\mcl{S}',\mcl{R}')$ such that $\mcl{R}' \subseteq\mcl{R}$
and $\mcl{S}' \subseteq \mcl{S}$ will be called a
\textbf{subnetwork}.  (In \cite{gagrani2024polyhedral}, a partial
network and subnetwork are referred to as subnetwork and motif,
respectively.)

    \subsubsection{Detailed balanced CRNs}
    \label{sec:prel_detailed_balanced}

    The reaction flux vector can be chosen to depend on the species concentrations $q$ to yield a dynamical system in the species concentration governed by the equations
    \begin{equation}
        \dot{q} = \St j(q).
    \end{equation}
    There are several canonical choices for parametrizing this dependence, each one admitting a different interpretation in terms of the phenomena they model. A parameterization $j(q)$ is called a \textbf{kinetic model} if $j_r(q)$ for a reaction $r:r^- \to r^+$ only depends on $r^-$ along with other minor technical requirements (Def.\ 2.1, \cite{vassena2024unstable}). 
    In this work, we only consider dynamics obtained under the \textbf{mass-action flux} assignment  
    \begin{equation}
        j_r \to J_r(q,k) =  k_r q^{r^-} \label{eq:MassAction},
    \end{equation}
    where, employing the multi-index notation, $q^{r^-} = \prod_{s \in \mcl{S}} q_s^{r^-_s}$. The assignment requires a vector of \textbf{rate constants} $k \in \mbb{R}^{\mcl{R}}_{\geq 0}$. Henceforth, we suppress the dependence on the rate constants, simply denoting the mass-action flux at concentration $q$ as $J(q)$ and the equations of mass-action kinetics as 
    \[\dot{q} = \St J(q).\] 

A CRN $\mcl{G} = (\mcl{S},\mcl{R})$ is \textbf{reversible} if for
every reaction $r: r^- \to r^+ \in \mcl{R}$, there exists the
\textbf{reverse reaction} $r^*: r^+ \to r^- \in \mcl{R}$.  Henceforth,
for a reaction $r$, we denote its reverse reaction and the rate
constant for the reverse reaction as $r^*$ and $k_r^*$, respectively.
Then, for any flux vector $j$, the reaction flux obtained by assigning
the flux from the reverse reaction $r^* \in \mcl{G}$ to the reaction
$r$ will be called the \textbf{reverse reaction flux}, and denoted
$j^*$ where
    \begin{equation}
        j^*_r = j_{r^*} \text{ for } r \in \mcl{G}. \label{eq:reverse_flux}
    \end{equation}
The reverse reaction flux under mass-action kinetics will be denoted
    \[J^*_r(q)=k_r^* q^{r^+},\] from which the species current for a
    reversible CRN under mass-action kinetics is then  
    \begin{align}
        \dot{q} &= \frac{1}{2}\sum_{r\in \mcl{G}} (\Delta r) (J_r(q)-J_r^*(q)) \nonumber\\
        &= \frac{1}{2}\sum_{r\in \mcl{G}} (\Delta r)(k_r q^{r^-} - k_r^* q^{r^+}). \label{eq:MAK_reversible}
    \end{align}
    A CRN is said to be \textbf{detailed-balanced}\footnote{\added{The detailed-balanced property is not to be confused with the local detailed-balanced property, see \cite{rao2018conservation}, which holds for any network where each reaction is reversible.}} if it admits a detailed balance equilibrium concentration $\eq \in \mbb{R}^{\mcl{S}}_{\geq 0}$ where the flux of each reaction and the reverse reaction balance out yielding
    \begin{equation}
      \Phi_r :=   k_r \eq^{r^-} = k_r^* \eq^{r^+}  \quad \forall r \in \mcl{R}, \label{eq:detailed_balance}
    \end{equation}
    where the factor of $1/2$ arises because we sum over all one-way reactions $r \in \mcl{R}$.
    Henceforth, we denote the \textbf{detailed balanced flux} in a reaction $r$ as $\Phi_r$.
    It is known (Ch.\ 14, \cite{feinberg2019foundations}) that for
    detailed-balanced systems, each stoichiometric compatibility class
    has exactly one equilibrium.  Thus for any positive $q$, there
    exists a unique $\eq$ in the same compatibility class. 
    Substituting the above in Eq.\ \ref{eq:MAK_reversible}, the species current of a detailed balanced CRN $\mcl{G}$ is given by
    \begin{equation}
        \dot{q} = \frac{1}{2}\sum_{r\in \mcl{G}} \Phi_r (\Delta r)\left[ \left(\frac{q}{\eq}\right)^{r^-} - \left(\frac{q}{\eq}\right)^{r^+}\right]. 
        \label{eq:detailed_MAK}
    \end{equation}

A relation from~\cite{schuster1989generalization}, known as
Wegscheider’s condition, gives that a reversible network admits a detailed balanced condition if and only if the rate constants satisfy
    \[ \ln\left(\frac{k_r}{k_r^*}\right) \in \text{Im}(\St).\]
    It follows that the rate constants of a detailed balanced CRN admit the thermodynamic parametrization
    \begin{equation}
    \begin{aligned}
        k_{r} &= e^{-(E_{r}^{\ddagger}-\mu_0 \cdot r^- )/RT} \\
        k_{r}^{*} &= e^{-(E_{r}^{\ddagger}-\mu_0 \cdot r^+ )/RT},
    \end{aligned}
    \label{eq:kin_const}
    \end{equation}
    where $\mu_0 \in \mbb{R}^{\mcl{S}}$ are the chemical energies of formation of the species, $E_r^\ddagger$ is the energy of the transition state, $T$ is the temperature, and $R$ is the molar gas constant. In this work, for all calculations, we set $RT = 1$. 

    % \NL{
    %     Yes looks good, maybe a minor thing in terms of notation: to prevent confusion between superscripts `T' and transpose operation we could use $E_r^t$ for the transition state energy
    % }\\
    % \PG{Good point. Do you wanna be fancier and use double dagger? $\ddagger$ Haha. If not, feel free to change it to $t$ as you suggested above}
    
    \subsubsection{Throughput currents and pathways}
    \label{sec:prelim_thruput}

In this work, we organize our study of open CRNs around species
exchange-currents with the environment that are kept constant as other
parameters are changed.  This condition is the Legendre
dual~\cite{Smith:Intr_Extr:20} to the frequently-used prescription of
\textit{chemostatting}, in which species concentrations (and hence
chemical potentials) are held
constant~\cite{polettini2014irreversible}.  In the sense that fixed
currents and concentrations could be mapped, in idealized
linear-response environments, to fixed chemical-potentials and their
(discrete) gradients, there is a correspondence between the Legendre
duality of currents and potentials, and the duality of Dirichlet and
Neumann boundary conditions for the solution of differential
equations~\cite{riley2006mathematical}.  Since the currents can be
arbitrary, the two treatments are interchangeable wherever the
Legendre duality is defined, and our fixed-current condition entails
no loss of generality. 

    Suppose a CRN $\mcl{G}$ is opened to operate at a fixed species current $\vext$, henceforth referred to as a \textbf{throughput current}. A reaction flux $j$ will be called \textbf{admissible} if the species current due to the reaction flux matches the throughput current, i.e.\ (using Eq.\ \ref{eq:stoich_CRN})
    \begin{equation}
       \vext = \St j. \label{eq:vext} 
    \end{equation}
In a reversible network, there are infinitely many admissible reaction
fluxes for a throughput current as for any admissible flux $j$, $j_r
\to j_r + \delta_r$ and $j^*_{r} \to j^*_{r} + \delta_r $ is also
admissible.  We therefore introduce the \textbf{net reaction flux} of
a reaction flux $\mfk{N}(j)$ as the difference between the flux
through a reaction and its reverse
    \[ \mfk{N}(j): \mfk{N}_r(j) = j_r - j^*_{r}.\]

    The support of the net reaction flux, denoted $\supp(\mfk{N})$ is the set of reactions where the net reaction flux is non-zero,
    \[ \supp(\mfk{N}(j)) = \{ r \in \mcl{G}: \mfk{N}_r(j) \neq 0  \}.\]
    Observe that the support of the net reaction flux is a partial network of the CRN
    \[ \supp(\mfk{N}(j)) \subseteq \mcl{G}.\]
    The support of the net reaction flux of an admissible flux for a throughput current is called a \textbf{pathway} for that throughput current. Equivalently, a partial network $\mcl{G}' \subseteq \mcl{G}$ is a pathway for a throughput current vector if there exists an admissible net reaction flux with support in $\mcl{G}'$
    \[ \mcl{G}' \text{ is a pathway for }\vext \iff \exists j: \St j = \vext \text{ and } \supp(\mfk{N}(j)) = \mcl{G}'.\]

    \subsubsection{Nonequilibrium Steady State (NESS)}
    \label{sec:prelim_NESS}
    Consider a strictly partial network $\mcl{G}' \subset \mcl{G}$. We define the \textbf{partial mass-action flux}, denoted by $\mcl{J}(\mcl{G}',q)$, as
        \begin{align}       
        \mcl{J}_r(\mcl{G}',q) &= \begin{cases}
            \qquad J_r(q) & \text{ if } r \in \mcl{G}'\\
            \sqrt{J_r(q) J^*_r(q)} &\text{ if } r \notin \mcl{G}'
        \end{cases}. \label{eq:partial_mass_action}
   \end{align}
    Since the partial mass-action flux assigns equal fluxes to forward and backward reactions in every reaction not in the partial network, the support of the net partial mass-action flux is the partial network
    \[ \supp(\mfk{N}(\mcl{J}(\mcl{G}',q))) = \mcl{G}',\]
    thus justifying its name. In App.\ \ref{app:variational_partial_MAK}, we prove that this assignment for detailed balanced CRNs minimizes the cost of \textit{blocking} reactions (defined in Sec.\ \ref{sec:cost_block}) not in the partial CRN.  

    \begin{comment}
    While this choice of defining a partial mass-action flux such that the support of the net flux is only in the partial network is not unique, it is well-motivated from CRN theory
    \footnote{
    In a reversible CRN, for a reaction flux flux $j$, the \textit{reversing force} 
    \[ f_r = \ln\left(\frac{j_r^*}{j_r}\right)\]
    reverses the reaction flux (see Eq.\ \ref{eq:reverse_flux}), i.e.\
    $ J^*_r = j_r e^{f_r}$.
    The flux at half the reversing force 
    \[j_r e^{f_r/2} = \sqrt{j_r J^*_r}\]
    is the flux in the partial mass-action flux in the reactions in the complement of the partial network and the dissipation potential \cite{kobayashi2022hessian} is minimum (identically zero) at this assignment.
    } and its utility will become clear in the next subsection.  
    \end{comment}

    For a throughput current, a concentration $q$ will be called a \textbf{nonequilibrium steady state (NESS)} if the mass-action or the partial mass-action flux at the concentration is admissible. In other words, a concentration $q$ is a NESS for a throughput current $\vext$ and partial network $\mcl{G}' \subseteq \mcl{G}$ if 
    \[ \vext = \St \mcl{J}(\mcl{G}',q). \]
    In general, each pathway for a throughput current may admit several or zero NESSs. In the remainder of this section, we will provide a formalism to thermodynamically rank the NESSs. 
    
    %------------------------------------------------------------------------------------------------------------------------------------
    \subsection{Thermodynamic cost of a pathway}
    \label{sec:thermo_cost}
    For a stochastic CRN $\mcl{G} = (\mcl{S},\mcl{R})$ at a population vector $n = V q$, let the probability that the reaction $r \in \mcl{G}$ fires $V \delta t j_r$ times, where $j \in \mathbb{R}_{\geq 0}^\mcl{R}$, be given as $\mbb{P}[V \delta t j | V q]$.
    Then, it is shown in \cite{lazarescu2019large} that, in the limit of large volume and small times step, $V \to \infty$ and $\delta t \to 0$, the probability follows a large deviation scaling 
    \begin{equation}
        \mbb{P}[V \delta t j | V q] \asymp e^{- \delta t \, V \mcl{D}(j||J(q))}, \label{eq:prob_LDT}
    \end{equation}
    with the rate function 
    \begin{equation}
        \mcl{D}(j||J(q)) = \sum_{r\in \mcl{G}} \left( j_r \ln \left( \frac{j_r}{J_r(q)}\right) - (j_r - J_r(q))\right). \label{eq:rate}
    \end{equation}
    Eq.~(\ref{eq:rate}) will be the starting point of our formalism for assigning thermodynamic costs to pathways.

    \subsubsection{Remarks on the rate function}
    $\mcl{D}(j||J(q))$ is the rate function of a Poisson distribution
    with a mean at the mass action reaction rates at concentration $q$
    of $J(q)$. Its appearance in the large-deviation function can be understood by recalling that a stochastic CRN consists of independent jump processes. Alternatively, $\mcl{D}(j||J(q))$ can also be seen as the exact Kullback-Leibler (KL) divergence between two Poisson distributions with mean at $j$ and $J(q)$. In this interpretation, it quantifies the difference between the Poisson distributions where $j$ is typical from that where the mass-action flux is typical. $\mcl{D}(j||J(q))$ is also known as a generalized KL divergence and satisfies  $ \mcl{D}(j_1||j_2)  \geq 0$
    for $j_1, j_2 \in \mathbb{R}_{\geq 0}^\mcl{R}$, where equality holds if and only if $j_1 = j_2$. 
    
\subsubsection{Maintenance cost}
\label{sec:thermo_cost_maintenence}

Consider a reversible CRN $\mcl{G}$ initialized with a concentration
$q$ for which the mass-action reaction flux $J(q)$ instantaneously
yields the throughput current $\vext$, i.e.
    \[ \mbb{S}J(q) = \vext.\] 
In the absence of external exchange, in a small-time step $\delta t$,
the species concentration will evolve to $q+\vext \delta t$ yielding a
species current $J(q+\vext \delta t)$.  To maintain the species
current at $\vext$, the species concentrations must be continuously
restored to $q$.

The probabilistic account of the deterministic mass-action flow is
that, by changing $q$ and redistributing internal energy to heat in
the bath, the system-plus-environment has moved to a less-improbable
distribution of configurations, losing the improbability of its
initial condition that drove the transition (in probability)
irreversibly.  The large-deviation measure of how much (log-)
improbability the system-plus-bath have lost in the transition is the
least improbability over all fluctuation events that could restore the
original configuration.  For a reversible stochastic CRN, from
Eq.\ \ref{eq:rate}, that log-fluctuation-improbability is bounded
below by the quantity
    \begin{align*}
        \dot{\Sigma}(\mcl{G},q) &= \min_{j:\mbb{S}j=-\vext} \mcl{D}(j||J(q)).
    \end{align*}
    Using Legendre-duality of the KL-divergence and Hamiltonian, the above quantity can be recast as \added{(see Eq.\ 4.20-4.27, \cite{chabane2021rarity})}
    \begin{align}
        \dot{\Sigma}(\mcl{G},q) &= \min_{j:\mbb{S}j=-\vext}
        \max_p \left(p^T \mbb{S}j - \sum_{r\in \mcl{G}}(e^{p^T \mbb{S}^r}-1)J(q)\right) \nonumber\\
        &= 
        \max_p \left(-p^T \vext - \sum_{r\in \mcl{G}}(e^{p^T \mbb{S}^r}-1)J(q)\right). \label{eq:excess_EPR}
    \end{align}
    We refer to the quantity $\dot{\Sigma}(\mcl{G},q)$ in
    Eq.\ \ref{eq:excess_EPR} as the \textbf{maintenance
      cost}\footnote{Also called the excess-entropy production
    (Eq.\ 42, \cite{kolchinsky2024generalized}).} as it is the minimum
    rate at which log-improbability must be continuously injected into
    the system to maintain a constant concentration $q$ and current
    $\vext$ through the CRN $\mcl{G}$.  Although the restoration of
    $q$ is performed by particle exchange with an environment and not
    by an internal fluctuation that re-absorbs heat, in the local
    equilibrium approximation the bound on the chemical work that must
    be delivered to compensate for dissipated heat is the same
    quantity (see~\cite{Wachtel:transduction:22}), permitting us to
    regard the driven external exchange as an ``injection'' of
    improbability (as popularized by
    Schr{\"{o}dinger~\cite{Schrodinger:WIL:92}).

    Observe that the reverse mass-action flux $J^*(q)$ always induces a species current of $-\vext$ and satisfies the equality
    \[-\vext = \mbb{S}J^*(q). \]
    It is known from Hamilton-Jacobi theory (Sec.\ 5,
    \cite{smith2020intrinsic}) that, for detailed-balanced systems, the minimizer over $j$ in Eq.\ \ref{eq:excess_EPR} is the the reverse mass-action flux $J^*(q)$. This result can be obtained by recognizing that escape paths \cite{gagrani2023action} for detailed-balanced systems are the time-reverses of their relaxation paths and thus, the species current along an escape path is exactly $-\vext$. Thus, the maintenance cost for detailed-balanced CRNs is the well-known \textbf{entropy production rate} (EPR)
    \begin{align}
        \dot{\Sigma}(\mcl{G},q)
        &= \mcl{D}(J^*(q)||J(q)) \nonumber\\
        &=\frac{1}{2}\sum_{r \in \mcl{G}}\ln\left(\frac{J_r^*(q)}{J_r(q)}\right)\left(J_r^*(q)-J_r(q)\right). 
        \label{eq:EPR}
    \end{align}
    Furthermore, for detailed-balanced CRNs, the above expression simplifies to (see Eq.\ A18, \cite{gagrani2025evolution})
    \begin{align}
        \dot{\Sigma}(\mcl{G},q)
        &= -\log\left(\frac{q}{\eq}\right)^T \vext.  \label{eq:DB_potential}
    \end{align}
    For non-detailed balanced systems, the maintenance cost is bounded
    above by the EPR and a detailed description of the relationship
    between the two quantities can be found in
    \cite{kolchinsky2024generalized}.  Since, in \deleted{this} this work, we
    restrict ourselves to detailed-balanced CRNs, we will use
    maintenance cost and EPR interchangeably.

    \subsubsection{Restriction cost of a pathway}
    \label{sec:cost_block}
    Consider an admissible reaction flux $j$ through a pathway $\mcl{G}' \subset \mcl{G}$ for some fixed throughput current. By definition, the net reaction flux must be zero in any reaction not in $\mcl{G}'$,
    \[ \mfk{N}_r(j) = j_r-j_r^* = 0 \quad \text{for } r \notin \mcl{G}', \]
requiring for these reactions that the flux and its reverse be
equal.  Henceforth, we refer to a reaction with zero net flux as
\textbf{blocked}.
    
    For a stochastic CRN at a concentration $q$, we know from Eq.\ \ref{eq:rate} that the generalized KL-divergence gives the log-improbability of observing a reaction flux. A simple calculation, shown in App.\ \ref{app:variational_partial_MAK}, yields that the least unlikely reaction flux required for blocking a reaction is 
    \[
    j_r = j^*_r =  \sqrt{J_r(q) J^*_r(q)} \quad \text{ for } r \notin \mcl{G}'.
    \]
    Thus, at a concentration $q$, the partial mass-action flux $\mcl{J}(\mcl{G}',q)$ defined in Eq.\ \ref{eq:partial_mass_action} is the least improbable reaction flux through the pathway $\mcl{G}'$.
    The rate function at the partial mass-action reaction flux evaluates to
    \begin{align}
    \mcl{D}(\mcl{J}(\mcl{G}',q)||J(q))
    &= \frac{1}{2}\sum_{r \notin \mcl{G}'} \left(\sqrt{J_r(q)}-\sqrt{J_r^*(q)}\right)^2 \label{eq:discover}\\
    &= \dot{\Delta}(\mcl{G}',q). \nonumber
    \end{align}

    We refer to $\dot{\delta}(r,q)$ as the \textbf{blocking cost}\footnote{Analogous to the
generalized KL divergence, $\dot{\delta}$ may be seen as a generalized
Hellinger distance \cite{ay2017information} defined over discrete
measures rather than probability distributions.} of a reaction $r$ at concentration $q$, where
    \begin{equation}
     \dot{\delta}(r,q):= \left(\sqrt{J_r(q)}-\sqrt{J_r^*(q)}\right)^2.\label{eq:block_cost_r}   
    \end{equation}
\added{We remark that, as proved in App.\ \ref{app:restriction_cost_bound}, the blocking cost of a reaction is never greater than its EPR.}
We define the \textbf{restriction cost} of a pathway to be
the sum of blocking costs of all the reactions not in the pathway,
    \begin{equation}
        \dot{\Delta}(\mcl{G}',q) = \sum_{r \notin \mcl{G}'} \dot{\delta}(r,q). \label{eq:discovery_decomp}
    \end{equation}
Like the maintenance cost, the restriction cost is a rate at which
improbability accumulates for the observation of an ongoing current
that would be atypical under the mass-action law.  Unlike the cost of
maintenance, which is ``paid'' by ``injection of improbability'' in
the form of chemical work along with particles from the environment,
neither matter nor energy is exchanged between the CRN and its
environment to accomplish restriction, and we do not need to stipulate
a particular mechanism of blocking.  Our ability, nonetheless, to
assign a cost to blocking from the large-deviation probabilities of
event prevention, may serve as a starting point for later study of the
system's interaction with specific mechanisms that accomplish
restriction.

    \subsubsection{Thermodynamic cost of a pathway}
    \label{sec:thermo_cost_thermo_cost}
   
    For a throughput current $\vext$, we define the \textbf{thermodynamic cost} of a pathway $\mcl{G}'$ in a CRN $\mcl{G}$ as 
    \begin{align}
        \chi(\mcl{G}') &= 
        \min_{q} \mcl{D}(\mcl{J}^*(\mcl{G}',q)||J(q)) \nonumber\\
        \text{subject to}&\qquad \vext = \St \mcl{J}(\mcl{G}',q),\nonumber\\
        & \qquad \mbb{L}^Tq = \mbb{L}^T \eq, \label{eq:cost_pathway}
    \end{align}
    where $\mbb{L}$ is the matrix of conservation laws defined in Eq.\ \ref{eq:conservation_law_matrix} and $\eq$ is the equilibrium concentration that fixes a stoichiometric compatibility class.
    The concentration that minimizes the cost is a NESS associated with the pathway, and a pathway will be called \textbf{infeasible} if it has no associated NESS. A pathway may also exhibit multiple local minima, leading to multiple NESSs, each with its associated cost.

    For a reversible $\mcl{G}' \subset \mcl{G}$, at any concentration $q$, it is straightforward to verify that 
    \begin{align}
        \mcl{D}(\mcl{J}^*(\mcl{G}',q)||J(q))
        &=
        \mcl{D}(\mcl{J}^*(\mcl{G}',q)||\mcl{J}(\mcl{G}',q)) + \mcl{D}(\mcl{J}(\mcl{G}',q)||J(q)) \nonumber\\
        &= \frac{1}{2}\sum_{r \in \mcl{G}'}\left(J_r^*(q)-J_r(q)\right) \ln\left(\frac{J_r^*(q)}{J_r(q)}\right)
        + \frac{1}{2}\sum_{r \notin \mcl{G}'} \left(\sqrt{J_r(q)}-\sqrt{J_r^*(q)}\right)^2 \nonumber\\
        &= \dot{\Sigma}(\mcl{G}',q) + \dot{\Delta}(\mcl{G}',q).
    \end{align}
    Thus, the cost of a pathway consists of two nonnegative contributions corresponding to its log-improbability of maintenance and restriction. The overdot, signifying time derivative, is a reminder that these costs are rates (of log-improbability) and must be paid at each instance to maintain the NESS.

%------------------------------------------------------------------------------------------------------------------------------------

\section{Nested detailed balanced pathways}
\label{sec:det_balanced}

In this section, we investigate the implications of our formalism from
Sec.\ \ref{sec:math_formalism} for assigning thermodynamic costs to
pathways in detailed-balanced CRNs.  Pathways $\mcl{G}'$ and $\mcl{G}$
will be called respectively a \textbf{nested} pathway and its
\textbf{embedding} pathway\footnote{Our choice of term makes reference
to the notion of an embedding space as the hosting space in
differential topology.} if both $\mcl{G}'$ and $\mcl{G}$ satisfy a
given throughput current and $\mcl{G}'$ is a strict partial network of
$\mcl{G}$.  For detailed-balanced CRNs in the linear response regime,
we develop a strict analogy with electrical circuits, demonstrating
that the cost of any nested pathway is strictly greater than the cost
of a pathway in which it is embedded (subject to certain technical
conditions explained later). For multimolecular CRNs, we illustrate
through an example that this trend may reverse for some fraction of
the NESSs far from equilibrium.

    \subsection{Linear response regime}
    \label{sec:uni_CRN_cost}

    Recall that, from Eq.\ \ref{eq:detailed_MAK}, the species current of a detailed balanced CRN $\mcl{G}$ is given by
    \begin{equation*}
        \dot{q} = \frac{1}{2}\sum_{r\in \mcl{G}} \Phi_r (\Delta r)\left[ \left(\frac{q}{\eq}\right)^{r^-} - \left(\frac{q}{\eq}\right)^{r^+}\right]. 
    \end{equation*}
    Denoting the vector of all ones of size $|\mcl{S}|$ by $\mbf{e} = [1,1,\ldots,1]^T$, a concentration $q$ will be said to be in the \textbf{linear response regime} if it satisfies
    \begin{equation}
        q \sim \eq \implies \frac{q-\eq}{\eq} \sim \mbf{0}. \label{eq:linear_response}
    \end{equation}
    In this regime, as can be verified by Taylor-expansion, the dynamics given by the above equation simplify to
    \begin{equation}
        \dot{q} = -\frac{1}{2}\sum_{r\in \mcl{G}} \Phi_r (\Delta r)(\Delta r)^T \fracqeqme . \label{eq:unimol_MAK} 
    \end{equation}

    A CRN where every reaction converts a single reactant species to a product species is called a unimolecular CRN. We remark that Eq.\ \ref{eq:unimol_MAK} is exact for unimolecular CRNs. For such a CRN, $q^{r^\pm} = r^\pm q$. Substituting it in Eq.\ \ref{eq:detailed_MAK}, the dynamics for a detailed balanced unimolecular CRN is given by the equation
    % The 
    \begin{align*}
        \dot{q} = -\frac{1}{2}\sum_{r\in \mcl{G}} \Phi_r (\Delta r)(\Delta r)^T \fracqeq.
    \end{align*}
    Observe that all unimolecular CRNs have the conservation law
    induced by the vector $\mbf{e}$, i.e.\, $\mbf{e}^T \dot{q} = 0=\mbf{e}^T \Delta r \, \forall r$ for any unimolecular CRN. Making use of the above observation, we can substitute $q/\eq$ with $q/\eq - \mbf{e}$, thus recovering Eq.\ \ref{eq:unimol_MAK}. 
    
    \subsubsection{Analogy with electrical circuits}
    The Ohm's law for electrical circuits consisting of only resistors and ideal current and voltage sources states
    \[ I = \frac{1}{R} V,\]
    where $I$ is the electrical current, $R$ is the effective resistance, $R^{-1}$ is the effective conductance, and $V$ is the potential difference between two points.
    To make analogy between CRNs and electrical circuits, we use the mapping
    \begin{align}
        I &\to \mcl{I} := \dot{q}, \nonumber\\
        V &\to \mcl{V}(q) := \fracqeqme, \nonumber\\
        \frac{1}{R} &\to \mbb{C}(\mcl{G}):= \frac{1}{2}\sum_{r\in \mcl{G}} \Phi_r (\Delta r)(\Delta r)^T,
        \label{eq:electrical_analogy}
    \end{align}
    and call $\mcl{I}$, $\mcl{V}(q)$, and $\mbb{C}(\mcl{G})$ as the species current vector, \textbf{species potential} vector, and \textbf{CRN conductance} matrix, respectively. Note that the species potential is the same as the \textit{chemical potential} \cite{rao2016nonequilibrium} in the linear response regime for detailed balanced systems which is defined as
    \begin{equation}
        \mu:= \ln\fracqeq \approx \fracqeqme. \label{eq:mu_chem_pot}
    \end{equation}
    Substituting Eq.\ \ref{eq:electrical_analogy} in Eq.\ \ref{eq:unimol_MAK}, we obtain for CRNs in the linear response regime
    \begin{equation}
        -\mcl{I} =  \mbb{C}(\mcl{G}) \mcl{V}(q), \label{eq:Ohm_CRN}
    \end{equation}
    where the additional minus sign stems from the convention that for CRNs a positive (negative) current means that species are flowing out of (into) the CRN which is opposite to that of conventional electric circuits.

    \subsubsection{Resistance of partial CRNs}
    A matrix $\mbb{M}$ is positive semidefinite if for any vector $v \in \mbb{R}^\mcl{S}$,
    \[ v^T \mbb{C} v \geq 0.\]
    For two positive semidefinite matrices $\mbb{M}_1$ and $\mbb{M}_2$, we say 
    \[ \mbb{M}_1 \geq \mbb{M}_2 \]
    if $\mbb{M}_1-\mbb{M}_2$ is positive semidefinite.
    The conductance matrix $\mbb{C}$ for any detailed-balanced CRN,
    constructed as a sum of dyadics in
    Eq.~\ref{eq:electrical_analogy}, is a symmetric positive
    semi-definite matrix.
%    The null space of $\mbb{C}$ is spanned by $\mbf{e}$, the vector of
   % all ones.
    Likewise, both the conductance matrix of any partial CRN $\mcl{G}'
    \subset \mcl{G}$ and its complement in $\mcl{G}$ are sums of
    dyadics, with $\bbcal{C}{G}{'}$ having fewer terms (indeed, a
    proper subset) than $\bbcal{C}{G}{}$; hence 
    \[ \bbcal{C}{G}{} \geq \bbcal{C}{G}{'}.\]

    Unlike a scalar electrical conductance, the conductance matrix for
    a CRN is singular, so we define the corresponding
    \textbf{resistance matrix} as its Moore-Penrose inverse
    (pseudoinverse) \cite{bernstein2009matrix} and denote it as
    \begin{equation}
        \mbb{R} := \mbb{C}^+ = (\mbb{C}^T \mbb{C})^{-1} \mbb{C}^T. \nonumber
    \end{equation}
    Using the above, we rewrite Eq.\ \ref{eq:Ohm_CRN} as
    \begin{equation}
        \mcl{V}(q) = - \bbcal{R}{G}{}\mcl{I} + \mbf{x} \label{eq:Ohm_CRN2}
    \end{equation} 
    where we use the freedom to vary $\mbf{x}$ within
    $\text{Im}(\mbb{R})^\perp$ to find a nonnegative $q$ in the same
    stoichiometric compatibility class as $\eq$. (If such a $q$ does
    not exist, then the species current is infeasible.)  Given a
    partial network $\mcl{G}' \subset \mcl{G}$, in
    App.\ \ref{app:resistance_nondecrease} we show that the difference
    between the resistance matrix of the partial network and network
    is positive semidefinite for any vector that is a feasible species
    current for $\mcl{G}'$,
    \begin{equation}
        v^T \left(\bbcal{R}{G}{'}-\bbcal{R}{G}{}\right)v \geq 0 \quad \text{ for }v \in \text{Im}(\bbcal{C}{G}{'}). \label{eq:R_semidef}
    \end{equation}
    As a slight abuse of notation, we denote this as 
    \begin{equation}
        \bbcal{R}{G}{} \leq \bbcal{R}{G}{'}, \label{eq:increasing_resistance}
    \end{equation}
    which is the analog of the result for electrical circuits that the effective resistance between two points never decreases if an intermediate resistor is removed.

    \subsection{Nondecreasing cost of nested pathways}
    \label{sec:non_dec_nested_path}
    Consider a detailed balanced multimolecular CRN in the linear regime or a unimolecular CRN $\mcl{G}$, a throughput current $\vext$, and two nested pathways $\mcl{G}_2 \subset \mcl{G}_1 \subset \mcl{G}$ for which the constraint is feasible. Furthermore, assume that $\mcl{G}/\mcl{G}_1$ is itself a pathway, i.e.\ the complement of $\mcl{G}_1$ in $\mcl{G}$ supports an admissible reaction flux for the throughput current $\vext$. In this subsection, we will show that the cost of the nested pathway is never less than the cost of the pathway in which it is embedded,
    \[ \chi(\mcl{G}_2) \geq \chi(\mcl{G}_1).\]

    Let the NESS concentrations at which the costs are evaluated be $q_1$ and $q_2$ for $\mcl{G}_1$ and $\mcl{G}_2$, respectively. Recall that the thermodynamic cost of a pathway is the sum of its maintenance cost and restriction cost evaluated at the NESSs
    \[\chi = \dot{\Sigma} + \dot{\Delta}.\]
    Let us denote the maintenance and restriction costs of $\mcl{G}_i$
    at $q_i$ as $\dot{\Sigma}_i$ and $\dot{\Delta}_i$, respectively,
    where $i \in \{1,2\}$. In what follows, we show that
    $\dot{\Sigma}_1 \leq \dot{\Sigma}_2$ and $\dot{\Delta}_1 \leq
    \dot{\Delta}_2$.  Since the differences of both summands are
    individually nonnegative, the nondecrease of $\chi$ under
    restriction follows. 

    \subsubsection{Nondecreasing maintenance cost}
    
    As shown in Eq.\ \ref{eq:DB_potential}, for a detailed balanced system, the expression of the maintenance cost (EPR) simplifies to:
    \begin{equation*}
        \dot{\Sigma} = -\ln{\fracqeq}^T \dot{q}. 
    \end{equation*}
    In the linear response regime, using Eq.\ \ref{eq:linear_response} and the mapping in Eq.\ \ref{eq:electrical_analogy}, we can write the EPR as
    \begin{equation*}
       \dot{\Sigma} = -\mcl{V}(q)^T \mcl{I}. %\label{eq:EPR_uni} 
    \end{equation*}
    Observe that the above equation is analogous to the formula for power $P$ dissipated in electrical circuits
    $P = V I.$
    Using Eq.\ \ref{eq:Ohm_CRN2}, we can further rewrite the above equation as
    \begin{equation}
        \dot{\Sigma} = \mcl{I}^T \mbb{R} \mcl{I}, \label{eq:EPR_uni} 
    \end{equation}
    which is the counterpart to the formula $P = I^2 R$ for electrical circuits. 
    
    Let us consider the difference of the EPRs due to $\mcl{G}_1$ and $\mcl{G}_2$. We denote the resistance matrix of $\mcl{G}_i$ as $\mbb{R}_i$, where $i \in \{1,2\}$. Then, we have
    \begin{align*}
        \dot{\Sigma}_1 - \dot{\Sigma}_2 &= \mcl{I}^T \mbb{R}_1 \mcl{I} - \mcl{I}^T \mbb{R}_2 \mcl{I}\\
        &= \mcl{I}^T (\mbb{R}_1 - \mbb{R}_2) \mcl{I}\\
        & \leq 0,
    \end{align*}
    where the last line follows from Eq.\ \ref{eq:increasing_resistance}. Thus, we have shown that $\dot{\Sigma}_1 \leq \dot{\Sigma}_2$.
    
\smallskip
\added{\noindent \textbf{Remark:}
  As we noted in the introduction, the non-decrease of dissipative
  entropy production is intuitive and familiar from electric circuit
  theory, where addition of parallel flow paths decreases
  whole-circuit resistance and thereby lowers voltage drop and
  delivered power across the input and output at fixed total flux.
  The CRN result is the stoichiometric generalization of the Kirchhoff
  result for scalar charge carriers.
}

    \subsubsection{Nondecreasing restriction cost}
    \label{sec:nondec_discovery}
    Using Eq.\ \ref{eq:discovery_decomp}, the restriction cost of $\mcl{G}_i$  
    is 
    \[ \dot{\Delta}_i = \sum_{r \in \mcl{G}^c_i} \dot{\delta}(r,q_i).\]
    Using Eq.\ \ref{eq:block_cost_r} for detailed balanced systems, the blocking cost of a reaction becomes
    \[ \dot{\delta}(r,q)= \Phi_r \left(\fracqeq^{r^-/2}-\fracqeq^{r^+/2}\right)^2.\]
    In the linear response regime, the expression simplifies to 
    \begin{align}
       \dot{\delta}(r,q)& =
       \Phi_r \left((\Delta r)^T\sqrt{\fracqeq}\right)^2
       \nonumber \\
       & \approx \frac{1}{4}\Phi_r \left((\Delta r)^T\fracqeqme\right)^2\nonumber\\
       & =  \mcl{V}(q)^T  (\Delta r)\frac{\Phi_r}{4} (\Delta r)^T  \mcl{V}(q) \nonumber \\
       & = \mcl{I}^T \mbb{R}  (\Delta r)\frac{\Phi_r}{4} (\Delta r)^T \mbb{R} \mcl{I}
       \label{eq:cost_block_uni}
    \end{align}
    where we have made use of the fact that $\mbb{R}$ is symmetric last line.

    For $i \in \{1,2\}$, let $\mcl{G}^c_i:=\mcl{G}/\mcl{G}_i$ denote the complement of $\mcl{G}_i$ in $\mcl{G}$. 
    Since $\mcl{G}_1 \supset \mcl{G}_2$, $\mcl{G}_1^c \subset \mcl{G}_2^c$. Thus,
    \[ \dot{\Delta}_2 = 
    \sum_{r \in \mcl{G}^c_1} \dot{\delta}(r,q_2) + 
    \sum_{r \in \mcl{G}^c_2/\mcl{G}^c_1} \dot{\delta}(r,q_2). 
    \]
    Using Eq.\ \ref{eq:cost_block_uni}, the total cost of blocking off all reactions in $\mcl{G}_1^c$ is
    \[ 
    \sum_{r \in \mcl{G}_1^c} \dot{\delta}(r,q_i) = \sum_{r \in
      \mcl{G}_1^c} \mcl{I}^T \mbb{R}_i  (\Delta r)\frac{\Phi_r}{4}
    (\Delta r)^T \mbb{R}_i \mcl{I} = \frac{1}{2} \mcl{I}^T \mbb{R}_i \mbb{C}(\mcl{G}_1^c) \mbb{R}_i \mcl{I},
    \]
    using the definition of the conductance matrix from Eq.\ \ref{eq:electrical_analogy}.
    Using $\mbb{R}_2 \geq \mbb{R}_1$ and the assumption that $\mcl{G}_1^c$ is itself a pathway for the throughput current $\mcl{I}$, we show in App.\ \ref{app:cost_nondecrease} that the difference in the cost of blocking of all reactions in $\mcl{G}_1^c$ between the nested subgraph $\mcl{G}_2$ and its embedding graph $\mcl{G}_1$ is never negative, i.e.,
    \begin{equation}
        \sum_{r \in \mcl{G}_1^c} \left(\dot{\delta}(r,q_2) - \dot{\delta}(r,q_1)\right) \geq 0. \label{eq:diff_cost_nonnegative}
    \end{equation}
    Clearly, $\sum_{r \in \mcl{G}^c_2/\mcl{G}^c_1} \dot{\delta}(r,q_2) \geq 0$. Thus, we have proved that $\dot{\Delta}_1 \leq \dot{\Delta}_2$.

    \subsection{\added{Possibility for opposite rankings at
        stable attractors and unstable saddle points}}
    \label{sec:higher_order}

   The species current vector field for multimolecular CRNs is typically non-injective \cite{banaji2010graph}. Thus, the same throughput current can generally admit multiple NESSs in multimolecular CRNs. For a pathway $\mcl{G}_1$ nested within a pathway $\mcl{G}$, i.e., $\mcl{G}_1 \subset \mcl{G}$, as shown in Sec.\ \ref{sec:non_dec_nested_path}, the close-to-equilibrium NESSs exhibit a strict ordering: the maintenance, restriction, and thermodynamic costs are all lower in $\mcl{G}$ than in $\mcl{G}_1$. However, for NESSs further from equilibrium, this relationship need not hold. We illustrate this breakdown of cost ordering far from equilibrium with an example in the remainder of the section. The implications of this phenomenon for biochemistry and the origins of life are discussed in Sec.\ \ref{sec:discussion}.

    \begin{figure}[t]
    \centering
    \begin{subfigure}{.48\textwidth}
      \centering
      \includegraphics[width=0.95\linewidth]{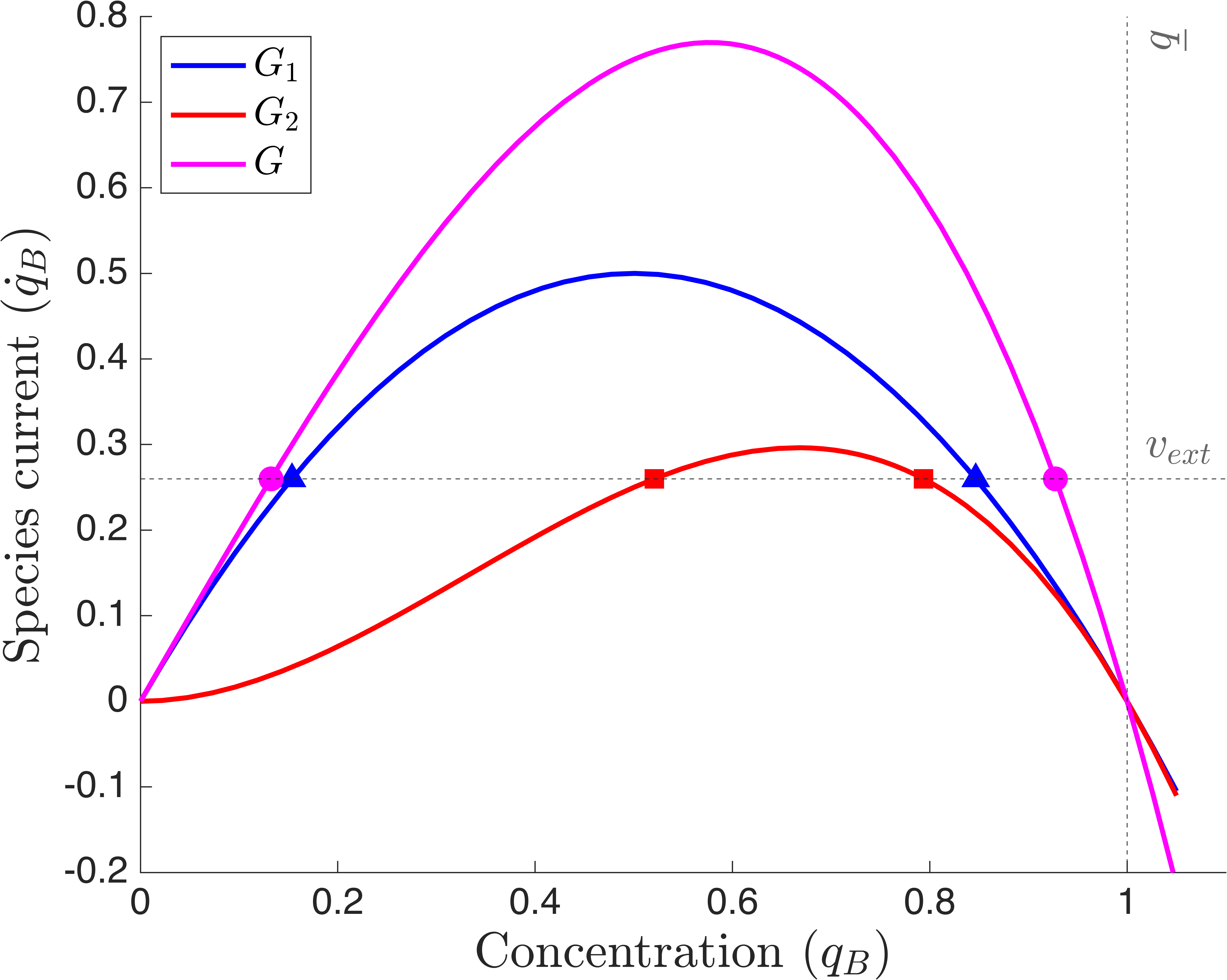}
      %\caption{A subfigure}
      %\label{fig:sub1}
    \end{subfigure}%
    \begin{subfigure}{.48\textwidth}
      \centering
      \includegraphics[width=0.95\linewidth]{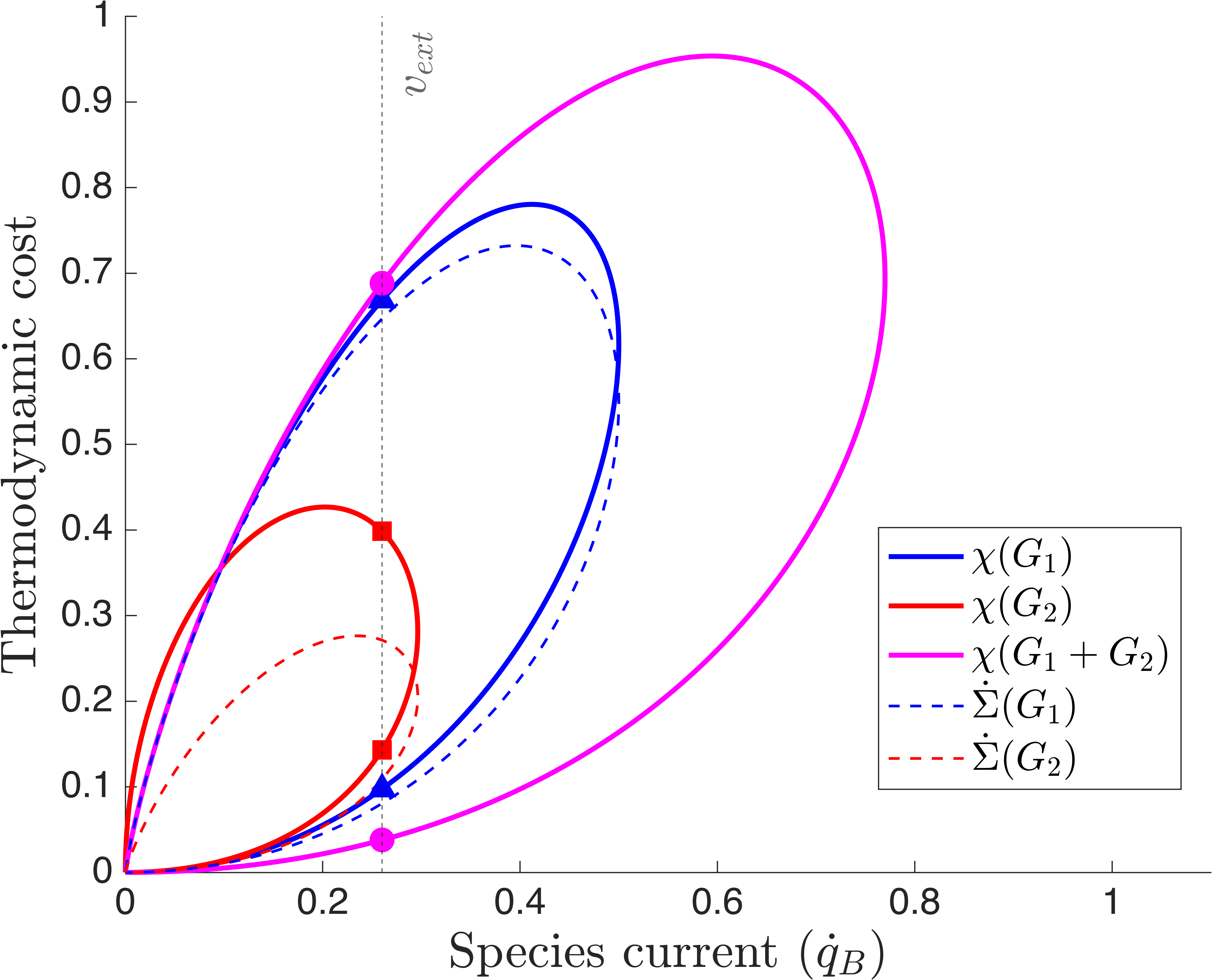}
%      \caption{A subfigure}
%      \label{fig:sub2}
    \end{subfigure}
    \caption{The velocity profile and the thermodynamic costs for the three multimolecular CRNs considered in Sec.\ \ref{sec:higher_order} are shown in the left and right panels, respectively. The NESSs are shown with markers and the maintenance cost of the nested pathways is also shown with dashed-curves.}
    \label{fig:conc_EPR_velocity}
    \end{figure}
        
    Consider the following three CRNs,
    \begin{align*}
        \mcl{G}_1 &= \{\ce{A + B <=>[1][1] 2B}\},\\
        \mcl{G}_2 &= \{\ce{A + 2B <=>[1][1] 3B}\},\\
        \mcl{G} &= \mcl{G}_1 \cup \mcl{G}_2.
    \end{align*}
    Clearly, $\mcl{G}_1,\mcl{G}_2 \subset \mcl{G}$. Consider the
    stoichiometric compatability class given by $q_A + q_B = 2$ in which $\eq_A = \eq_B = 1$. For a throughput current $\vext = [-0.26, 0.26]^T$ such that $A$ and $B$ are flowing into and out of the the system, respectively, the velocity vs.\ concentration and thermodynamic costs ($\chi$) vs.\ velocity plots are shown in Fig.\ \ref{fig:conc_EPR_velocity}. The EPR $\dot{\Sigma}$ for different pathways is also shown. It can be seen that the throughput current admits two NESSs for each pathway (shown with markers at the appropriate interesections).
    
    %\NL{(a stability analysis of the possible NESSs can be found in App.~\ref{app:stability})}.
    While the thermodynamic cost and EPR are both smaller for the
    embedding pathway $\mcl{G}$ than for the nested pathways
    $\mcl{G}_1$ and $\mcl{G}_2$, this trend reverses for the NESSs
    further away.  This behavior is specific to our choice of $\vext$
    and is does not hold for arbitrary throughput currents, as can be
    seen from regions in the figure where the thermodynamic cost of
    $\mcl{G}_2$ exceeds that of $\mcl{G}$ at both NESSs.  The
    existence of the counterexample shows that no single monotonicity
    relation follows from restriction for costs of NESSs that are
    far-from-equilibrium.

%------------------------------------------------------------------------------------------------------------------------------------

    %----------------------------------------------------------------------------------
    \section{Applications}
    \label{sec:methods}
    In this section, we consider several toy models to elucidate different aspects of our formalism. In Sec.\ \ref{sec:unimol}, we study two unimolecular CRNs involving four and five species. The four-species model features two nested pathways, and we present symmetric, asymmetric, and strongly asymmetric choices of reaction rates to demonstrate that the maintenance cost, $\dot{\Sigma}$, and the thermodynamic cost, $\chi$, always increase in nested pathways. While this result strictly holds, the different rate constant assignments illustrate thermodynamic choices through which one pathway can be made to dominate over the other. The five-species model, which contains six nested pathways, is used to demonstrate the non-decrease of restriction costs for nested pathways within nested pathways.

    In Sec.\ \ref{sec:multi}, we examine a multimolecular example of competing autocatalytic pathways \cite{andersen2021defining,blokhuis2020universal,gagrani2024polyhedral}. Under a given throughput current, these networks are shown to possess multiple non-equilibrium steady states (NESSs). While NESSs near the detailed-balanced attractor behave similarly to unimolecular CRNs, the thermodynamic costs at NESSs of nested pathways further away from the detailed-balanced attractor are found to be lower than the corresponding costs of the embedding pathway. The implications of this finding are discussed further in Sec.\ \ref{sec:discussion}.

%\added{\textbf{Praful and Nino: check this, to be sure it is
%    what was actually done.}}
In the models below, several different ways of blocking reactions can
result in the elimination of flow from the complement to a nested
pathway, and these may either disconnect subsets of species from the
nested pathway entirely, or place them in equilibrium with one or more
internal species in the nested pathway. In the examples that follow, we adopt the convention that if the flow through an external species is eliminated by blocking all reactions to and from it, we assign its concentration to the value within the stoichiometric compatibility class that minimizes the total blocking cost. Any other choice—such as setting the concentration to its Gibbs equilibrium value—would leave the qualitative ordering of the results unchanged, as it can only increase the total cost of the nested pathway.

\subsection{Unimolecular CRNs}
\label{sec:unimol}

        \subsubsection{Four-species model}
        
        \begin{figure}[!htb]
            \centering
            \begin{subfigure}[c]{0.2\textwidth}
                \centering
                \includegraphics[width=\textwidth]{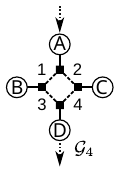}
                \caption{}
                \label{fig:abcd_crn}
            \end{subfigure}
            \hfil
            \begin{subfigure}[c]{0.2\textwidth}
                \centering
                \includegraphics[width=\textwidth]{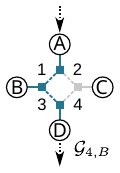}
                \caption{}
                \label{fig:abcd_sub1}
            \end{subfigure}
            \hfil
            \begin{subfigure}[c]{0.2\textwidth}
                \centering
                \includegraphics[width=\textwidth]{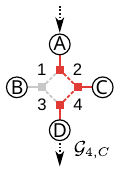}
                \caption{}
                \label{fig:abcd_sub2}
            \end{subfigure}
            \caption{
                Representation of the possible reaction pathways in the four species model that support $\vext=(-1,0,0,1)$:
                \textbf{(a)} full CRN $\mcl{G}_{4}$,
                \textbf{(b)} subgraph $\mcl{G}_{4,B}$,
                \textbf{(c)} subgraph $\mcl{G}_{4,C}$.
                Edge-labels represent the index of the reaction in Eq.~\ref{eq:ABCD_react}.
            }
            \label{fig:abcd}
        \end{figure}

        The first example is a simple system with four species (\ce{A}, \ce{B}, \ce{C}, \ce{D}) and four reversible reactions:

            \begin{equation}
                \begin{aligned}
                \ce{A &<=>[k_{1}][k^{*}_{1}] B <=>[k_{3}][k^{*}_{3}] D}\\
                \ce{A &<=>[k_{2}][k^{*}_{2}] C <=>[k_{4}][k^{*}_{4}] D}.\\
                \end{aligned}
                \label{eq:ABCD_react}
            \end{equation}
            The resulting CRN $\mcl{G}_{4}$ is visualized in Fig.~\ref{fig:abcd_crn}\footnote{
                Note that for simplicity the edges representing the reactions are drawn undirected as all reactions are assumed to be reversible.
            }.
            The stoichiometric matrix $\mbb{S}_{4}$ associated with this CRN is given as:
            \[
                \mbb{S}_{4}=\begin{pNiceMatrix}[r,first-col,first-row]
                  &r_{1}&r_{1}^{*}&r_{2}&r_{2}^{*}&r_{3}&r_{3}^{*}&r_{4}&r_{4}^{*}\\
                A &-1& 1&-1& 1& 0& 0& 0& 0\\
                B & 1&-1& 0& 0&-1& 1& 0& 0\\
                C & 0& 0& 1&-1& 0& 0&-1& 1\\
                D & 0& 0& 0& 0& 1&-1& 1&-1
                \end{pNiceMatrix}.
            \]
            From $\mbb{S}_{4}$ one can find that the CRN has one
            conserved quantity, generated by $\mbb{L}^{T}=(1,1,1,1)$ (i.e. the sum of all species concentrations is conserved).

            In what follows, we use the following setup. Species \ce{A} and \ce{D} are designated as in- and out-flowing species respectively, while the species \ce{B} and \ce{C} serve as intermediates, i.e.: $\vext=(-1,0,0,1)$. As shown in Fig.~\ref{fig:abcd}, in addition to the complete graph $\mcl{G}_{4}$, there are two additional partial networks that can support $\vext$: $\mcl{G}_{4,B}$ and $\mcl{G}_{4,C}$ representing the reaction pathway going only through the intermediate \ce{B} (see Fig.~\ref{fig:abcd_sub1}) and \ce{C} (see Fig.~\ref{fig:abcd_sub2}), respectively. The reaction rate constants $k_{r}$ and $k_{r}^{*}$, which are necessary to calculate the mass-action fluxes $j_{r}$ and $j_{r}^{*}$, can be calculated using Eq.~\ref{eq:kin_const} with the parameters $E^{\ddagger}$ and $\mu_{0}$ that follow from the thermodynamic landscape of the system (compare also Fig.~\ref{fig:motivation}). Explicitly, we have:            
            \begin{align*}
                k_{1} &= e^{-(E_{1}^{\ddagger}-\mu_0^A )},& k_{1}^{*} &= e^{-(E_{1}^{\ddagger}-\mu_0^B )}, 
            \end{align*}
        and so on. Any choice of values for $E_{r}^{\ddagger}$ and $\mu_{0}$ is thermodynamically consistent\footnote{
            For real CRNs, these values would actually come from the appropriate computational tools and/or experimental data.
        }. 
        
        In this subsection, different choices for the energy landscape of the system are going to be set
        resulting in different values for the rate constants of forward- and backward reactions $k_{r},k_{r}^{*}$. Then, the optimization problem defined in Eq.~\ref{eq:cost_pathway} is solved to calculate the costs of the full-CRN, $\chi(\mcl{G}_{4})$, as well as the two nested pathways, $\chi(\mcl{G}_{4,B})$ and $\chi(\mcl{G}_{4,C})$ for each choice
        (see App.~\ref{app:four_species} for a detailed description of the implementation).
        Through this, we study the effect that different energy landscapes, i.e. the choices of kinetic parameters, have on the costs of a reaction pathway $\chi(\mcl{G})$. To ensure comparability between the different results, the stoichiometric compatibility class defined in Eq.~\ref{eq:cost_pathway} is set to $\mbb{L}^{T}\eq=50.0$

         \begin{figure}[!htb]
            \centering
            \begin{subfigure}[c]{0.46\textwidth}
                \centering
                \includegraphics[width=\linewidth]{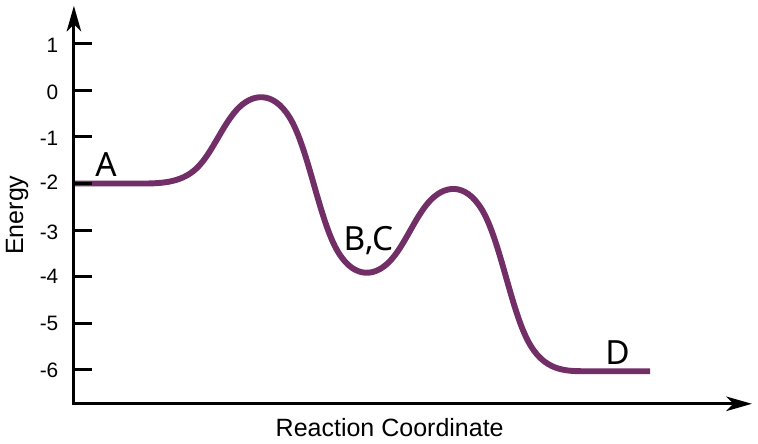}
                \caption{}
                \label{fig:abcd_lndscp1}
            \end{subfigure}
            \hspace{2ex}
            \begin{subfigure}[c]{0.46\textwidth}
                \centering
                \includegraphics[width=\linewidth]{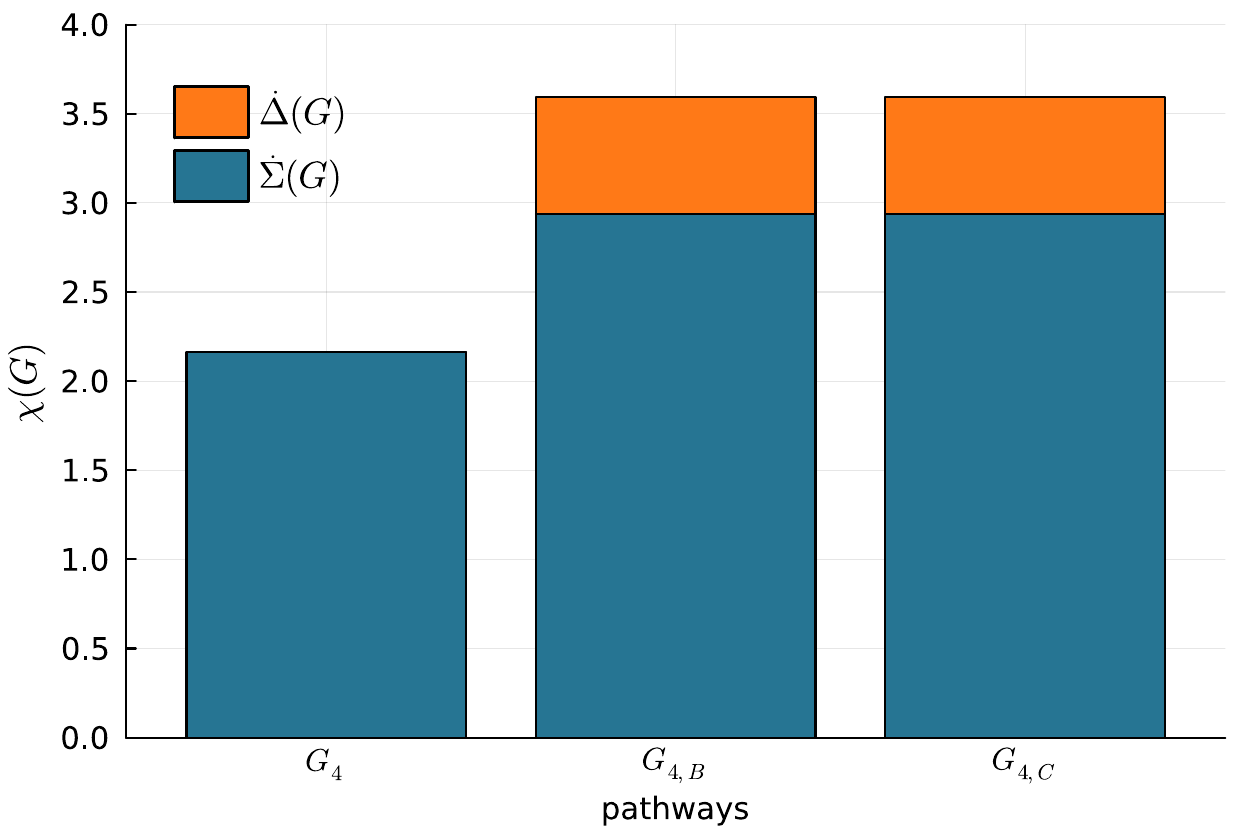}
                \caption{}
                \label{fig:abcd_cost1}
            \end{subfigure}
            
            \begin{subfigure}[c]{\textwidth}
                \centering
                \includegraphics[width=\linewidth]{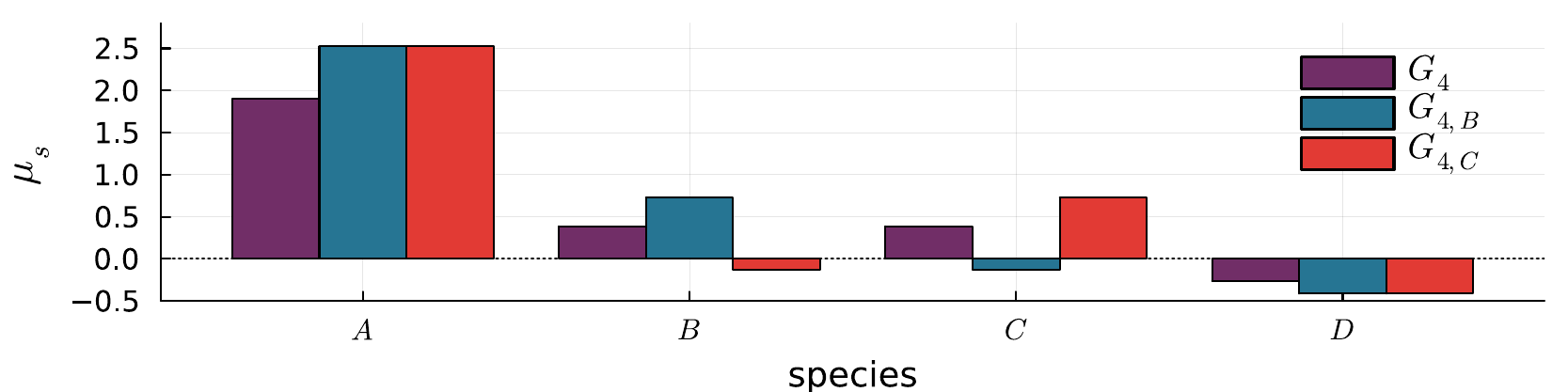}
                \caption{}
                \label{fig:abcd_conc1}
            \end{subfigure}
            \label{abcd1}
            \caption{
                Collection of the results in the four species model for the symmetric energy landscape with parameter settings: $\mu_{0}=(-2.0,-4.0,-4.0,-6.0)$, $E^{\ddagger}=(0.0,0.0,-2.0,-2.0)$.
                \textbf{(a)} Visualization of  the energy landscape.
                \textbf{(b)} Bar plot of the cost of the full CRN and the two reaction pathways: the bar height corresponds to thermodynamic cost $\chi(\mcl{G})$ while the bar composition corresponds to the maintenance cost $\dot{\Sigma}(\mcl{G})$ (blue) and the restriction cost $\dot{\Delta}(\mcl{G})$ (orange).
                \textbf{(c)} Bar plot of the chemical potential $\mu$ (Eq.\ \ref{eq:mu_chem_pot}) of each species for $\mcl{G}_{4}, \mcl{G}_{4,B}$ and $\mcl{G}_{4,C}$ (bar colors are the same as in Fig.~\ref{fig:abcd}).
            }
        \end{figure}

            \paragraph{Symmetric energy landscape: }The energies of formation $\mu_{0}$ are set such that the forward reactions are favored and \ce{B} and \ce{C} are set to have the same formation energy, i.e.: $\mu_{0}^{A}>\mu_{0}^{B}=\mu_{0}^{C}>\mu_{0}^{D}$ (shown in Fig.~\ref{fig:abcd_lndscp1}). The transition state energies $E^{\ddagger}$ are set such that the kinetic barriers are equal for all reactions. The result of this \textit{symmetric} assignment is that the rate constants $k_{r},k_{r}^{*}$ are the same in $\mcl{G}_{4,B}$ and $\mcl{G}_{4,C}$.

            The costs and concentration of NESSs for this model are shown, respectively, in Fig.~\ref{fig:abcd_cost1} and Fig.~\ref{fig:abcd_conc1}. It can be seen that the thermodynamic costs for $\mcl{G}_{4,B}$ and $\mcl{G}_{4,C}$ are the same, and they are both greater than $\mcl{G}_{4}$. Additionally, the EPR in $\mcl{G}_{4,B}$ and $\mcl{G}_{4,C}$ is also greater than that of $\mcl{G}_{4}$. Observe that the chemical potentials $\mu$ (Eq.\ \ref{eq:mu_chem_pot}) of $A$ and $D$ are larger in magnitude at the NESSs for the nested pathways. This can be interpreted as the system having to deviate further from the equilibrium to support the same current through a nested pathway, consequently having a higher EPR.

            \begin{figure}[!htb]
                \centering
                \begin{subfigure}[c]{0.46\textwidth}
                    \centering
                    \includegraphics[width=\linewidth]{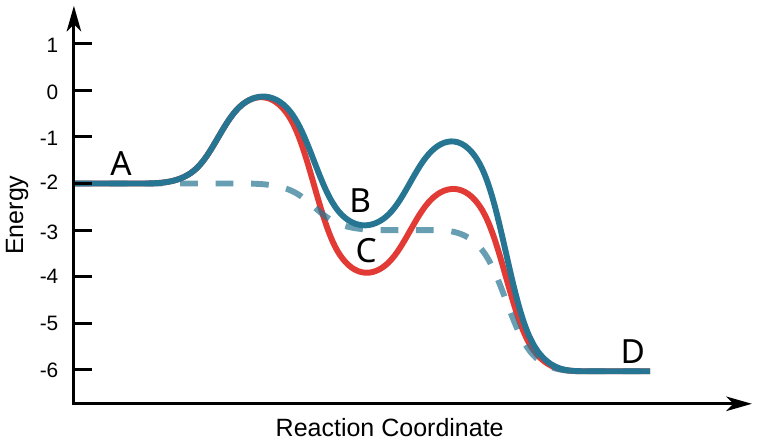}
                    \caption{}
                    \label{fig:abcd_lndscp2}
                \end{subfigure}
                \hspace{2ex}
                \begin{subfigure}[c]{0.46\textwidth}
                    \centering
                    \includegraphics[width=\linewidth]{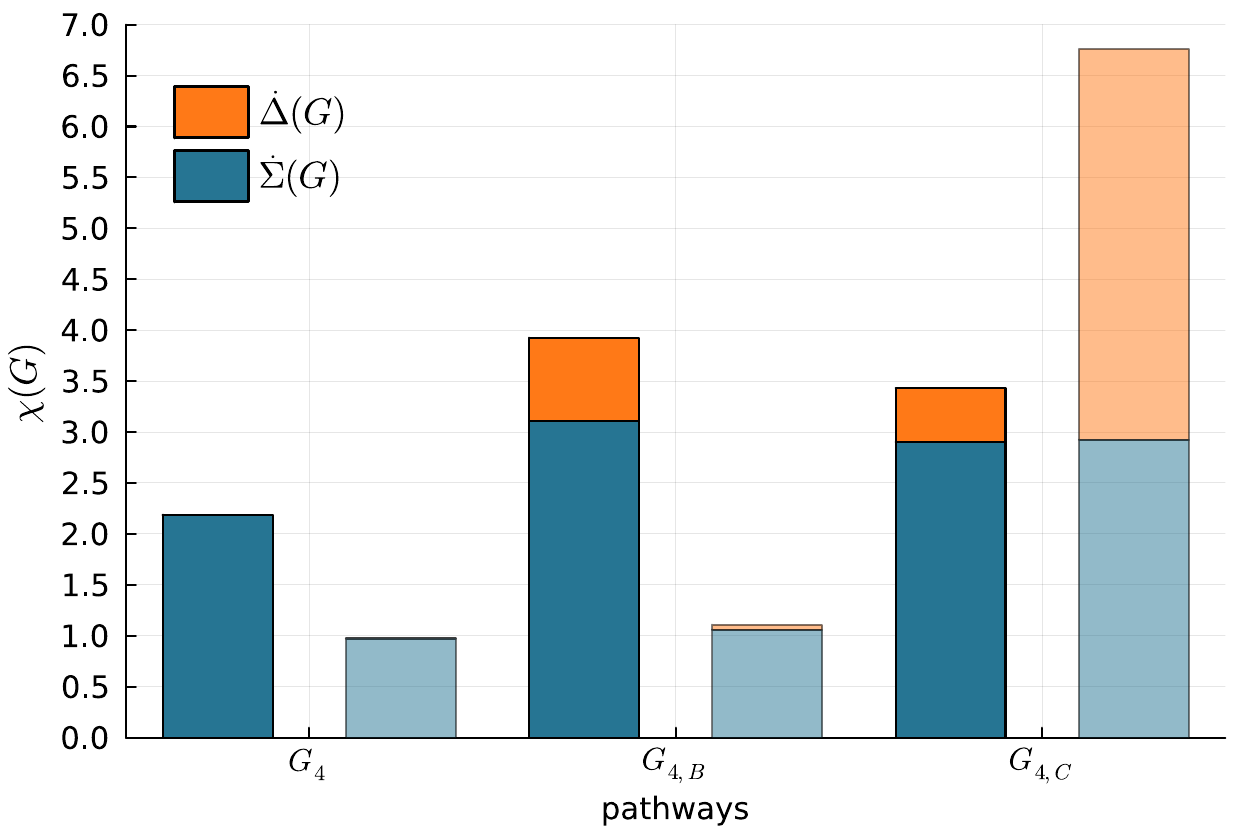}
                    \caption{}
                    \label{fig:abcd_cost2}
                \end{subfigure}
                
                \begin{subfigure}[c]{\textwidth}
                    \centering
                    \includegraphics[width=\linewidth]{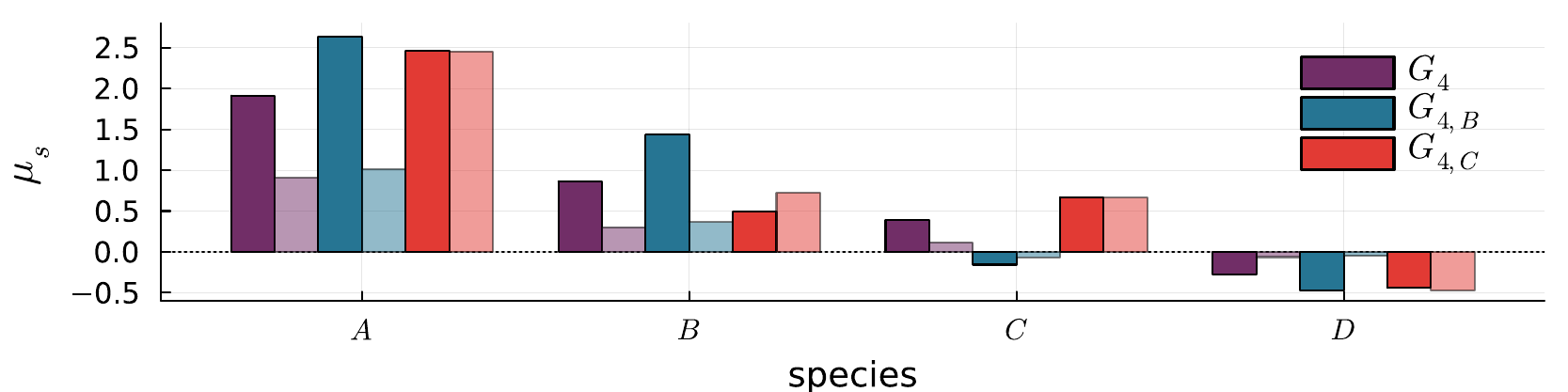}
                    \caption{}
                    \label{fig:abcd_conc2}
                \end{subfigure}
                \label{abcd2}
                \caption{
                    Collection of the results in the four species model for the asymmetric energy landscape with parameter settings: $\mu_{0}=(-2.0,-3.0,-4.0,-6.0)$, $E_{1}^{\ddagger}=(0.0,0.0,-1.0,-2.0)$;
                    and for the modified landscape: $E_{2}^{\ddagger}=(-2.0,0.0,-3.0,-2.0)$ ($\mu_{0}$ is the same).
                    \textbf{(a)} Visualization of the energy landscape (solid line) and the modified landscape (dashed line).
                    \textbf{(b)} Bar plot of the costs of the full CRN and the two reaction pathways (same bar composition as in Fig.~\ref{fig:abcd_cost1}).
                    \textbf{(c)} Bar plot of the chemical potential $\mu$ (Eq.\ \ref{eq:mu_chem_pot}) of each species for $\mcl{G}_{4}, \mcl{G}_{4,B}$ and $\mcl{G}_{4,C}$ (same color coding as in Fig.~\ref{fig:abcd_conc1}).
                    For both bar plots the opaque bars correspond to the initial energy landscape while the clear bars correspond to the results for the modified energy landscape.
                }
             \end{figure}
            
            \paragraph{Asymmetric energy landscape:} For this landscape (shown with solid curves in Fig.~\ref{fig:abcd_lndscp2}), the energies of formation $\mu_{0}$ are set such that the forward directions are favored and species \ce{B} is set to have a higher formational energy than species \ce{C}, i.e.: $\mu_{0}^{A}>\mu_{0}^{B}>\mu_{0}^{C}>\mu_{0}^{D}$. The transition state energies $E^{\ddagger}$ are set such that the kinetic barriers in the forward direction are equal for all reactions.\footnote{\added{Note that the Arrhenius law for
  reaction rates is a separation-of-scales property, between the
  elementary sampling frequencies (thermal prefactors) that establish
  the dimensions of rate and scale as powers (specifically, linear) of
  temperature~\cite{Eyring:ToRP:41}, and non-dimensional multipliers
  for first-passage time over reaction barriers that scale as
  exponentials of inverse temperature, an essential singularity with
  respect to polynomial expansions~\cite{Cardy:Instantons:78}; see
  also~\cite{Coleman:AoS:85}, Ch.~7.  The zero-barrier limit marks
  the dissolution of this separation of scales and describes processes
  such as simple diffusion.  Its use in a CRN model has the
  interpretation of enzymes that have evolved rates comparable to the
  diffusion rates for their substrates.  We adopt it here to furnish a
  reaction model parametrized by formation free energies of species,
  an equilibrium property that in real systems is generic, in contrast
  to kinetic parameters that are highly sensitive to \textit{ad hoc}
  molecular details and are most meaningfully addressed in relation to
  case-specific evolutionary or regulatory questions.  }} 
  The result of this assignment is that the rate constants $k_{r},k_{r}^{*}$ are different between $\mcl{G}_{4,B}$ and $\mcl{G}_{4,C}$. This energy landscape is then modified (blue dashed curve in Fig.~\ref{fig:abcd_lndscp2}) by lowering the transition state energies of the reactions in the $\mcl{G}_{4,B}$ pathway such that the kinetic barriers in the forward direction are effectively zero. An effect like this could be, for example, due to perfect \textit{catalysis} through specific types of enzymes.

        The thermodynamic costs for the different pathways in this model are shown in Fig.~\ref{fig:abcd_cost2}. It can be seen that the cost of $\mcl{G}_{4,B}$ is slightly higher than the cost of $\mcl{G}_{4,C}$, and they are both greater than $\mcl{G}_{4}$. For the modified landscape, however, the cost of $\mcl{G}_{4,B}$ is significantly lower than the cost of $\mcl{G}_{4,C}$, $\chi(\mcl{G}_{4,B})$, and is almost the same as the cost of the complete network $\chi(\mcl{G}_{4})$. Moreover, notice that the costs in the modified landscape, are much lower than the costs in the original landscape.

            \begin{figure}
                \centering
                \begin{subfigure}[c]{0.46\textwidth}
                    \centering
                    \includegraphics[width=\linewidth]{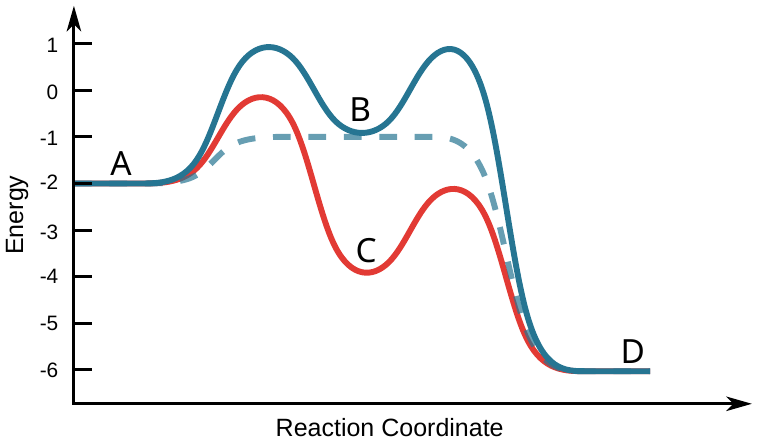}
                    \caption{}
                    \label{fig:abcd_lndscp3}
                \end{subfigure}
                \hspace{2ex}
                \begin{subfigure}[c]{0.46\textwidth}
                    \centering
                    \includegraphics[width=\linewidth]{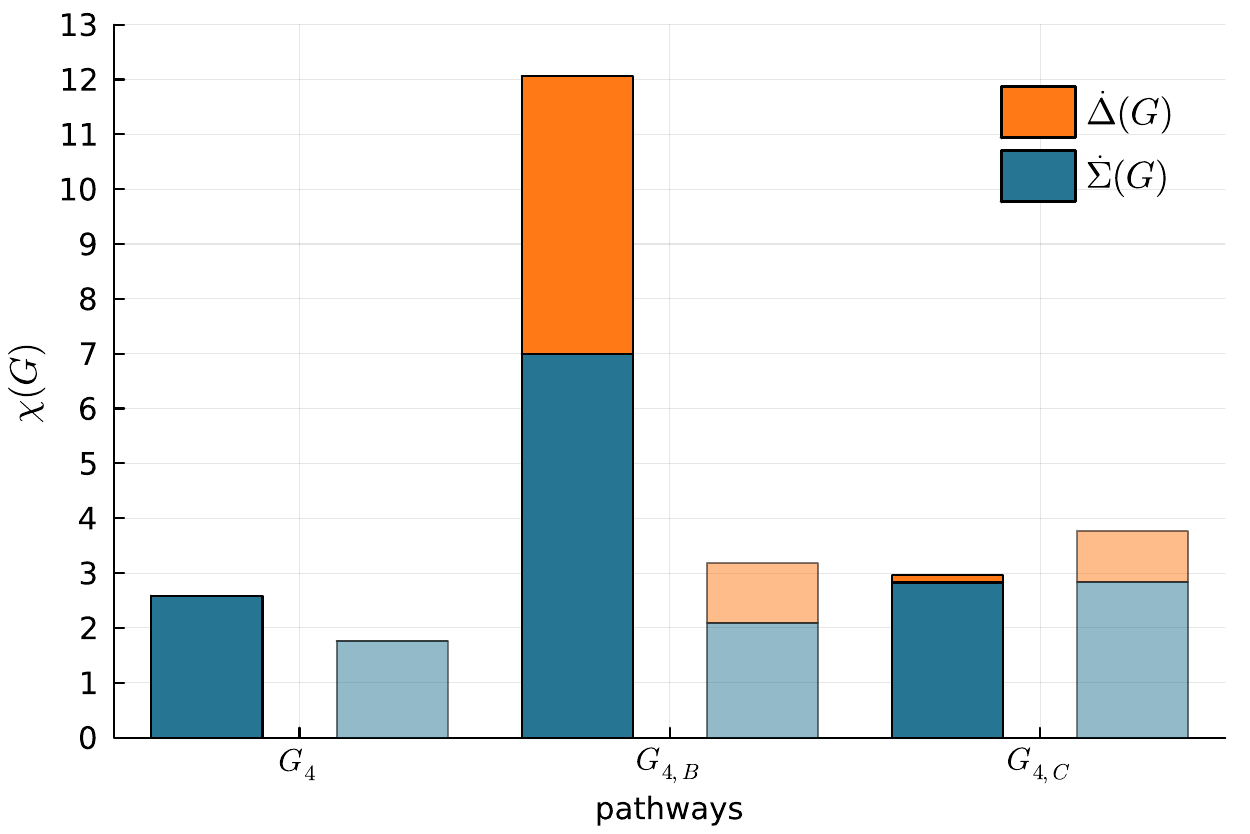}
                    \caption{}
                    \label{fig:abcd_cost3}
                \end{subfigure}
                
                \begin{subfigure}[c]{\textwidth}
                    \centering
                    \includegraphics[width=\linewidth]{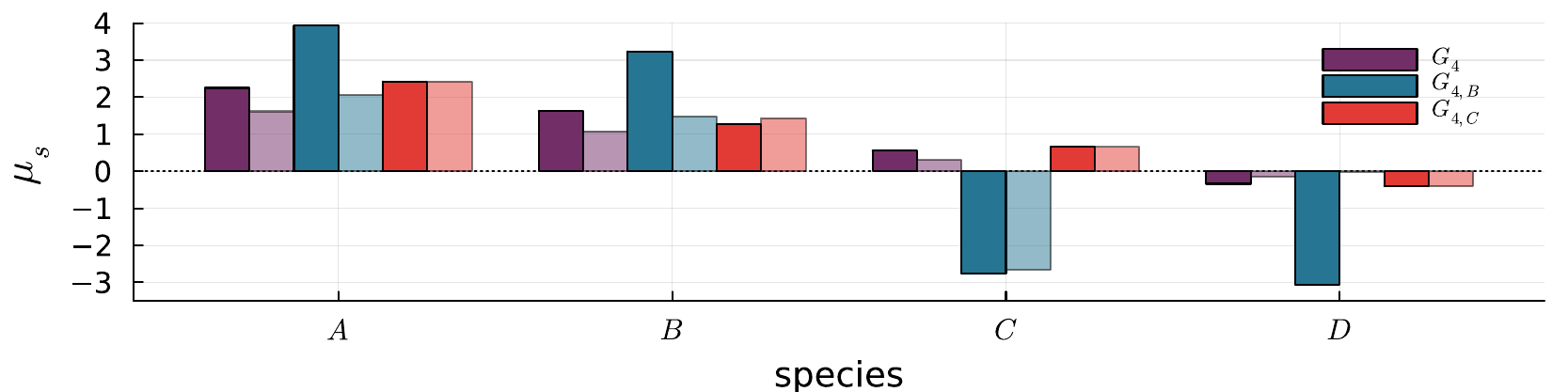}
                    \caption{}
                    \label{fig:abcd_conc3}
                \end{subfigure}
                \caption{
                    Collection of the results in the four species model for the strong asymmetric energy landscape with parameter settings: $\mu_{0}=(-2.0,-1.0,-4.0,-6.0)$, $E_{1}^{\ddagger}=(1.0,0.0,1.0,-2.0)$;
                    and for the modified landscape: $E_{2}^{\ddagger}=(-1.0,0.0,-1.0,-2.0)$ ($\mu_{0}$ is the same).
                    \textbf{(a)} Visualization of the energy landscape (solid line) and the modified landscape (dashed line).
                    \textbf{(b)} Bar plot of the cost of the full CRN and the two reaction pathways.
                    \textbf{(c)} Bar plot of the chemical potential $\mu$ (Eq.\ \ref{eq:mu_chem_pot}) of each species for $\mcl{G}_{4}, \mcl{G}_{4,B}$ and $\mcl{G}_{4,C}$.
                    As in Fig.~\ref{fig:abcd_lndscp2}, the opaque bars correspond to the initial energy landscape while the clear bars correspond to the results for the modified energy landscape.
                }
                \label{fig:abcd3}
            \end{figure}

            \paragraph{Strongly asymmetric energy landscape:} As shown in Fig.~\ref{fig:abcd_lndscp3}, the energies of formation $\mu_{0}$ are set such that species \ce{B} is set to have a higher formational energy than both \ce{A} and \ce{C}, i.e.: $\mu_{0}^{B}>\mu_{0}^{A}>\mu_{0}^{C}>\mu_{0}^{D}$. Thus, while the formation of $D$ is favored, the reaction \ce{A -> B} is favored in the reverse direction. This energy landscape is again modified by lowering the transition state energies of the reactions in the $\mcl{G}_{4,B}$ pathway such that 
            % the kinetic barriers in the forward direction are effectively zero
            the kinetic barrier of the \ce{A -> B} reaction in the reverse direction and the kinetic barrier of the \ce{B -> D} reaction in the forward direction (i.e. the two favored directions)
            are effectively zero (blue dashed curve in Fig.~\ref{fig:abcd_lndscp3}). As explained earlier, this can be seen as a particular pathway being catalyzed. 

            As shown in Fig.~\ref{fig:abcd_cost3}, the cost of $\mcl{G}_{4,B}$ is significantly higher than the cost of $\mcl{G}_{4,C}$ ($\chi(\mcl{G_{4,C}})$ which is almost the same as $\chi(\mcl{G}_{4})$). However, for the modified network, the cost of $\mcl{G}_{4,B}$ is once again lower than the cost of $\mcl{G}_{4,C}$.

            \paragraph{Comment on catalysis: }
            The two asymmetric models above demonstrate that in a network with multiple pathways, catalyzing a pathway can significantly reduce the thermodynamic cost of the NESS associated with that pathway. Furthermore, the thermodynamic cost remains largely unchanged if all uncatalyzed pathways are blocked. As our examples illustrate, this cost reduction occurs regardless of whether the uncatalyzed reaction is favored. It can also be seen that the cost of blocking the catalyzed pathway is significantly higher than the cost of blocking its uncatalyzed counterpart. We hypothesize that these findings are general and discuss their implications for the specificity of biological pathways in Sec.\ \ref{sec:discussion}.

            \begin{comment}
                
            the following settings are chosen:
            \begin{itemize}
                \item 
                \item the transition state energies $E^{\ddagger}$ are set such that the kinetic barriers in the forward direction are equal for all reactions
                \item the result of this ``strongly asymmetric'' energy landscape is that the rate constants $k_{r},k_{r}^{*}$ are partially different between $\mcl{G}_{4,B}$ and $\mcl{G}_{4,C}$
            \end{itemize}
            The results of this setting for the energy landscape are the following:
            \begin{itemize}
                \item as shown in Fig.~\ref{fig:abcd_cost3}, the cost of $\mcl{G}_{4,B}$ is significantly higher than the cost of $\mcl{G}_{4,C}$ ($\chi(\mcl{G_{4,C}})$ is almost the same as $\chi(\mcl{G}_{4})$)
                \item additionally, Fig.~\ref{fig:abcd_conc2} shows the chemical potentials of each species within the different pathways
            \end{itemize}
            
            The results of this modified energy landscape are the following:
            \begin{itemize}
                \item as shown in Fig.~\ref{fig:abcd_cost3}, the cost of $\mcl{G}_{4,B}$ is now lower than the cost of $\mcl{G}_{4,C}$ (albeit not as pronounced).
                \item Fig.~\ref{fig:abcd_conc3} shows the chemical potentials of each species within the different pathways for this modified landscape
            \end{itemize}

        \end{comment}

        \subsubsection{Five species model}

            \begin{figure}[!htb]
                \centering
                \begin{subfigure}[c]{0.3\textwidth}
                    \centering
                    \includegraphics[width=\textwidth]{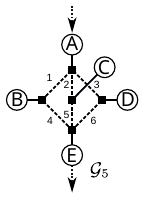}
                    \caption{}
                    \label{fig:abcde_crn}
                \end{subfigure}
                \hfil
                \begin{subfigure}[c]{0.5\textwidth}
                    \centering
                    \includegraphics[width=\textwidth]{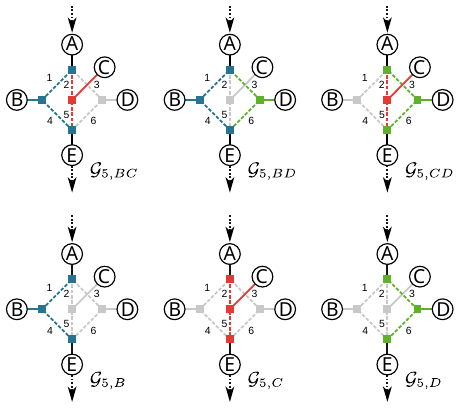}
                    \caption{}
                    \label{fig:abcde_subs}
                \end{subfigure}
                \caption{
                    Representation of the possible reaction pathways in the five species model that support $\vext=(-1,0,0,0,1)$:
                    \textbf{(a)} full CRN $\mcl{G}_{5}$,
                    \textbf{(b)} collection of all subgraphs of $\mcl{G}_{5}$ that support $\vext$.
                    Edge-labels represent the index of the reaction in Eq.~\ref{eq:abcde_react}.
                }
                \label{fig:abcde}
            \end{figure}
            In this subsection, we give an example of the result derived in Sec.\ \ref{sec:nondec_discovery} that the restriction cost of a pathway nested inside another pathway is always higher than the restriction cost of the embedding pathway if the complement of the embedding pathway is also a pathway. 
            Consider the CRN $\mcl{G}_{5}$ visualized in Fig.~\ref{fig:abcde_crn} with five species and 6 reversible reactions:
            \begin{equation}
                \begin{aligned}
                \ce{A &<=>[k_1][k_1^{*}] B <=>[k_4][k_4^{*}] E}\\
                \ce{A &<=>[k_2][k_2^{*}] C <=>[k_5][k_5^{*}] E}\\
                \ce{A &<=>[k_3][k_3^{*}] D <=>[k_6][k_6^{*}] E}.
                \end{aligned}
                \label{eq:abcde_react}
            \end{equation}
\begin{comment}
    
            The stoichiometric matrix $\mbb{S}_{5}$ associated with this CRN is given as:
            \[
                \mbb{S}_{5}=\begin{pNiceMatrix}[r,first-col,first-row]
                  &r_{1}&r_{1}^{*}&r_{2}&r_{2}^{*}&r_{3}&r_{3}^{*}&r_{4}&r_{4}^{*}&r_{5}&r_{5}^{*}&r_{6}&r_{6}^{*}\\
                A &-1& 1&-1& 1&-1& 1& 0& 0& 0& 0& 0& 0\\
                B & 1&-1& 0& 0& 0& 0&-1& 1& 0& 0& 0& 0\\
                C & 0& 0& 1&-1& 0& 0& 0& 0&-1& 1& 0& 0\\
                D & 0& 0& 0& 0& 1&-1& 0& 0& 0& 0&-1& 1\\
                E & 0& 0& 0& 0& 0& 0& 1&-1& 1&-1& 1&-1
                \end{pNiceMatrix}.
            \]
            From $\mbb{S}_{5}$ one can find that the CRN has one conservation law: $\mbb{L}^{T}=(1,1,1,1,1)$ (like in the four species model the sum of all species concentrations is conserved). 
\end{comment}

            We designate species \ce{A} and \ce{E} as the in- and out-flowing species, respectively, while the species \ce{B}, \ce{C} and \ce{D} serve as intermediates, i.e.: $\vext=(-1,0,0,0,1)^T$ in the basis $(\ce{A},\ce{B},\ce{C},\ce{D},\ce{E})$. As shown in Fig.~\ref{fig:abcde_subs}, along with $\mcl{G}_{5}$, six additional pathways can support $\vext$. Three pathways require blocking out two reactions of the CRN (upper row of Fig.~\ref{fig:abcde_subs}) and three pathways require blocking out four reactions of the CRN (lower row of Fig.~\ref{fig:abcde_subs}). A close inspection of Fig.~\ref{fig:abcde_subs} reveals that the following relationships between these pathways hold: 
            \begin{align*}
            \mcl{G}_{5,B}\subset\mcl{G}_{5,BC},&\quad \mcl{G}_{5,B}\subset \mcl{G}_{5,BD},\\ \mcl{G}_{5,C}\subset\mcl{G}_{5,BC},&\quad \mcl{G}_{5,C}\subset\mcl{G}_{5,CD},\\ \mcl{G}_{5,D}\subset\mcl{G}_{5,BD},
            &\quad \mcl{G}_{5,D}\subset
            \mcl{G}_{5,CD}.
            \end{align*}
            Finally, observe that the complements $\mcl{G}_{5}/\mcl{G}_{5,BC},\mcl{G}_{5}/\mcl{G}_{5,BD},\mcl{G}_{5}/\mcl{G}_{5,CD}$ also support $\vext$. i.e. are again pathways\footnote{
                    In fact, $\mcl{G}_{5}/\mcl{G}_{5,BC}=\mcl{G}_{5,D},\,\mcl{G}_{5}/\mcl{G}_{5,BD}=\mcl{G}_{5,C}$ and $\mcl{G}_{5}/\mcl{G}_{5,CD}=\mcl{G}_{5,B}$
            }

            %will demonstrate the nondecrease of thermodynamic cost of nested pathways 
            %This means that the reaction system is a good model to test out the theorem of increasing cost on nested subgraphs from Sec. 3.2.

            \begin{figure}[!htb]
                \centering
                \begin{subfigure}[c]{0.45\textwidth}
                    \centering
                    \includegraphics[width=\textwidth]{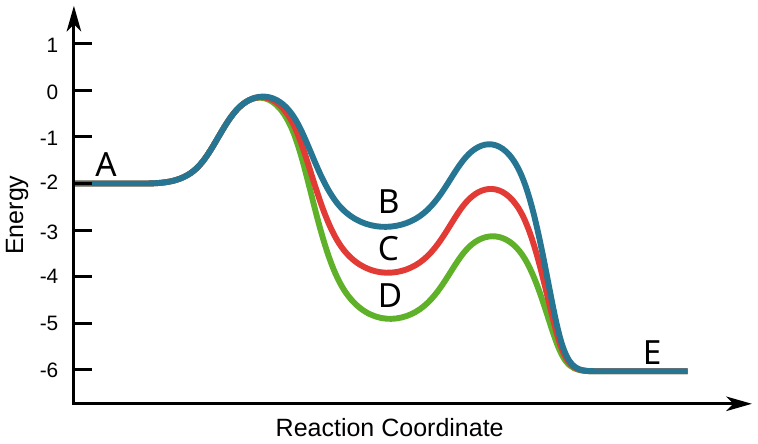}
                    \caption{}
                    \label{fig:abcde_lndscp1}
                \end{subfigure}
                \hfil
                \begin{subfigure}[c]{0.45\textwidth}
                    \centering
                    \includegraphics[width=\textwidth]{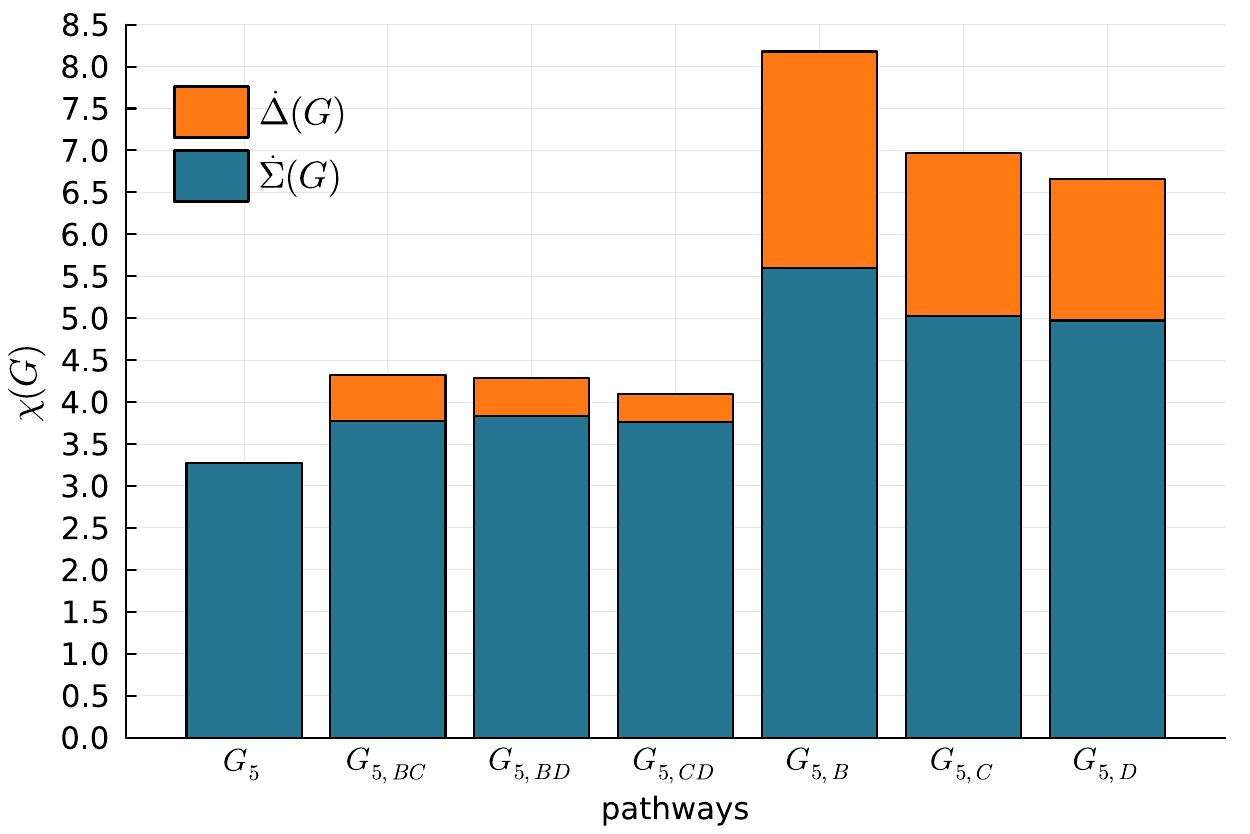}
                    \caption{}
                    \label{fig:abcde_cost1}
                \end{subfigure}
                \caption{
                    Results in the five species model for the energy landscape with parameter settings: $\mu_{0}=(-2.0,-3.0,-4.0,-5.0,-6.0), E^{\ddagger}=(0.0,0.0,0.0,-1.0,-2.0,-3.0)$.
                    \textbf{(a)} Visualization of the energy landscape.
                    \textbf{(b)} Bar plot of the cost for the possible reaction pathways: the bar height corresponds to thermodynamic cost $\chi(\mcl{G})$ while the bar composition corresponds to the maintenance cost $\dot{\Sigma}(\mcl{G})$ (blue) and the restriction cost $\dot{\Delta}(\mcl{G})$ (orange).
                    Reaction pathways, (sub)-graphs are grouped on the x-axis by increasing number of blocked out reactions.
                }
                \label{fig:abcde_res}
            \end{figure}
        
            Similar to the four-species model, an asymmetric energy landscape (shown in Fig.~\ref{fig:abcde_lndscp1}) is chosen such that the energies of formation $\mu_{0}$ the forward reactions are favored: $\mu_{0}^{A}>\mu_{0}^{B}>\mu_{0}^{C}>\mu_{0}^{D}>\mu_{0}^{E}.$
            The transition state energies $E^{\ddagger}$ are set such that the kinetic barriers in the forward direction are equal for all reactions.
            The stoichiometric compatibility class defined in Eq.~\ref{eq:cost_pathway} is set to $\mbb{L}^{T}\eq=20.0$ and the optimization problem defined in Eq.~\ref{eq:cost_pathway} is solved to calculate the cost of the full CRN $\chi(\mcl{G}_{5})$ as well as all its nested pathways.

            The results from the calculations are summarized in Fig.~\ref{fig:abcde_cost1}. It can be seen that the costs for the pathways that require blocking out two reactions are, in general, lower than their counterparts for the pathways that require blocking out four reactions, i.e. $\chi(\mcl{G}_{5,ij})>\chi(\mcl{G}_{5,i})>\chi(\mcl{G}_{5})$, $\dot{\Sigma}(\mcl{G}_{5,ij})>\dot{\Sigma}(\mcl{G}_{5,i})>\dot{\Sigma}(\mcl{G}_{5})$, and $\dot{\Delta}(\mcl{G}_{5,ij})>\dot{\Delta}(\mcl{G}_{5,i})>\dot{\Delta}(\mcl{G}_{5})$, for $i,j \in \{B,C,D\}$. \deleted{$\chi(\mcl{G}_{5,ij})<\chi(\mcl{G}_{5,i})<\chi(\mcl{G}_{5})$, $\dot{\Sigma}(\mcl{G}_{5,ij})<\dot{\Sigma}(\mcl{G}_{5,i})<\dot{\Sigma}(\mcl{G}_{5})$, and $\dot{\Delta}(\mcl{G}_{5,ij})<\dot{\Delta}(\mcl{G}_{5,i})<\dot{\Delta}(\mcl{G}_{5})$, for $i,j \in {B,C,D}$.} In particular, this verifies the results proved in Sec.\ \ref{sec:non_dec_nested_path}.

    \subsection{Multimolecular CRNs: competing autocatalytic cycles}
    \label{sec:multi}
    
            \begin{figure}[!htb]
                \centering
                \begin{subfigure}[c]{0.32\textwidth}
                    \centering
                    \includegraphics[width=\textwidth]{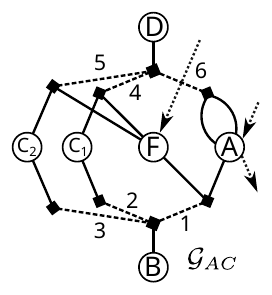}
                    \caption{}
                    \label{fig:ACcomp_crn}
                \end{subfigure}
                \hfill
                \begin{subfigure}[c]{0.32\textwidth}
                    \centering
                    \includegraphics[width=\textwidth]{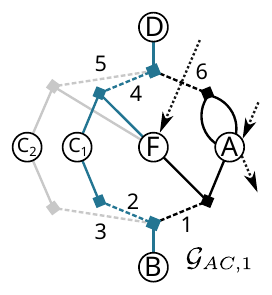}
                    \caption{}
                    \label{fig:ACcomp_sub1}
                \end{subfigure}
                \hfill
                \begin{subfigure}[c]{0.32\textwidth}
                    \centering
                    \includegraphics[width=\textwidth]{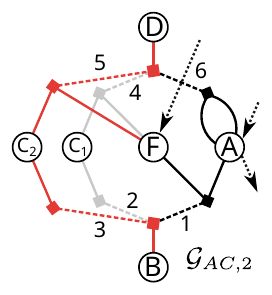}
                    \caption{}
                    \label{fig:ACcomp_sub2}
                \end{subfigure}
                \caption{
                    Representation of the different possible reaction pathways in the competing AC-cycles model:
                    \textbf{(a)} full CRN,
                    \textbf{(b)} sub-CRN representing cycle-1,
                    \textbf{(c)} sub-CRN representing cycle-2.
                }
                \label{fig:ACcomp}
            \end{figure}
            
    In this subsection, we study a CRN composed of two competing autocatalytic (AC) cycles: 
            \begin{equation}
                \begin{alignedat}{3}
                \ce{F + A &<=>[k_1][k_1^{*}] B} &\qquad \ce{F + C1 &<=>[k_4][k_4^{*}] D}\\
                \ce{B &<=>[k_2][k_2^{*}] C1} & \ce{F + C2 &<=>[k_5][k_5^{*}] D}\\
                \ce{B &<=>[k_3][k_3^{*}] C2} & \ce{D &<=>[k_6][k_6^{*}] 2A}.\\
                \end{alignedat}
                \label{eq:ACcomp_react}
            \end{equation}
        The CRN $\mcl{G}_{AC}$ that is formed by this system of reaction equations is visualized\footnote{
                Note: as we are now dealing with multimolecular reactions we use the K{\"o}nig-represntation for a CRN
                \cite{zykov:1974,andersen2021defining}. As above, the edges are drawn undirected because all reactions are assumed to be reversible.
            } in Fig.~\ref{fig:ACcomp_crn}.
            The stoichiometric matrix $\mbb{S}_{AC}$ associated with this CRN is given as:
            \[
                \mbb{S}_{AC}=\begin{pNiceMatrix}[r,first-col,first-row]                  &r_{1}&r_{1}^{*}&r_{2}&r_{2}^{*}&r_{3}&r_{3}^{*}&r_{4}&r_{4}^{*}&r_{5}&r_{5}^{*}&r_{6}&r_{6}^{*}\\
                F     &-1& 1& 0& 0& 0& 0&-1& 1&-1& 1& 0& 0\\
                A     &-1& 1& 0& 0& 0& 0& 0& 0& 0& 0& 2&-2\\
                B     & 1&-1&-1& 1&-1& 1& 0& 0& 0& 0& 0& 0\\
                C_{1} & 0& 0& 1&-1& 0& 0&-1& 1& 0& 0& 0& 0\\
                C_{2} & 0& 0& 0& 0& 1&-1& 0& 0&-1& 1& 0& 0\\
                D     & 0& 0& 0& 0& 0& 0& 1&-1& 1&-1&-1& 1
                \end{pNiceMatrix}.
            \]
            From $\mbb{S}_{AC}$ one can find that the CRN has one conservation law: $\mbb{L}^{T}=(1,2,3,3,3,4)$.

            For an analysis of NESSs, \ce{F} (``food source'') is designated as the in-flowing species, \ce{A} (the auto-catalyst) is designated as the out-flowing species, and the remaining species serve as internal species, i.e.: $\vext=(-2,1,0,0,0,0)$. As shown in Fig.~\ref{fig:ACcomp}, in addition to $\mcl{G}_{AC}$ there are two nested pathways that can support $\vext$: $\mcl{G}_{AC,1}$ and $\mcl{G}_{AC,2}$ representing the pathway going through the intermediate \ce{C1} and \ce{C2}, respectively, (see Fig.~\ref{fig:ACcomp_sub1} and Fig.~\ref{fig:ACcomp_sub2}).
            for simplicity the reaction rate constants are all set to: \[k_{r}=k_{r}^{*}=1.0\quad \forall r.\]
            Solving the optimization problem defined in Eq.~\ref{eq:cost_pathway} allows us to calculate the NESSs and their associated costs of the full CRN $\chi(\mcl{G}_{AC})$ as well as the two reaction pathways $\chi(\mcl{G}_{AC,1})$ and $\chi(\mcl{G}_{AC,2})$. The stoichiometric compatibility class defined in Eq.~\ref{eq:cost_pathway} is set to $\mbb{L}^{T}\eq=20.0$.

            \begin{figure}[!htb]
                \centering
                \begin{subfigure}[c]{0.45\textwidth}
                    \centering
                    \includegraphics[width=\textwidth]{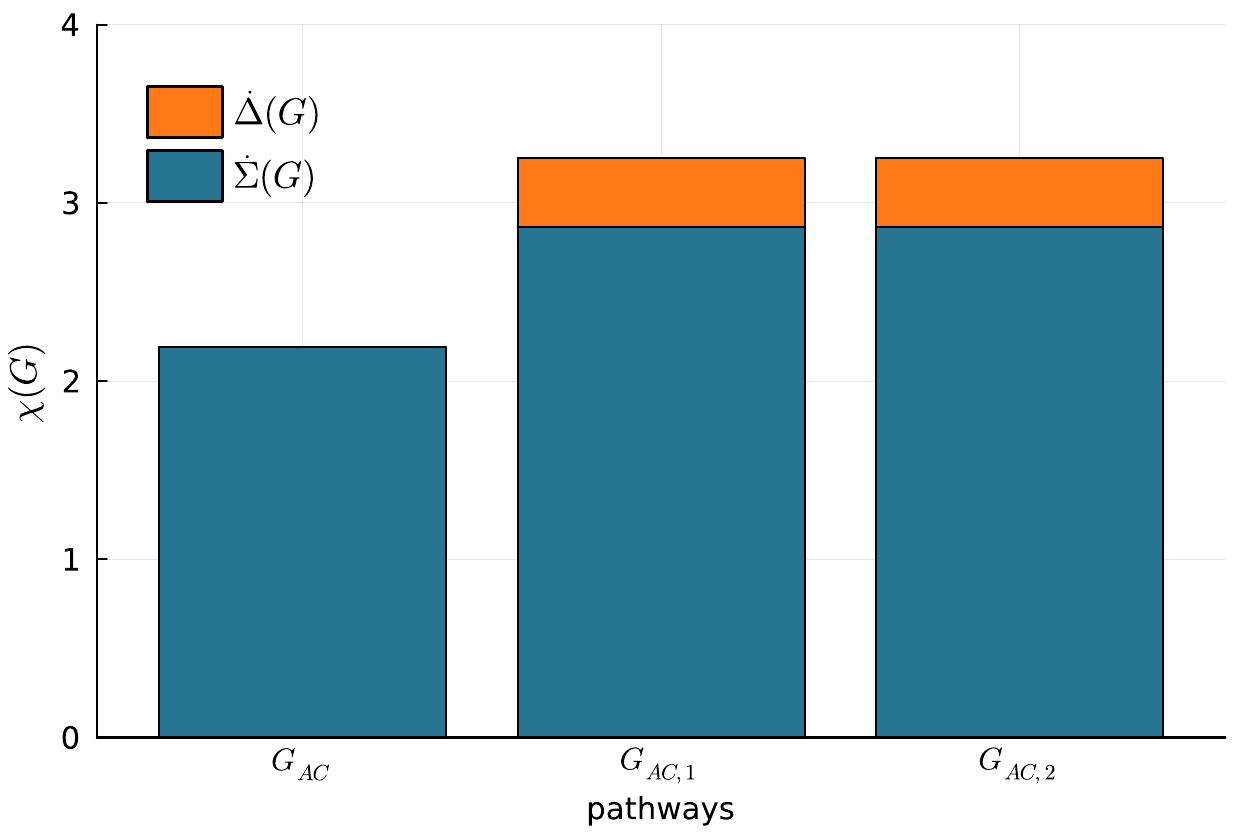}
                    \caption{}
                    \label{fig:ACcomp_cost1}
                \end{subfigure}
                \hfil
                \begin{subfigure}[c]{0.45\textwidth}
                    \centering
                    \includegraphics[width=\textwidth]{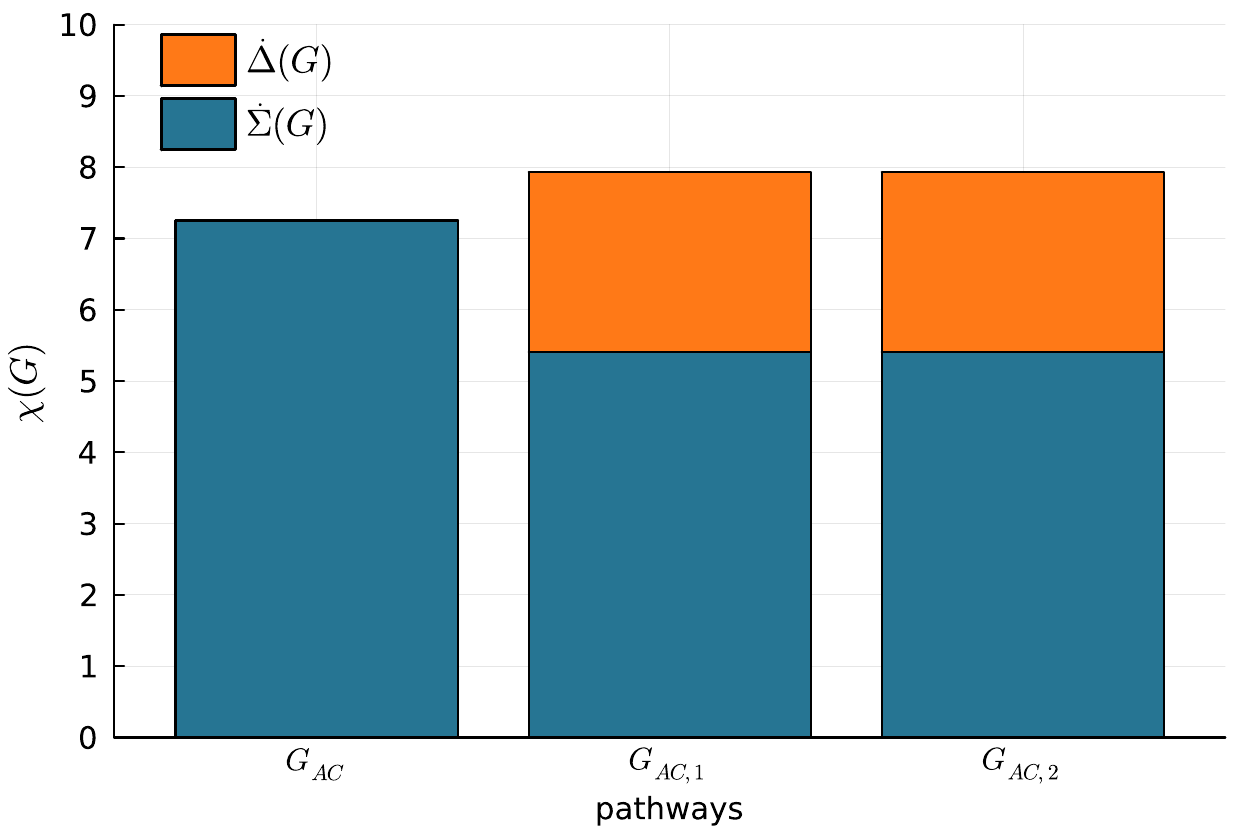}
                    \caption{}
                    \label{fig:ACcomp_cost2}
                \end{subfigure}
                
                \begin{subfigure}[c]{0.45\textwidth}
                    \centering
                    \includegraphics[width=\textwidth]{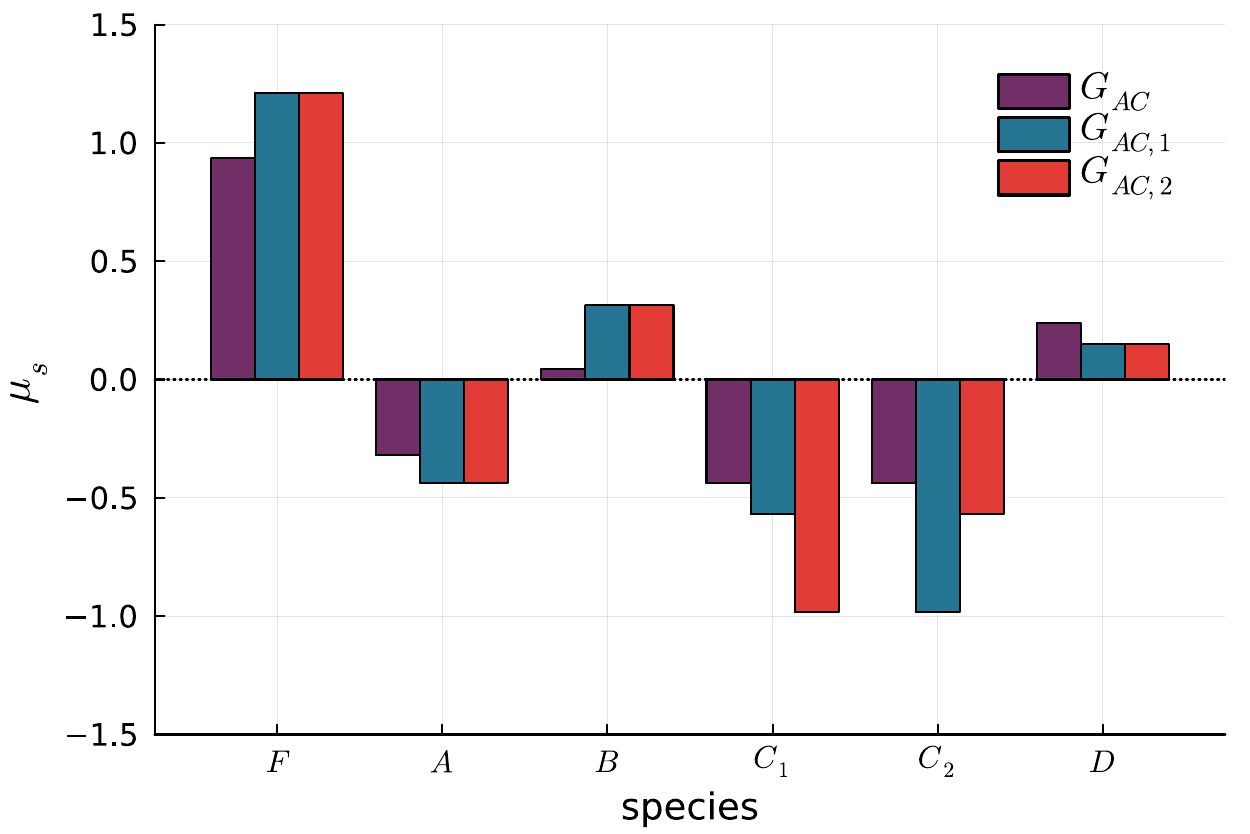}
                    \caption{}
                    \label{fig:ACcomp_conc1}
                \end{subfigure}
                \hfil
                \begin{subfigure}[c]{0.45\textwidth}
                    \centering
                    \includegraphics[width=\textwidth]{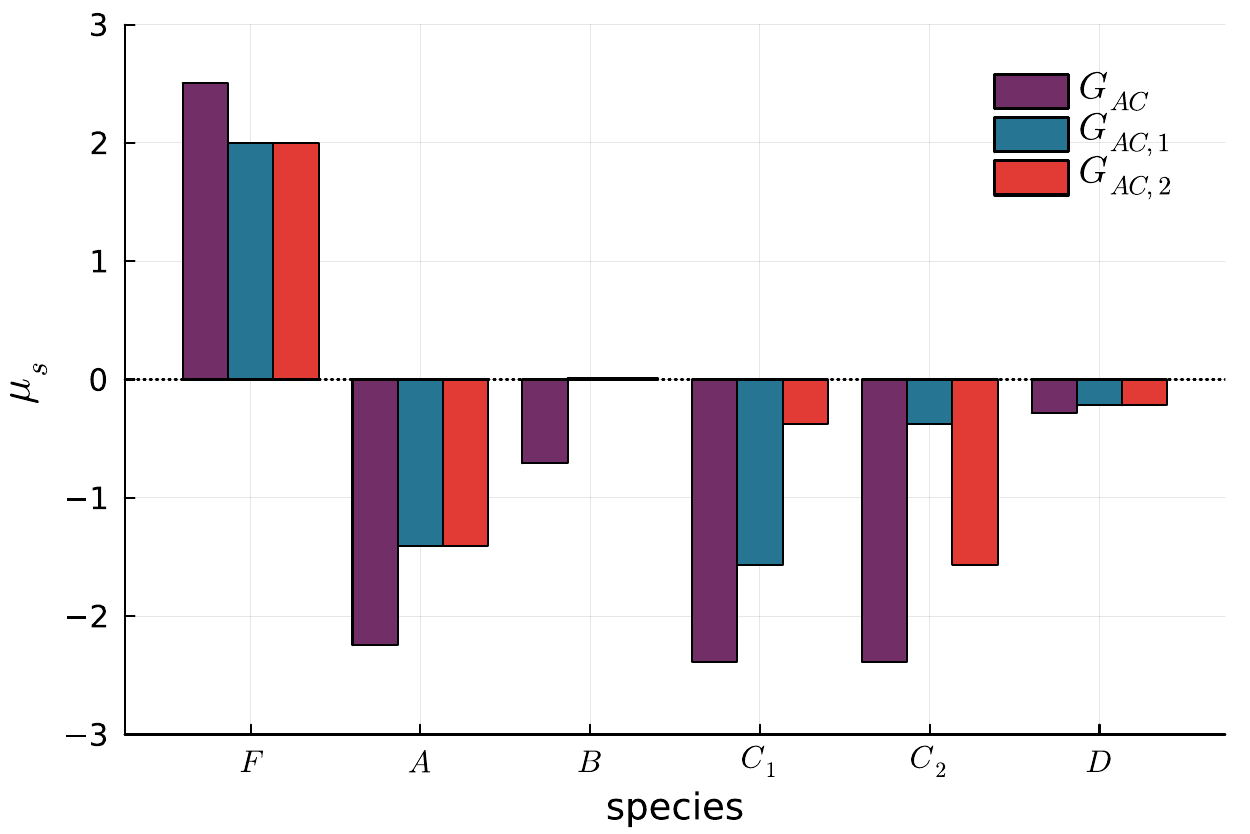}
                    \caption{}
                    \label{fig:ACcomp_conc2}
                \end{subfigure}
                \caption{
                    Results in the competing autocatalytic cycles model.
                    \textbf{(a)} Bar plot of the cost for the possible reaction pathways for the NESS closer to the detailed balance equilibrium \added{$\eq$} \deleted{eq}.
                    \textbf{(b)} Bar plot of the cost for the possible reaction pathways for the NESS away from the detailed balance equilibrium.
                    For both figures: the bar height corresponds to thermodynamic cost $\chi(\mcl{G})$ while the bar composition corresponds to the maintenance cost $\dot{\Sigma}(\mcl{G})$ (blue) and the discovery cost $\dot{\Delta}(\mcl{G})$ (orange).
                    \textbf{(c)} Bar plot of the chemical potential $\mu$ (Eq.\ \ref{eq:mu_chem_pot}) of each species for $\mcl{G}_{AC}, \mcl{G}_{AC,1}$ and $\mcl{G}_{AC,2}$ for the NESS closer to the detailed balance equilibrium \added{$\eq$} \deleted{eq}.
                    \textbf{(d)} Bar plot of the chemical potentials for the NESS away from the detailed balance equilibrium.
                    }
                \label{fig:ACcomp_res}
            \end{figure}

        The results of our analysis are shown in Fig.~\ref{fig:ACcomp_res} and are as follows. We find that, similar to the example in Sec.\ \ref{sec:higher_order}, each pathway admits two NESSs. For the NESSs closer to the detailed balance equilibrium $\eq$ of the system, both the thermodynamic cost as well as the EPR of the nested pathways is higher than that of the full CRN, i.e. $\chi(\mcl{G}_{AC})<\chi(\mcl{G}_{AC,1}),\chi(\mcl{G}_{AC,2})$ \deleted{$\chi(\mcl{G}_{AC,1})$} and $\dot{\Sigma}(\mcl{G}_{AC})<\dot{\Sigma}(\mcl{G}_{AC,1}),\dot{\Sigma}(\mcl{G}_{AC,2})$ \deleted{$\dot{\Sigma}(\mcl{G}_{AC,1})$} ( see Fig.~\ref{fig:ACcomp_cost1}). On the other hand, for the NESSs away from the detailed balance equilibrium, while the cost of the nested pathways is still higher than for the full CRN, the EPR of the subgraphs is lower, i.e. $\chi(\mcl{G}_{AC})<\chi(\mcl{G}_{AC,1}),\chi(\mcl{G}_{AC,2})$ \deleted{$\chi(\mcl{G}_{AC,1})$} and $\dot{\Sigma}(\mcl{G}_{AC})>\dot{\Sigma}(\mcl{G}_{AC,1}),\dot{\Sigma}(\mcl{G}_{AC,2})$ \deleted{$\dot{\Sigma}(\mcl{G}_{AC,2})$} (see Fig.~\ref{fig:ACcomp_cost2}).

\section{Discussion}
\label{sec:discussion}

The versatility of CRNs as a model class within and outside chemistry,
and the feature that makes a concept of pathway ranking well-defined,
is \emph{compositionality}: subnetworks may be aggregated to produce
embedding networks in which any subnetwork is part of the environment
for the others, and in which ecological concepts such as competition,
mutualism, or cooperativity are defined~\cite{peng2020ecological}.
Distinct pathways may correspond to alternative metabolic sequences in
a cell or organism~\cite{Palsson:systems_bio:06}, or to alternative
populations of trophic niches by species in an
ecosystem~\cite{Dunne:foodwebstats:02,Sterner:eco_stoich:02}.
Because the abstraction does not change between the parts and the
whole, CRNs define a stoichiometric notion of open subnetworks, which
we have used here to implement general constant-current boundary
conditions, and which offer a systematic way to refine environmental
boundary constraints~\cite{smith2024rules} beyond an open-system
prescription based only on the conservation laws of a subnetwork, like
the one used (under the definition of \textit{chemical
  transformations}) to study free-energy transduction
in~\cite{Wachtel:transduction:22}.

The topological compositionality of stoichiometric graphs is joined in
our treatment with a second equally important compositionality: that
of partition functions in statistical ensembles.  Unlike probability
densities (which must be re-normalized at each change in the scale at
which a system is described), partition functions simply
nest~\cite{Smith:Intr_Extr:20}, when the accessible state spaces
available to microvariables conditioned on fixed values of
higher-scale constraints or control parameters are subsumed in
ensembles over the values of those conditioning variables.  This
nesting gives rise to relations such as the chain rule for entropies,
and the relations of \added{Hartley
  entropies~\cite{Hartley:information:28} (the log-partition functions
  of ensembles, see~\cite{Smith:Intr_Extr:20})} to relative entropies
in their ensembles.  Nesting makes it coherent for us to assign cost
rate-functions to the operation of stochastic processes at a
microscale, which retain their definitions as conditional information
summands within information (log-improbability) measures for larger
ensembles that can be constructed to model the process of design
selection or feedback control, such as a biological population
process~\cite{Smith:evo_games:15} selecting over alternative enzyme
specificities for the metabolisms of the member organisms.

Making use of both aspects of compositionality here, and generalizing
the role of the Gibbs free energy (divided by $k_B T$) from
equilibrium thermodynamics to path ensembles, much as is done in
stochastic
thermodynamics~\cite{Jarzynski:neq_FE_diffs:97,Crooks:NE_work_relns:99},
we use the large-deviation rate function for an atypical current
relative to the mass-action flux to define a continuously-accruing
improbability cost for a current anchored in a throughput requirement
and a restriction constraint.  The NESSs associated with a pathway
correspond to the points where the cost function achieves a local
minimum.  This cost function further decomposes into two components:
the maintenance cost, which measures the minimum log-improbability of
maintaining the NESS (and here coincides with an entropy-production
rate), and the restriction cost, which quantifies the minimum
log-improbability of \added{forfeiting all events through
  any} reactions not included in the pathway.  Our cost function is
well-defined for reversible CRNs operating under any kinetic
framework, and is not restricted to mass-action kinetics.

By recasting mass-action kinetics in the framework of Ohm's law, we
have demonstrated that for small driving-departures from equilibrium,
the resistance of a CRN decreases as the number of pathways supporting
the same throughput current increases.  The maintenance cost
corresponds to the power dissipated by an electrical
network,\footnote{Ohmic networks are, of course, an instance of CRNs
in the linear-response regime.  The space-filling electric field
permits distributed voltage drop, maintaining the linear-response
relation between current and the voltage gradient, and electrons at
different spatial nodes in a network are effectively distinct species,
a device for modeling spatial compartmentation in CRNs more
generally.} and is always greater for a nested pathway than for its
embedding pathway.  The restriction cost, being proportional to the
EPR in the linear regime, is then likewise non-decreasing under
recursive nesting.

We illustrate these findings using four- and five-species unimolecular
CRNs.  While the cost of a nested pathway is always bounded below by
the cost of its embedding pathway, our examples demonstrate that this
lower bound can be approached closely.  For a CRN composed of two
non-overlapping nested pathways, if the thermodynamic landscape of
formation energies makes one pathway costlier than the other,
introducing an effective catalyst that collapses the reaction barriers
can make the unfavorable pathway favorable.  The cost of the catalyzed
pathway can be made to approach that of the embedding pathway by
increasing barrier heights or inhibiting the alternative pathway.
Thus, catalysts and inhibitors can significantly alter pathway costs,
effectively overriding the original uncatalyzed energy landscape.

Alternatively, the topology of the network and the indirectness of a
pathway can cause the elimination of some reactions to be less
impactful on network resistance than others, a result that was shown
in~\cite{smith2024rules} to single out the Calvin-Benson cycle for carbon
fixation as the least costly to implement among those using comparable
reaction mechanisms~\cite{andersen:2019}.  For a throughput
function such as sugar-group shuffling in a carbon-fixation pathway,
which gates all other cellular processes, the relative costs among
alternative pathways translate directly into selective advantages in
conventional terms of free-energy savings.

Multimolecular CRNs, pervasive in studies of biological systems, can
exhibit
\added{behavior at unstable saddle fixed points differeng
  from their behavior near stable attractors.}
We have shown that among NESSs operating on
far-from-equilibrium branches of the concentration-current curve, the
thermodynamic cost and the maintanence cost
\added{to drive a nested pathway near its unstable fixed
  point can be lower than their counterparts for its hosting pathway.
  We do not currently know the parameter conditions under which this
  reversal can arise, and we have not shown that it is restricted only
  to unstable fixed points, though we conjecture that the latter is
  the case.}

To conclude we mention possible avenues for future research.

More can be done to understand CRNs through the mapping to electric
circuits~\cite{avanzini2023circuit,raux2024thermodynamic} and to use
stoichiometric methods to inform circuit
theory~\cite{cardelli2020electric}.  In electrical network theory, the
non-decrease of power dissipation in embedding circuits has been
established by counting spanning
trees~\cite{loebl2010discrete}.  Similar graph-theoretic concepts have
recently been developed for CRNs~\cite{dal2023geometry}.  Applying
these methods to the ranking of maintenance costs for nested pathways
may offer useful mathematical and physical insights.

An obvious next step is to use pathway ranking to integrate properties
of reaction mechanism and individual reactions into measures of
system-level function, along the lines of~\cite{smith2024rules}, to study
likelihoods for complex organosynthesis in geochemistry and the
transition to biochemistry and early pathway evolution.
Following early indications~\cite{Huber:Ace_CoA:97,Cody:pyruvate:00,Cody:ACF:01,Cody:carbonyl_insert:04} that
redox-driven and mineral-catalyzed order in planetary-surface organic
geochemistry could resemble parts of
universal~\cite{Smith:universality:04} core metabolic and
carbon-fixation pathways, a cascade of more recent
results~\cite{Keller:2014,Patel:2015,Keller:2016,Muchowska:2017,Varma:metals_ACA:18,Kitadai:geo_CO:18,Muchowska:2019,Kitadai:met_HT_protomet:19,Preiner:geo_WL:20,Stubbs:2020,Krishnamurthy:2023,Beyazay:NiFe_CO2:23,Song:alk_earth_CO2_fixn:24,Song:CO2_fixn_heterocat:24}
have begun to systematize the complexity and specificity with which
naturally-occurring catalysts and driving conditions can synthesize
organic molecules from inorganic feedstocks.  It has been natural to
instantiate both
hypothesized~\cite{Eschenmoser:HCN:07,Andersen:Eschenmoser_HCN:15} and
demonstrated synthetic
pathways~\cite{Oro:adenine:60,Andersen:HCN_graphs:13,Ricardo:borate:04,Andersen:generic_strat:14}
within rule-based graph-grammars, to discover the combinatorial
diversity that follows from assuming reaction mechanisms, beyond the
single or few pathway completions envisioned by the originating
authors.

The step that integrates these isolated results into system models
begins with our construction here: to rank pathway performance in
terms of both maintaining and restricting flows as all mechanisms
interact through stoichiometry.  The results may assist modeling and
prediction for laboratory systems, and suggest selection criteria as
environmentally-scaffolded pathways gradually came under control of
first captured, and then self-generated, catalysts by an emerging
biosphere.  A question of conceptual as well as practical interest is
whether there are quantitative relations between differences in
restriction cost and the evolutionary innovations that produce them,
analogous to the obvious relation between reducing dissipation of
chemical work and re-routing that work to self-maintain
systems~\cite{Smith:refrig:03}.

    \section*{Acknowledgements}    
    PG and ES were partially funded by the National Science
    Foundation, Division of Environmental Biology (Grant No:
    DEB-2218817). PG is supported by JST CREST (JPMJCR2011) and JSPS
    grant No: 25H01365.  ES was supported in part by National
    Aeronautics and Space Agency Grant no.~80NSSC24K0344 and in part
    by the Earth-Life Science Institute. NL and CF were supported by
    the Novo Nordisk Foundation grant reference NNF21OC0066551.

    \appendix

    \section{Mathematical appendix}

    \subsection{Variational characterization of the partial-mass action flux assignment}
    \label{app:variational_partial_MAK}
    Consider a CRN with a single reversible reaction $r$.
    To block $r$, the forward and reverse fluxes must be the same. Let $j_r = j_r^* = j$. The rate function for such an assignment is
    \begin{align*}
        \mcl{D}(j||J(q)) &= j \ln\left(\frac{j^2}{J(q)J^*(q)}\right) - (2j - J(q) - J^*(q)). 
    \end{align*}
    The $j^*$ that minimizes the cost satisfies
    \begin{align*}
        \pdv{\mcl{D}(j||J(q))}{j} &= \ln\left(\frac{(j^*)^2}{J(q)J^*(q)}\right)\\
        &= 0,
    \end{align*}
    yielding
    \[ j^* = \sqrt{J(q)J^*(q)}.\]
    The rate function at $j^*$ is given by
    \[\mcl{D}(j^*||J(q)) = - (2j^* - J(q) - J^*(q)) = (\sqrt{J(q)} -  \sqrt{J^*(q)})^2. \]

    \subsection{\added{Restriction cost of a reaction is bounded above by its entropy production rate}}
    \label{app:restriction_cost_bound}
    \added{
    Recall from Eqs.\ \ref{eq:EPR} and \ref{eq:block_cost_r}, the entropy production rate for a reaction $\dot{\Sigma}(r,q)$ is
    \[ \dot{\Sigma}(r,q) = \ln\left(\frac{J_r^*(q)}{J_r(q)}\right)\left(J_r^*(q)-J_r(q)\right)\]
    and
    the blocking cost $\dot{\delta}(r,q)$ is
    \[      \dot{\delta}(r,q)= \left(\sqrt{J_r(q)}-\sqrt{J_r^*(q)}\right)^2.
    \]
    Using 
    \[\ln(x)\geq 1 - \frac{1}{x},\]
    we get
    \begin{align*}
        \dot{\Sigma}(r,q) &=2 \ln\left(\frac{\sqrt{J_r^*(q)}}{\sqrt{J_r(q)}}\right)\left(J_r^*(q)-J_r(q)\right)\\
        &\geq
        2 \left(1 - \frac{\sqrt{J_r(q)}}{\sqrt{J_r^*(q)}}\right) 
        \left(J_r^*(q)-J_r(q)\right)\\
        &= 
        2 \left(1 + \frac{\sqrt{J_r(q)}}{\sqrt{J_r^*(q)}}\right) 
        \left(\sqrt{J_r^*(q)}-\sqrt{J_r(q)}\right)^2\\
        & \geq 2  \left(\sqrt{J_r^*(q)}-\sqrt{J_r(q)}\right)^2\\
        & \geq \dot{\delta}(r,q),
    \end{align*}
    as was to be shown.
    }
    
    \subsection{Resistance in a nested pathway never decreases}
    \label{app:resistance_nondecrease}

    Fact $8.20.11$ in \cite{bernstein2009matrix} states that, for any two positive semidefinite matrices $A,B$, if two of the following statements hold, then the remaining statement also holds:
    \begin{enumerate}
        \item $A \leq B$.
        \item $B^+ \leq A^+$.
       \item $\text{rank } A = \text{rank }B$.
    \end{enumerate}

    Consider two nested graphs $\mcl{G}' \subset \mcl{G}$ and their conductance matrices $\bbcal{C}{G}{}$ and $\bbcal{C}{G}{'}$. We know that
    \begin{align}
        \bbcal{C}{G}{} \geq \bbcal{C}{G}{'}. \label{eq:assumption}
    \end{align}
    \deleted{
    Consider a projector matrix $P$ such that 
    $\text{Im}\left(P \bbcal{C}{G}{}\right) = \text{Im}\left(\bbcal{C}{G}{'}\right)$.
    Clearly, 
    $P \bbcal{C}{G}{} \geq \bbcal{C}{G}{'}$.
    Then, using the fact above on $P \bbcal{C}{G}{}$ and $\bbcal{C}{G}{'}$, we get
    $v^T(\bbcal{C}{G}{'}^+ - \bbcal{C}{G}{}^+P^+)v \leq 0 \text{ for all }v$.
    }
    \added{
    For every normalized eigenvector of $\bbcal{C}{G}{'}$ with a positive eigenvalue $e_i$, define 
    \[ P = \sum_i e_i e_i^T. \]
    Then $P$ is a projector matrix such that 
    \[ \text{Im}\left(P \bbcal{C}{G}{}P^T\right) = \text{Im}\left(\bbcal{C}{G}{'}\right).\]
    Also, using Eq.\ \ref{eq:assumption}, clearly 
    \[ P \bbcal{C}{G}{} P^T \geq \bbcal{C}{G}{'}.\]
    Then, using the fact above on $P \bbcal{C}{G}{} P^T$ and $\bbcal{C}{G}{'}$, we get
    \[ v^T(\bbcal{C}{G}{'}^+ - (P^+)^T\bbcal{C}{G}{}^+P^+)v \leq 0 \text{ for all }v. \]}
    Notice that for $v \in \text{Im}(\bbcal{C}{G}{'})$, 
    \[ Pv = v = P^+ v.\]
    Thus, 
    \[ v^T \left(\bbcal{C}{G}{'}^+ - \bbcal{C}{G}{}^+\right)v \geq 0 \quad \text{ for }v \in \text{Im}(\bbcal{C}{G}{'}).\]
    Finally, using 
    \[ \mbb{C}^+ = \mbb{R},\]
    we get the desired relation between the resistance matrices of two nested graphs shown in Eq.\ \ref{eq:R_semidef}.
    
    \subsection{Cost of blocking a pathway in a nested graph never decreases}
    \label{app:cost_nondecrease}
        
    For reaction $r \in \mcl{G}_1^c$ blocked in both $\mcl{G}_1$ and $\mcl{G}_2$, the difference in their costs of blocking is
    \begin{align}
     \sum_{r \in \mcl{G}_1^c}
     \dot{\delta}(r,q_2) - \dot{\delta}(r,q_1) 
    &= \mcl{I}^T \mbb{R}_2 \mbb{C}(\mcl{G}_1^c) \mbb{R}_2 \mcl{I} 
    - 
    \mcl{I}^T \mbb{R}_1 \mbb{C}(\mcl{G}_1^c) \mbb{R}_1 \mcl{I} \nonumber\\
    &= \mcl{I}^T (\mbb{R}_2-\mbb{R}_1) \mbb{C}(\mcl{G}_1^c) (\mbb{R}_2 + \mbb{R}_1) \mcl{I}.  \label{eq:diff_cost_pf}
    \end{align}
    By assumption, $\mcl{G}_1^c$ is a pathway for the throughput current $\mcl{I}$, i.e.\ there exists a vector $\mbf{x}$ such that 
    \[ \mbb{C}(\mcl{G}_1^c) \mbf{x} = \mcl{I}.\]
    Substituting this in the last line of Eq.\ \ref{eq:diff_cost_pf}, we get 
    \begin{align*}
    &\sum_{r \in \mcl{G}_1^c}
     \dot{\delta}(r,q_2) - \dot{\delta}(r,q_1) \\
     =& 
     \mbf{x}^T \mbb{C}(\mcl{G}_1^c) (\mbb{R}_2-\mbb{R}_1) \mbb{C}(\mcl{G}_1^c) (\mbb{R}_2 + \mbb{R}_1) \mbb{C}(\mcl{G}_1^c)\mbf{x}\\
     =& 
      \mbf{x}^T \mbb{C}(\mcl{G}_1^c) (\mbb{R}_2-\mbb{R}_1) \mbb{C}(\mcl{G}_1^c)
      \mbb{C}^+(\mcl{G}_1^c) \mbb{C}(\mcl{G}_1^c)(\mbb{R}_2 + \mbb{R}_1) \mbb{C}(\mcl{G}_1^c)\mbf{x},
    \end{align*}
    where $\mbb{C}\mbb{C}^+\mbb{C} = \mbb{C}$ is used in the final line. Defining matrix $\mbb{F} := \mbf{x}\mbf{x}^T$ and recognizing that $\mbb{F}$ is invertible, the last line can be re-expressed as 
    \begin{align*}
    \sum_{r \in \mcl{G}_1^c}
     \dot{\delta}(r,q_2) - \dot{\delta}(r,q_1) 
     =& 
     \mbf{x}^T \left[\mbb{C}(\mcl{G}_1^c) (\mbb{R}_2-\mbb{R}_1) \mbb{C}(\mcl{G}_1^c)\right]\mbf{x} \\
    &\times \mbf{x}^T \left[\mbb{F}^{-1}
      \mbb{C}^+(\mcl{G}_1^c) \mbb{F}^{-1}\right] \mbf{x} \\
      &\times \mbf{x}^T \left[\mbb{C}(\mcl{G}_1^c)(\mbb{R}_2 + \mbb{R}_1) \mbb{C}(\mcl{G}_1^c)\right]\mbf{x}.
      \end{align*}
     \cite{707780} shows that if the product of three positive semidefinite matrices is symmetric, it is also positive semidefinite. It is easy to verify that the term in each square bracket is symmetric, and thus positive semidefinite. Thus, $\sum_{r \in \mcl{G}_1^c}
     \dot{\delta}(r,q_2) - \dot{\delta}(r,q_1) \geq 0$, as was to be shown.  

    \section{Details of implementation}

        %\NL{
        In general all the optimization problems for the toy chemistry models presented in Sec.~\ref{sec:methods} were solved by implementing them using the \textit{Julia} programming language \cite{bezanson:2017}.
        More precisely, the JuMP package \cite{lubin:2023} was used together with the IPOPT solver \cite{wachter:2006}.
        The code for the implementation can be found on the following public repository:
        In the following we provide a detailed description of the four species model.
        
        \subsection{Four-species model}
        \label{app:four_species}
            % \begin{equation}
            %     \begin{aligned}
            %     -k_{1}\,q_{A}+k_{1}^{*}\,q_{B}-k_{2}\,q_{A}+k_{2}^{*}\,q_{C}&=-1\\
            %     k_{1}\,q_{A}-k_{1}^{*}\,q_{B}-k_{3}\,q_{B}+k_{3}^{*}\,q_{D}&=0\\
            %     k_{2}\,q_{A}-k_{2}^{*}\,q_{C}-k_{4}\,q_{C}+k_{4}^{*}\,q_{D}&=0\\
            %     k_{3}\,q_{B}-k_{3}^{*}\,q_{D}+k_{4}\,q_{C}-k_{4}^{*}\,q_{D}&=1
            %     \end{aligned}
            % \end{equation}
            % \begin{equation}
            %     q_{A}+q_{B}+q_{C}+q_{D}=50
            % \end{equation}
            Using Eq.~\ref{eq:MassAction} together with Eq.~\ref{eq:EPR} we obtain the EPR for the full CRN $\mcl{G_{4}}$ as:
            \begin{equation}
                \begin{aligned}
                \dot{\Sigma}(\mcl{G_{4})} = &\left(k_{1}^{*}\,q_{B}-k_{1}\,q_{A}\right)\ln\left(\frac{k_{1}^{*}\,q_{B}}{k_{1}\,q_{A}}\right) + \left(k_{2}^{*}\,q_{C}-k_{2}\,q_{A}\right)\ln\left(\frac{k_{2}^{*}\,q_{C}}{k_{2}\,q_{A}}\right)\\[1ex]
                &\left(k_{3}^{*}\,q_{D}-k_{3}\,q_{B}\right)\ln\left(\frac{k_{3}^{*}\,q_{D}}{k_{3}\,q_{B}}\right) + \left(k_{4}^{*}\,q_{D}-k_{4}\,q_{C}\right)\ln\left(\frac{k_{4}^{*}\,q_{D}}{k_{4}\,q_{C}}\right)
                \end{aligned}
            \end{equation}
            As there are no reactions to be blocked out in the full CRN we obtain the restriction cost simplifies to: $\dot{\Delta}(\mcl{G}_{4})=0$
            Using the stoichiometric matrix $\mbb{S_{4}}$ and the conservation law $\mbb{L}$ together with the throughput current $\vext$ and the stoichiometric compatibility class $\mbb{L}^{T}\eq$ from the main text, we obtain the optimization problem from Eq.~\ref{eq:cost_pathway} for the thermodynamic cost of the full CRN as:
            \begin{equation}
                \begin{tabular}{rrl}
                &$\chi(\mcl{G}_{4})=\min_{q}\,\dot{\Sigma}(\mcl{G}_{4})$&\\[1ex]
                subject to:& $-k_{1}\,q_{A}+k_{1}^{*}\,q_{B}-k_{2}\,q_{A}+k_{2}^{*}\,q_{C}$&$=-1$\\
                & $k_{1}\,q_{A}-k_{1}^{*}\,q_{B}-k_{3}\,q_{B}+k_{3}^{*}\,q_{D}$&$=0$\\
                & $k_{2}\,q_{A}-k_{2}^{*}\,q_{C}-k_{4}\,q_{C}+k_{4}^{*}\,q_{D}$&$=0$\\
                & $k_{3}\,q_{B}-k_{3}^{*}\,q_{D}+k_{4}\,q_{C}-k_{4}^{*}\,q_{D}$&$=1$\\
                & $q_{A}+q_{B}+q_{C}+q_{D}$&$=50$
                \end{tabular}
            \end{equation}
            % \begin{equation}
            %     \begin{aligned}
            %     -k_{1}\,q_{A}+k_{1}^{*}\,q_{B}&=-1\\
            %     k_{1}\,q_{A}-k_{1}^{*}\,q_{B}-k_{3}\,q_{B}+k_{3}^{*}\,q_{D}&=0\\
            %     k_{3}\,q_{B}-k_{3}^{*}\,q_{D}&=1
            %     \end{aligned}
            % \end{equation}
            Likewise, for the pathway $\mcl{G_{4,B}}$ we obtain the EPR as:
            \begin{equation}
                \dot{\Sigma}(\mcl{G_{4,B})} = \left(k_{1}^{*}\,q_{B}-k_{1}\,q_{A}\right)\ln\left(\frac{k_{1}^{*}\,q_{B}}{k_{1}\,q_{A}}\right) + \left(k_{3}^{*}\,q_{D}-k_{3}\,q_{B}\right)\ln\left(\frac{k_{3}^{*}\,q_{D}}{k_{3}\,q_{B}}\right)
            \end{equation}
            while the restriction cost from Eq.~\ref{eq:discover} is:
            \begin{equation}
                \dot{\Delta}(\mcl{G_{4,B}}) = \left(\sqrt{k_{2}\,q_{A}}-\sqrt{k_{2}^{*}\,q_{C}}\right)^2 + \left(\sqrt{k_{4}\,q_{C}}-\sqrt{k_{4}^{*}\,q_{D}}\right)^2
            \end{equation}
            The optimization problem for the thermodynamic cost of the pathway $\mcl{G}_{4,B}$ is:
            \begin{equation}
                \begin{aligned}
                \chi(\mcl{G}_{4,B})=\min_{q}\left(\dot{\Sigma}(\mcl{G}_{4,B})+\dot{\Delta}(\mcl{G}_{4,B})\right)\\[1ex]
                \begin{tabular}{rrl}
                subject to:& $-k_{1}\,q_{A}+k_{1}^{*}\,q_{B}$&$=-1$\\
                & $k_{1}\,q_{A}-k_{1}^{*}\,q_{B}-k_{3}\,q_{B}+k_{3}^{*}\,q_{D}$&$=0$\\
                & $k_{3}\,q_{B}-k_{3}^{*}\,q_{D}$&$=1$\\
                & $q_{A}+q_{B}+q_{C}+q_{D}$&$=50$
                \end{tabular}
                \end{aligned}
            \end{equation}
            (equations that cancel to zero due to the definition of the partial mass-action flux from Eq.~\ref{eq:partial_mass_action} were left out).
            Finally, for the reaction pathway $\mcl{G_{4,C}}$ the EPR is:
            % \begin{equation}
            %     \begin{aligned}
            %     -k_{2}\,q_{A}+k_{2}^{*}\,q_{C}&=-1\\
            %     k_{2}\,q_{A}-k_{2}^{*}\,q_{C}-k_{4}\,q_{C}+k_{4}^{*}\,q_{D}&=0\\
            %     k_{4}\,q_{C}-k_{4}^{*}\,q_{D}&=1
            %     \end{aligned}
            % \end{equation}
            \begin{equation}
                \dot{\Sigma}(\mcl{G_{4,C})} = \left(k_{2}^{*}\,q_{C}-k_{2}\,q_{A}\right)\ln\left(\frac{k_{2}^{*}\,q_{C}}{k_{2}\,q_{A}}\right) + \left(k_{4}^{*}\,q_{D}-k_{4}\,q_{C}\right)\ln\left(\frac{k_{4}^{*}\,q_{D}}{k_{4}\,q_{C}}\right)
            \end{equation}
            and the restriction cost is:
            \begin{equation}
                \dot{\Delta}(\mcl{G_{4,C}}) = \left(\sqrt{k_{1}\,q_{A}}-\sqrt{k_{1}^{*}\,q_{B}}\right)^2 + \left(\sqrt{k_{3}\,q_{B}}-\sqrt{k_{3}^{*}\,q_{D}}\right)^2
            \end{equation}
            The optimization problem for the thermodynamic cost of the pathway $\mcl{G}_{4,C}$ is:
            \begin{equation}
                \begin{aligned}
                \chi(\mcl{G}_{4,C})=\min_{q}\left(\dot{\Sigma}(\mcl{G}_{4,C})+\dot{\Delta}(\mcl{G}_{4,C})\right)\\[1ex]
                \begin{tabular}{rrl}
                subject to:& $-k_{2}\,q_{A}+k_{2}^{*}\,q_{C}$&$=-1$\\
                & $k_{2}\,q_{A}-k_{2}^{*}\,q_{C}-k_{4}\,q_{C}+k_{4}^{*}\,q_{D}$&$=0$\\
                & $k_{4}\,q_{C}-k_{4}^{*}\,q_{D}$&$=1$\\
                & $q_{A}+q_{B}+q_{C}+q_{D}$&$=50$
                \end{tabular}
                \end{aligned}
            \end{equation}
            These optimization problems are then solved using the above mention software packages for the different settings of $k_{r},k_{r}^{*}$ that follow from Eq.~\ref{eq:kin_const} and the different choices for $E^{\ddagger}$ and $\mu_{0}$ discussed in the main text.
            An analogous approach was used for the five species model as well as the competing autocatalytic cycles model.
            The source code for all models can be found here: \cite{gitrepo}.

    %----------------------------------------------------------------------------------
    % \bibliographystyle{plain} % We choose the "plain" reference style
    \bibliographystyle{unsrt}
    \bibliography{refs/refs} % Entries are in the refs.bib file

\begin{thebibliography}{100}

\bibitem{Khersonsky:promiscuity:10}
Olga Khersonsky and Dan~S. Tawfik.
\newblock Enzyme promiscuity: A mechanistic and evolutionary perspective.
\newblock {\em Annu.~Rev.~Biochem.}, 79:471--505, 2010.

\bibitem{Khersonsky:enzyme_families:11}
Olga Khersonsky, Sergey Malitsky, Ilana Rogachev, and Dan~S. Tawfik.
\newblock Role of chemistry versus substrate binding in recruiting promiscuous enzyme functions.
\newblock {\em Biochem.}, 50:2683--2690, 2011.

\bibitem{Zhu:CBB_control:24}
Xin-Guang Zhu, Haim Treves, and Honglong Zhao.
\newblock Mechanisms controlling metabolite concentrations of the calvin benson cycle.
\newblock {\em Sem.~Cell Dev.~Biol.}, 155:3--9, 2024.

\bibitem{Orgel:cycles:08}
Leslie~E. Orgel.
\newblock The implausibility of metabolic cycles on the early earth.
\newblock {\em PLoS Biology}, 6:e18, 2008.

\bibitem{Gudelj:coli_syntrophy:16}
Ivana Gudelj, Margie Kinnersley, Peter Rashkov, Karen Schmidt, and Frank Rosenzweig.
\newblock Stability of cross-feeding polymorphisms in microbial communities.
\newblock {\em PLoS Comp.~Biol.}, 12:e1005269, 2016.

\bibitem{Kimura:sel_info:61}
Motoo Kimura.
\newblock Natural selection as the process of accumulating genetic information in adaptive evolution.
\newblock {\em Genet. Res., Camb}, 2:127--140, 1961.

\bibitem{Iwasa:free_fitness:88}
Yoh Iwasa.
\newblock Free fitness that always increases in evolution.
\newblock 135:265--281, 1988.

\bibitem{Mustonen:fitness_flux:10}
Ville Mustonen and Michael L{\"{a}}ssig.
\newblock Fitness flux and ubiquity of adaptive evolution.
\newblock {\em Proc.~Nat.~Acad.~Sci.~USA}, 107:4248--4253, 2010.

\bibitem{Fermi:TD:56}
Enrico Fermi.
\newblock Thermodynamics, 1956.

\bibitem{Smith:gaps:25}
Eric Smith and Alicja Kubica.
\newblock Science of the gaps.
\newblock {\em Phil.~Trans.~R.~Soc.~B}, 380:20240282, 2025.
\newblock https://doi.org/10.1098/rstb.2024.0282.

\bibitem{Odum:max_power:55}
Howard~T. Odum and Richard~C. Pinkerton.
\newblock Time's speed regulator: the optimum efficiency for maximum output in physical and biological systems.
\newblock {\em Am.~Sci.}, 43:331--343, 1955.

\bibitem{vanDenBroeck:eff_max_power:05}
C.~Van~den Broeck.
\newblock Thermodynamic efficiency at maximum power.
\newblock {\em Phys.~Rev.~Lett.}, 95:190602, 2005.

\bibitem{Esposito:copol_eff:10}
Massimiliano Esposito, Katja Lindenberg, and Christian Van~den Broeck.
\newblock {Extracting chemical energy by growing disorder: efficiency at maximum power}.
\newblock {\em J.~Stat.~Mech.}, doi:10.1088/1742-5468/2010/01/P01008:1--11, 2010.

\bibitem{Allahverdyan:engines:13}
Armen~E. Allahverdyan, Karen~V. Hovhannisyan, Alexey~V. Melkikh, and Sasun~G. Gevorkian.
\newblock Carnot cycle at finite power: Attainability of maximal efficiency.
\newblock {\em Phys.~Rev.~Lett.}, 111:050601, 2013.

\bibitem{Peliti:ST_intro:21}
Luca Peliti and Simone Pigolotti.
\newblock {\em Stochastic Thermodynamics: An Introduction}.
\newblock Princeton U.~Press, Princeton, NJ, 2021.

\bibitem{Smith:three_level:24}
Eric Smith, Harrison~B. Smith, and Jakob~Lykke Andersen.
\newblock Rules, hypergraphs, and probabilities: the three-level analysis of chemical reaction systems and other stochastic stoichiometric population processes.
\newblock {\em PLoS Compl.~Syst.}, 1:e0000022, 2024.
\newblock https://doi.org/10.1371/journal.pcsy.0000022.

\bibitem{Horn:mass_action:72}
Friedrich Josef~Maria Horn and Roy Jackson.
\newblock General mass action kinetics.
\newblock {\em Arch.~Rat.~Mech.~Anal}, 47:81--116, 1972.

\bibitem{Feinberg:notes:79}
Martin Feinberg.
\newblock Lectures on chemical reaction networks.
\newblock lecture notes, 1979.
\newblock https://crnt.osu.edu/LecturesOnReactionNetworks.

\bibitem{Feinberg:def_01:87}
Martin Feinberg.
\newblock {Chemical reaction network structure and the stability of complex isothermal reactors -- I. The deficiency zero and deficiency one theorems}.
\newblock {\em Chem.~Enc.~Sci.}, 42:2229--2268, 1987.

\bibitem{schilling1998underlying}
Christophe~H Schilling and Bernhard~O Palsson.
\newblock The underlying pathway structure of biochemical reaction networks.
\newblock {\em Proceedings of the National Academy of Sciences}, 95(8):4193--4198, 1998.

\bibitem{Palsson:systems_bio:06}
Bernhard~O. Palsson.
\newblock {\em Systems Biology}.
\newblock Cambridge U.~Press, Cambridge, MA, 2006.

\bibitem{BarEven12}
Arren Bar-Even, Avi Flamholz, Elad Noor, and Ron Milo.
\newblock Thermodynamic constraints shape the structure of carbon fixation pathways.
\newblock {\em Biochimica et Biophysica Acta (BBA) - Bioenergetics}, 1817(9):1646 -- 1659, 2012.

\bibitem{Sterner:eco_stoich:02}
Robert~W. Sterner and James~J. Elser.
\newblock {\em Ecological Stoichiometry: The Biology of Elements From Molecules to the Biosphere}.
\newblock Princeton U. Press, Princeton, NJ, 2002.

\bibitem{peng2020ecological}
Zhen Peng, Alex~M Plum, Praful Gagrani, and David~A Baum.
\newblock An ecological framework for the analysis of prebiotic chemical reaction networks.
\newblock {\em Journal of theoretical biology}, 507:110451, 2020.

\bibitem{avram2024advancing}
Florin Avram, Rim Adenane, and Mircea Neagu.
\newblock Advancing mathematical epidemiology and chemical reaction network theory via synergies between them.
\newblock {\em Entropy}, 26(11):936, 2024.

\bibitem{Smith:evo_games:15}
Eric Smith and Supriya Krishnamurthy.
\newblock {\em Symmetry and Collective Fluctuations in Evolutionary Games}.
\newblock IOP Press, Bristol, 2015.

\bibitem{cuevas2023modular}
Bruno Cuevas-Zuvir{\'\i}a, Evrim Fer, Zachary~R Adam, and Bet{\"u}l Ka{\c{c}}ar.
\newblock The modular biochemical reaction network structure of cellular translation.
\newblock {\em npj Systems Biology and Applications}, 9(1):52, 2023.

\bibitem{Smith:BFI_Lifecycles:23}
Eric Smith.
\newblock Beyond fitness: The nature of selection acting through the constructive steps of lifecycles.
\newblock {\em Evolution}, 77:1967--1986, 2023.
\newblock https://academic.oup.com/evolut/article-abstract/77/9/1967/7158801.

\bibitem{Smith:BFII_Information:24}
Eric Smith.
\newblock Beyond fitness: The information imparted in population states by selection throughout lifecycles.
\newblock {\em Theor.~Popul.~Biol.}, 157:86--117, 2024.
\newblock https://www.sciencedirect.com/science/article/pii/S0040580924000364.

\bibitem{Polettini:open_CNs_I:14}
Matteo Polettini and Massimiliano Esposito.
\newblock {Irreversible thermodynamics of open chemical networks. I. Emergent cycles and broken conservation laws}.
\newblock {\em J.~Chem.~Phys.}, 141:024117, 2014.

\bibitem{rao2016nonequilibrium}
Riccardo Rao and Massimiliano Esposito.
\newblock Nonequilibrium thermodynamics of chemical reaction networks: Wisdom from stochastic thermodynamics.
\newblock {\em Physical Review X}, 6(4):041064, 2016.

\bibitem{Wachtel:transduction:22}
Artur Wachtel, Riccardo Rao, and Massimiliano Esposito.
\newblock Free-energy transduction in chemical reaction networks: from enzymes to metabolism.
\newblock {\em J.~Chem.~Phys.}, 157:024109, 2022.
\newblock https://arxiv.org/pdf/2202.01316.pdf.

\bibitem{baez2017compositional}
John~C Baez and Blake~S Pollard.
\newblock A compositional framework for reaction networks.
\newblock {\em Reviews in Mathematical Physics}, 29(09):1750028, 2017.

\bibitem{anderesen:2016}
Jakob~L. Andersen, Christoph Flamm, Daniel Merkle, and Peter~F. Stadler.
\newblock A software package for chemically inspired graph transformation.
\newblock In Rachid Echahed and Mark Minas, editors, {\em Graph Transformation (ICGT 2016)}, volume 9761 of {\em LNCS}, pages 73--88, Cham, 2016. Springer.

\bibitem{andersen:2019}
Jakob~L. Andersen, Christoph Flamm, Daniel Merkle, and Peter~F. Stadler.
\newblock Chemical transformation motifs—modelling pathways as integer hyperflows.
\newblock {\em IEEE/ACM Transactions on Computational Biology and Bioinformatics}, 16(2):510--523, 2019.

\bibitem{barato2015formal}
Andre~C Barato and Raphael Chetrite.
\newblock A formal view on level 2.5 large deviations and fluctuation relations.
\newblock {\em Journal of Statistical Physics}, 160:1154--1172, 2015.

\bibitem{touchette2009large}
Hugo Touchette.
\newblock The large deviation approach to statistical mechanics.
\newblock {\em Physics Reports}, 478(1-3):1--69, 2009.

\bibitem{peliti2021stochastic}
Luca Peliti and Simone Pigolotti.
\newblock {\em Stochastic thermodynamics: an introduction}.
\newblock Princeton University Press, 2021.

\bibitem{lazarescu2019large}
Alexandre Lazarescu, Tommaso Cossetto, Gianmaria Falasco, and Massimiliano Esposito.
\newblock Large deviations and dynamical phase transitions in stochastic chemical networks.
\newblock {\em The Journal of Chemical Physics}, 151(6), 2019.

\bibitem{Harris:fluct_thms:07}
R.~J. Harris and G.~M. Sch{\"{u}}tz.
\newblock Fluctuation theorems for stochastic dynamics.
\newblock {\em J.~Stat.~Mech.}, page P07020, 2007.
\newblock doi:10.1088/1742-5468/2007/07/P07020.

\bibitem{Esposito:fluct_theorems:10}
Massimiliano Esposito and Christian Van~den Broeck.
\newblock Three detailed fluctuation theorems.
\newblock {\em Phys.~Rev.~Lett.}, 104:090601, 2010.

\bibitem{Seifert:stoch_thermo_rev:12}
Udo Seifert.
\newblock Stochastic thermodynamics, fluctuation theorems, and molecular machines.
\newblock {\em Rep.~Prog.~Phys.}, 75:126001, 2012.
\newblock arXiv:1205.4176v1.

\bibitem{Jarzynski:neq_FE_diffs:97}
C.~Jarzynski.
\newblock Nonequilibrium equality for free energy differences.
\newblock {\em Phys.~Rev.~Lett.}, 78:2690--2693, 1997.

\bibitem{Crooks:NE_work_relns:99}
Gavin~E. Crooks.
\newblock Entropy production fluctuation theorem and the nonequilibrium work relation for free energy differences.
\newblock {\em Phys.~Rev.~E}, 6:2721--2726, 1999.

\bibitem{kolchinsky2021work}
Artemy Kolchinsky and David~H Wolpert.
\newblock Work, entropy production, and thermodynamics of information under protocol constraints.
\newblock {\em Physical Review X}, 11(4):041024, 2021.

\bibitem{plenio2001physics}
Martin~B Plenio and Vincenzo Vitelli.
\newblock The physics of forgetting: Landauer's erasure principle and information theory.
\newblock {\em Contemporary physics}, 42(1):25--60, 2001.

\bibitem{Nicolis:fluct_NEQ:71}
G.~Nicolis and I.~Prigogine.
\newblock Fluctuations in nonequlibrium systems.
\newblock 68:2102--2107, 1971.

\bibitem{Glansdorff:structure:71}
P.~Glansdorff and I.~Prigogine.
\newblock {\em Thermodynamic Theory of Structure, Stability, and Fluctuations}.
\newblock Wiley, New York, 1971.

\bibitem{Prigogine:MT:98}
Dilip Kondepudi and Ilya Prigogine.
\newblock {\em Modern Thermodynamics: From Heat Engines to Dissipative Structures}.
\newblock Wiley, New York, 1998.

\bibitem{Oono:nonlin_world:13}
Yoshitsugu Oono.
\newblock {\em The Nonlinear World: Conceptual Analysis and Phenomenology}.
\newblock Springer, New York, 2013.

\bibitem{Jaynes:MEPP:80}
E.~T. Jaynes.
\newblock The minimum entropy production principle.
\newblock {\em Annu.~Rev.~Phys.~Chem.}, 31:579--601, 1980.
\newblock reprinted in~\cite{Jaynes:Papers:83}.

\bibitem{Presse:max_cal:13}
{Press\'{e}, Steve and Ghosh, Kingshuk and Lee, Julian and Dill, Ken A.}
\newblock The principles of maximum entropy and maximum caliber in statistical physics.
\newblock 85:1115--1141, 2013.

\bibitem{Yang:NEQ_net_flows:25}
Ying-Jen Yang and Ken~A. Dill.
\newblock A principled basis for nonequilibrium network flows.
\newblock 2025.
\newblock https://arxiv.org/abs/2410.17495.

\bibitem{feinberg2019foundations}
Martin Feinberg.
\newblock {\em Foundations of chemical reaction network theory}.
\newblock Springer, 2019.

\bibitem{dai2023hypergraph}
Qionghai Dai and Yue Gao.
\newblock {\em Hypergraph Computation}.
\newblock Springer Nature, 2023.

\bibitem{gagrani2024polyhedral}
Praful Gagrani, Victor Blanco, Eric Smith, and David Baum.
\newblock Polyhedral geometry and combinatorics of an autocatalytic ecosystem.
\newblock {\em Journal of Mathematical Chemistry}, 62(5):1012--1078, 2024.

\bibitem{vassena2024unstable}
Nicola Vassena and Peter~F Stadler.
\newblock Unstable cores are the source of instability in chemical reaction networks.
\newblock {\em Proceedings of the Royal Society A}, 480(2285):20230694, 2024.

\bibitem{rao2018conservation}
Riccardo Rao and Massimiliano Esposito.
\newblock Conservation laws shape dissipation.
\newblock {\em New Journal of Physics}, 20(2):023007, 2018.

\bibitem{schuster1989generalization}
Stefan Schuster and Ronny Schuster.
\newblock A generalization of wegscheider's condition. implications for properties of steady states and for quasi-steady-state approximation.
\newblock {\em Journal of Mathematical Chemistry}, 3(1):25--42, 1989.

\bibitem{Smith:Intr_Extr:20}
Eric Smith.
\newblock Intrinsic and extrinsic thermodynamics for stochastic population processes with multi-level large-deviation structure.
\newblock {\em Entropy}, 22:1137, 2020.

\bibitem{polettini2014irreversible}
Matteo Polettini and Massimiliano Esposito.
\newblock Irreversible thermodynamics of open chemical networks. i. emergent cycles and broken conservation laws.
\newblock {\em The Journal of chemical physics}, 141(2), 2014.

\bibitem{riley2006mathematical}
Kenneth~Franklin Riley, Michael~Paul Hobson, and Stephen~John Bence.
\newblock {\em Mathematical methods for physics and engineering: a comprehensive guide}.
\newblock Cambridge university press, 2006.

\bibitem{chabane2021rarity}
Lydia Chabane.
\newblock {\em From rarity to typicality: the improbable journey of a large deviation}.
\newblock PhD thesis, Universit{\'e} Paris-Saclay, 2021.

\bibitem{kolchinsky2024generalized}
Artemy Kolchinsky, Andreas Dechant, Kohei Yoshimura, and Sosuke Ito.
\newblock Generalized free energy and excess entropy production for active systems.
\newblock {\em arXiv preprint arXiv:2412.08432}, 2024.

\bibitem{Schrodinger:WIL:92}
E.~Schr{\"{o}}dinger.
\newblock {\em What is Life?: The Physical Aspect of the Living Cell}.
\newblock Cambridge U.~Press, New York, 1992.

\bibitem{smith2020intrinsic}
Eric Smith.
\newblock Intrinsic and extrinsic thermodynamics for stochastic population processes with multi-level large-deviation structure.
\newblock {\em Entropy}, 22(10):1137, 2020.

\bibitem{gagrani2023action}
Praful Gagrani and Eric Smith.
\newblock Action functional gradient descent algorithm for estimating escape paths in stochastic chemical reaction networks.
\newblock {\em Physical Review E}, 107(3):034305, 2023.

\bibitem{gagrani2025evolution}
Praful Gagrani and David Baum.
\newblock Evolution of complexity and the transition to biochemical life.
\newblock {\em Physical Review E}, 111(6):064403, 2025.

\bibitem{ay2017information}
Nihat Ay, J{\"u}rgen Jost, H{\^o}ng V{\^a}n~L{\^e}, and Lorenz Schwachh{\"o}fer.
\newblock {\em Information geometry}, volume~64.
\newblock Springer, 2017.

\bibitem{bernstein2009matrix}
Dennis~S Bernstein.
\newblock {\em Matrix mathematics: theory, facts, and formulas}.
\newblock Princeton university press, 2009.

\bibitem{banaji2010graph}
Murad Banaji and Gheorghe Craciun.
\newblock Graph-theoretic criteria for injectivity and unique equilibria in general chemical reaction systems.
\newblock {\em Advances in Applied Mathematics}, 44(2):168--184, 2010.

\bibitem{andersen2021defining}
Jakob~L. Andersen, Christoph Flamm, Daniel Merkle, and Peter~F. Stadler.
\newblock Defining autocatalysis in chemical reaction networks.
\newblock {\em Journal of Systems Chemistry}, 8:121--133, 2021.

\bibitem{blokhuis2020universal}
Alex Blokhuis, David Lacoste, and Philippe Nghe.
\newblock Universal motifs and the diversity of autocatalytic systems.
\newblock {\em Proceedings of the National Academy of Sciences}, 117(41):25230--25236, 2020.

\bibitem{Eyring:ToRP:41}
S.~Glasstone, K.~J. Laidler, and H.~Eyring.
\newblock {\em The Theory of Rate Processes}.
\newblock McGraw Hill, New York, 1941.

\bibitem{Cardy:Instantons:78}
J.~L. Cardy.
\newblock Electron localisation in disordered systems and classical solutions in ginzburg-landau field theory.
\newblock {\em J.~Phys.~C}, 11:L321 -- L328, 1987.

\bibitem{Coleman:AoS:85}
Sidney Coleman.
\newblock {\em Aspects of Symmetry}.
\newblock Cambridge, New York, 1985.

\bibitem{zykov:1974}
Alexander~Aleksandrovich Zykov.
\newblock Hypergraphs.
\newblock {\em Russian Mathematical Surveys}, 29(6):89, 1974.

\bibitem{Dunne:foodwebstats:02}
Jennifer~A. Dunne, Richard~J. Williams, and Neo~D. Martinez.
\newblock Food-web structure and network theory: The role of connectance and size.
\newblock {\em Proc.~Nat.~Acad.~Sci.~USA}, 99:12917--12922, 2002.

\bibitem{smith2024rules}
Eric Smith, Harrison~B Smith, and Jakob~Lykke Andersen.
\newblock Rules, hypergraphs, and probabilities: the three-level analysis of chemical reaction systems and other stochastic stoichiometric population processes.
\newblock {\em PLOS Complex Systems}, 1(4):e0000022, 2024.

\bibitem{Hartley:information:28}
R.~V.~L. Hartley.
\newblock Transmission of information.
\newblock {\em Bell system technical journal}, July:535--563, 1928.

\bibitem{avanzini2023circuit}
Francesco Avanzini, Nahuel Freitas, and Massimiliano Esposito.
\newblock Circuit theory for chemical reaction networks.
\newblock {\em Physical Review X}, 13(2):021041, 2023.

\bibitem{raux2024thermodynamic}
Paul Raux, Christophe Goupil, and Gatien Verley.
\newblock Thermodynamic circuits: Association of devices in stationary nonequilibrium.
\newblock {\em Physical Review E}, 110(1):014134, 2024.

\bibitem{cardelli2020electric}
Luca Cardelli, Mirco Tribastone, and Max Tschaikowski.
\newblock From electric circuits to chemical networks.
\newblock {\em Natural Computing}, 19:237--248, 2020.

\bibitem{loebl2010discrete}
Martin Loebl.
\newblock {\em Discrete mathematics in statistical physics}.
\newblock Springer, 2010.

\bibitem{dal2023geometry}
Sara Dal~Cengio, Vivien Lecomte, and Matteo Polettini.
\newblock Geometry of nonequilibrium reaction networks.
\newblock {\em Physical Review X}, 13(2):021040, 2023.

\bibitem{Huber:Ace_CoA:97}
Claudia Huber and {W\"{a}chtersh\"{a}user, G\"{u}nter}.
\newblock Activated acetic acid by carbon fixation on (fe,ni)s under primordial conditions.
\newblock {\em Science}, 276:245--247, 1997.

\bibitem{Cody:pyruvate:00}
George~D. Cody, Nabil~Z. Boctor, Timothy~R. Filley, Robert~M. Hazen, James~H. Scott, Anurag Sharma, and Hatten S.~Jr. Yoder.
\newblock Primordial carbonylated iron-sulfur compounds and the synthesis of pyruvate.
\newblock {\em Science}, 289:1337--1340, 2000.

\bibitem{Cody:ACF:01}
G.~D. Cody, N.~Z. Boctor, R.~M. Hazen, J.~A. Brandes, H.~J. Morowitz, and Jr. Yoder, H.~S.
\newblock Geochemical roots of autotrophic carbon fixation: hydrothermal experiments in the system citric acid, ${\mbox{h}}_2 \mbox{O}$-$\left( \pm \mbox{FeS} \right)$ $\left( \pm \mbox{NiS} \right)$.
\newblock {\em Geochimica et Cosmochimica Acta.}, 65:3557--3576, 2001.

\bibitem{Cody:carbonyl_insert:04}
G.~D. Cody, N.~Z. Boctor, J.~A. Brandes, T.~E. Filley, R.~M. Hazen, and Jr. Yoder, H.~S.
\newblock Assaying the catalytic potential of transition metal sulfides for abiotic carbon fixation.
\newblock {\em Geochim.~Cosmochim.~Acta}, 68:2185--2196, 2004.

\bibitem{Smith:universality:04}
Eric Smith and Harold~J. Morowitz.
\newblock Universality in intermediary metabolism.
\newblock {\em Proc.~Nat.~Acad.~Sci.~USA}, 101:13168--13173, 2004.
\newblock PMID: 15340153.

\bibitem{Keller:2014}
Markus~A. Keller, Alexandra~V. Turchyn, and Markus Ralser.
\newblock Non-enzymatic glycolysis and pentose phosphate pathway-like reactions in a plausible a rchean ocean.
\newblock {\em Molecular systems biology}, 10(4):725, 2014.

\bibitem{Patel:2015}
Bhavesh~H Patel, Claudia Percivalle, Dougal~J Ritson, Colm~D Duffy, and John~D Sutherland.
\newblock Common origins of rna, protein and lipid precursors in a cyanosulfidic protometabolism.
\newblock {\em Nature chemistry}, 7(4):301--307, 2015.

\bibitem{Keller:2016}
Markus~A. Keller, Andre Zylstra, Cecilia Castro, Alexandra~V. Turchyn, Julian~L. Griffin, and Markus Ralser.
\newblock Conditional iron and ph-dependent activity of a non-enzymatic glycolysis and pentose phosphate pathway.
\newblock {\em Science advances}, 2(1):e1501235, 2016.

\bibitem{Muchowska:2017}
Kamila~B. Muchowska, Sreejith~J. Varma, Elodie Chevallot-Beroux, Lucas Lethuillier-Karl, Guang Li, and Joseph Moran.
\newblock Metals promote sequences of the reverse krebs cycle.
\newblock {\em Nature ecology \& evolution}, 1(11):1716--1721, 2017.

\bibitem{Varma:metals_ACA:18}
Sreejith~J. Varma, Kamila~B. Muchowska, Paul Chatelain, and Joseph Moran.
\newblock {Native iron reduces ${\rm CO}_2$ to intermediates and end-products of the acetyl-CoA pathway}.
\newblock {\em Nat.~Ecol.~Evol.}, 2:1019--1024, 2018.

\bibitem{Kitadai:geo_CO:18}
Norio Kitadai, Ryuhei Nakamura, Masahiro Yamamoto, Ken Takai, Yamei Li, Akira Yamaguchi, Alexis Gilbert, Yuichiro Ueno, Naohiro Yoshida, and Yoshi Oono.
\newblock {Geoelectrochemical CO production: Implications for the autotrophic origin of life}.
\newblock {\em Sci.~Adv.}, 4:eaao7265, 2018.

\bibitem{Muchowska:2019}
Kamila~B Muchowska, Sreejith~J Varma, and Joseph Moran.
\newblock Synthesis and breakdown of universal metabolic precursors promoted by iron.
\newblock {\em Nature}, 569(7754):104--107, 2019.

\bibitem{Kitadai:met_HT_protomet:19}
Norio Kitadai, Ryuhei Nakamura, Masahiro Yamamoto, Ken Takai, Naohiro Yoshida, and Yoshi Oono.
\newblock Metals likely promoted protometabolism in early ocean alkaline hydrothermal systems.
\newblock {\em Sci.~Adv.}, 5:eaav7848, 2019.

\bibitem{Preiner:geo_WL:20}
Martina Preiner, Kensuke Igarashi, Kamila~B. Muchowska, Mingquan Yu, Sreejith~J. Varma, Karl Kleinermanns, Masaru~K. Nobu, Yoichi Kamagata, Harun T{\"{u}}ys{\"{u}}z, Joseph Moran, and William~F. Martin.
\newblock A hydrogen-dependent geochemical analogue of primordial carbon and energy metabolism.
\newblock {\em Nature Ecol.~Evol.}, 4:534--542, 2020.

\bibitem{Stubbs:2020}
R~Trent Stubbs, Mahipal Yadav, Ramanarayanan Krishnamurthy, and Greg Springsteen.
\newblock A plausible metal-free ancestral analogue of the krebs cycle composed entirely of $\alpha$-ketoacids.
\newblock {\em Nature chemistry}, 12(11):1016--1022, 2020.

\bibitem{Krishnamurthy:2023}
Ramanarayanan Krishnamurthy and Charles~L Liotta.
\newblock The potential of glyoxylate as a prebiotic source molecule and a reactant in protometabolic pathways—the glyoxylose reaction.
\newblock {\em Chem}, 9(4):784--797, 2023.

\bibitem{Beyazay:NiFe_CO2:23}
Tu{\u{g}}{\c{c}}e Beyazay, Cristina Ochoa-Hern{\'{a}}ndez, Youngdong Song, Kendra~S. Belthle, William~F. Martin, and Harun T{\"{u}}ys{\"{u}}z.
\newblock Influence of composition of nickel-iron nanoparticles for abiotic ${\rm co}_2$ conversion to early prebiotic organics.
\newblock {\em Angew.~Chem.~Int.~Ed.}, 62:e202218189, 2023.

\bibitem{Song:alk_earth_CO2_fixn:24}
Youngdong Song, Tu{\u{g}}{\c{c}}e Beyazay, and Harun T{\"{u}}ys{\"{u}}z.
\newblock {Effect of Alkali- and Alkaline-Earth-Metal Promoters on Silica-Supported Co--Fe Alloy for Autocatalytic ${\rm CO}_2$ Fixation}.
\newblock {\em Angew.~Chem.~Int.~Ed.}, 63:e202316110, 2024.

\bibitem{Song:CO2_fixn_heterocat:24}
Youngdong Song and Harun T{\"{u}}ys{\"{u}}z.
\newblock {${\rm CO}_2$ Fixation to Prebiotic Intermediates over Heterogeneous Catalysts}.
\newblock {\em Acc.~Chem.~Res.}, 57:2038--2047, 2024.

\bibitem{Eschenmoser:HCN:07}
Albert Eschenmoser.
\newblock On a hypothetical generational relationship between hcn and constituents of the reductive citric acid cycle.
\newblock {\em Chem.~Biodivers.}, 4:554--573, 2007.

\bibitem{Andersen:Eschenmoser_HCN:15}
Jakob~L. Andersen, Christoph Flamm, Daniel Merkle, and Peter~F. Stadler.
\newblock {\textit{In silico} support for Eschenmoser's glyoxylate scenario}.
\newblock {\em Israeli J.~Chem.}, 55:919--933, 2015.

\bibitem{Oro:adenine:60}
J.~Or{\'{o}} and A.~Kimball.
\newblock Synthesis of adenine from ammonium cyanide.
\newblock {\em Biochem.~Biophys.~Res.~Commun.}, 2:407--412, 1960.

\bibitem{Andersen:HCN_graphs:13}
Jakob~L. Andersen, Tommy Andersen, Christoph Flamm, Martin~M. Hanczyc, Daniel Merkle, and Peter~F. Stadler.
\newblock Navigating the chemical space of hcn polymerization and hydrolysis: Guiding graph grammars by mass spectrometry data.
\newblock {\em Entropy}, 15:4066--4083, 2013.

\bibitem{Ricardo:borate:04}
A.~Ricardo, M.~A. Carrigan, A.~N. Olcott, and S.~A. Benner.
\newblock Borate minerals stabilize ribose.
\newblock {\em Science}, 303:196, 2004.

\bibitem{Andersen:generic_strat:14}
Jakob~L. Andersen, Christoph Flamm, Daniel Merkle, and Peter~F. Stadler.
\newblock Generic strategies for chemical space exploration.
\newblock {\em Int.~J.~Comput.~Biol.~Drug Des.}, 7:225--258, 2014.

\bibitem{Smith:refrig:03}
Eric Smith.
\newblock Self-organization from structural refrigeration.
\newblock {\em Phys.~Rev.~E}, 68:046114, 2003.
\newblock PMID: 14683009.

\bibitem{707780}
user104254 (https://math.stackexchange.com/users/104254/user104254).
\newblock Is the product of $3$ positive semidefinite matrices positive semidefinite?
\newblock Mathematics Stack Exchange.
\newblock URL:https://math.stackexchange.com/q/707780 (version: 2023-05-23).

\bibitem{bezanson:2017}
Jeff Bezanson, Alan Edelman, Stefan Karpinski, and Viral~B. Shah.
\newblock Julia: {A} fresh approach to numerical computing.
\newblock {\em SIAM review}, 59(1):65--98, 2017.

\bibitem{lubin:2023}
Miles Lubin, Oscar Dowson, Joaquim~Dias Garcia, Joey Huchette, Beno{\^\i}t Legat, and Juan~Pablo Vielma.
\newblock {JuMP} 1.0: {R}ecent improvements to a modeling language for mathematical optimization.
\newblock {\em Mathematical Programming Computation}, 15(3):581--589, 2023.

\bibitem{wachter:2006}
Andreas W{\"a}chter and Lorenz~T. Biegler.
\newblock On the implementation of an interior-point filter line-search algorithm for large-scale nonlinear programming.
\newblock {\em Mathematical Programming}, 106:25--57, 2006.

\bibitem{gitrepo}
Nino Lauber.
\newblock Ranking pathways.
\newblock Available at: \url{https://github.com/ninolauber/ranking-pathways-paper}, 2025.

\end{thebibliography}
    %----------------------------------------------------------------------------------

\end{document}